\documentclass[%
 reprint,
nofootinbib,
 amsmath,amssymb,
 aps
]{revtex4-1}
\usepackage{color}
\usepackage{gensymb}
\usepackage{graphicx}
\usepackage{dcolumn}
\usepackage{bm}
\usepackage{amsmath}
\usepackage{mathtools} 
\usepackage{crossreftools}
\makeatletter
\newcommand{\optionaldesc}[2]{%
  \phantomsection
  #1\protected@edef\@currentlabel{#1}\label{#2}%
}
\makeatother
\usepackage{hyperref}
\usepackage{aas_macros}
\newcommand{\RNum}[1]{\uppercase\expandafter{\romannumeral #1\relax}}
\newcommand{\Msun}{\mathrm{M}_{\odot}}

\usepackage{gensymb}
\usepackage{soul}
\begin{document}

\newcommand{\sa}{\left<s\right>}
\title{Anisotropic mass segregation: two-component mean-field model}
\author{Hanxi Wang}
\email[Email: ]{hanxi.wang@physics.ox.ac.uk}
\affiliation{Department of Physics, Astrophysics, University of Oxford, Denys Wilkinson Building, Keble Road, Oxford, OX1 3RH, UK}

\author{Bence Kocsis}

\affiliation{Rudolf Peierls Centre for Theoretical Physics, University of Oxford, Clarendon Laboratory, Parks Road, Oxford, OX1 3PU, UK}
\affiliation{St Hugh’s College, University of Oxford, St Margaret’s Rd, Oxford, OX2 6LE, UK}

\date{\today}
\begin{abstract}

Galactic nuclei, the densest stellar environments in the Universe, exhibit a complex geometrical structure. The stars orbiting the central supermassive black hole follow a mass segregated distribution both in the radial distance from the center and in the inclination angle of the orbital planes. The latter distribution may represent the equilibrium state of vector resonant relaxation (VRR). In this paper, we build simple models to understand the equilibrium distribution found previously in numerical simulations. Using the method of maximising the total entropy and the quadrupole mean-field approximation, we determine the equilibrium distribution of axisymmetric two-component gravitating systems with two distinct masses, semimajor axes, and eccentricities. We also examine the limiting case when one of the components dominates over the total energy and angular momentum, approximately acting as a heat bath, which may represent the surrounding astrophysical environment such as the tidal perturbation from the galaxy, a massive perturber, a gas torus, or a nearby stellar system. Remarkably, the bodies above a critical mass in the subdominant component condense into a disk in a ubiquitous way. We identify the system parameters where the transition is smooth and where it is discontinuous. The latter cases exhibit a phase transition between an ordered disk-like state and a disordered nearly spherical distribution both in the canonical and in the microcanonical ensembles for these long-range interacting systems. 
\end{abstract}
\maketitle

\section{Introduction}

Supermassive black holes (SMBH) are commonly observed at the centers of galaxies \cite{supmassivBH}. The strong gravity of the SMBH influences the dynamics of the nuclear star cluster; these are the densest environments of the Universe \cite{Neumayer+2020}. At the center of the Milky Way, the lighter and older stars are observed to be distributed almost spherically while the younger and more massive stars form a more complicated anisotropic distribution including a coeval warped stellar disk, the so-called clockwise disk, and a counterrotating structure \cite{Bartko_2009,yelda2014,2022genzel}. It is difficult to explain the observed distribution with in-situ star formation because of the presence of strong tidal forces. Another possibility is that the anisotropy represents a dynamical equilibrium where the more massive objects segregate in a counter-rotating disk where objects orbit in both directions \cite{Szolgyen_Kocsis2018,Gruzinov+2020,magnan2021orbital,Mathe:2022azz}.

An effective way to study the equilibrium state of self-gravitating systems is statistical mechanics, which has been explored for a long time in this context \cite{1967lyn,1968lyn}. However, because of the long-range nature of gravity, energy is non-extensive which makes many results of statistical mechanics inapplicable \cite{Campa2014}. Another complication is that the uniform distribution on the energy hypersurface required by the ergodic hypothesis may not hold due to the unbounded nature of phase space and energy. All these issues make it challenging to construct the statistical mechanics of gravitating stellar systems. Fortunately however, the complications may be circumvented in dense stellar systems due to the existence of a timescale-hierarchy, which corresponds to the rate of change of the different orbital elements of objects in the system. This leads to an equilibrium distribution in certain bounded subsets of phase space \cite{RAUCH1996149,Roupas2020,kocsis2011,Touma_Tremaine2014,BarOr_Alexander2014,Sridhar_Touma2016,Roupas_2017,Gruzinov+2020,Touma+2019,Tremaine2020,Tremaine2020b,Levin2022}. 
In particular, in the mean field potential dominated by the SMBH and perturbed by a spherical star cluster, the orbital motion and apsidal in-plane precession are much faster than the diffusion of the orientation of the orbital planes (i.e. the argument of node and the $z$-component of angular momentum, or the  direction of the angular momentum normal vector). The diffusion of the orientation of the orbital planes is in turn much faster than the diffusion of the eccentricity and semimajor axis. The dominant mechanism that drives the dynamics of orbital orientations is vector resonant relaxation (VRR) \cite{RAUCH1996149,kocsis2011}, where the torque between stellar orbits averaged over the orbital period and the apsidal precession period accumulate coherently in time for extended periods while the semi-major axis and eccentricity (hence magnitude of angular momentum and energy) of each orbit is fixed. The corresponding VRR time scale is shorter than the age of the stars, and this leads to the diffusion of angular momentum vectors toward the VRR equilibrium \cite{Kocsis_Tremaine2015,Fouvry+2019}, much faster than two-body relaxation which drives a general gravitating system to thermal equilibrium. This hierarchy allows one to average the gravitational interaction over the faster processes and freeze the slowly changing orbital elements to obtain the corresponding VRR equilibrium of angular momentum.

The equilibrium distribution and phase diagram of VRR has been mapped out in the mean-field-theory approximation in the special case of a one-component system with the same semimajor axis and eccentricity for all stars \cite{Roupas_2017,takacs2018,Roupas2020}, which showed that the system exhibits a first order phase transition from a disk-like configuration to a nearly isotropic ordered state in the canonical ensemble.\footnote{Similar results hold for the so-called scalar resonant relaxation equilibria attained on even longer timescales \cite{Touma_Tremaine2014,BarOr_Alexander2014,Sridhar_Touma2016,Gruzinov+2020,Touma+2019,Tremaine2020,Tremaine2020b}.} The microcanonical ensemble for an isolated multi-radius and multi-mass system was obtained in Ref.~\cite{kocsis2011} for circular orbits and the thin disk limit showing that the disk oscillates in independent normal modes each being at the same temperature, but typically not in equipartition if the cluster is rotating\footnote{Equipartition holds in the center-of-mass corotating frame, as expected.} (see also \cite{Hunter_Toomre1969,Nelson_Tremaine1995,Ulubay-Siddiki+2009,Batygin18}). Numerical studies using Monte Carlo Markov Chain, mean field theory, and $N$-body simulations also showed that for multimass models the more massive components generally tend to settle in more flattened configurations while low mass components assume a nearly isotropic state \cite{Szolgyen_Kocsis2018,magnan2021orbital,Mathe:2022azz,taras_2022_discs}. The massive components assume a disk-like state even if the initial configuration has a very low amount of anisotropy of order a per cent \cite{Mathe:2022azz,magnan2021orbital}. This suggests that stellar mass black holes, which are typically more massive than typical main sequence stars, may efficiently settle into disks in dense star clusters, which may boost the collision rate between these objects and contribute to the observed gravitational wave events \cite{Szolgyen_Kocsis2018,Rasskazov_Kocsis2019,Samsing+2022}. 

The goal of this paper is to construct toy mean-field models to give a qualitative understanding of the orbital inclination, mass, and semimajor axis dependence of the VRR equilibrium states in multimass models. We examine the possible equilibrium distributions which may be applied to stellar systems or a small population of massive stars, or intermediate mass black holes (IMBHs) and determine how the surrounding astrophysical environment such as the tidal perturbation from the galaxy, a massive perturber, a gas torus, or a nearby stellar system may affect the equilibrium distribution.

We examine the interesting limiting case when one of the components with some given mass and semimajor axis dominates the energy and angular momentum of the system, which acts as a heat bath for the cluster. The equilibria may be found analytically in this case, which we compare with the exact calculation of a two-component mean field model. We show that these multicomponent systems exhibit a phase transition in both the canonical and the microcanonical ensembles analogous to the nematic-isotropic phase transition in liquid crystals, which is not possible in the case of a one-component system \cite{Roupas_2017}. We determine the critical minimum stellar mass where the distribution collapses to a disk-like state as a function of semimajor axis and the level of anisotropy of the dominant component. 

The physical origin of this analogy between liquid crystals comes and stellar systems is explained by the similarity between the Coulomb and the orbit-averaged Newtonian interactions, and the similarity in the geometry of the interacting objects; i.e. the liquid crystal molecules which are axisymmetric and the stellar orbits which rapidly cover axisymmetric disks due to the rapid eccentric orbital motion and in-plane apsidal precession. This correspondence manifests a similarity between the mean field Hamiltonian of the Maier-Saupe model of liquid crystal \cite{maier_saupe,plischke2006} and  the orbit averaged gravitational Hamiltonian of VRR \citep{Kocsis_Tremaine2015,Roupas_2017}. The interaction energy is minimized in the configuration where the orientation of these axisymmetric objects align. This leads to the alignment or antialignment of the axisymmetric molecules' orientation vectors at low temperatures, called the nematic phase, while there is a first order phase transition at a certain temperature to an isotropic orientation distribution. 
Similarly, in the nematic phase, the VRR interactions in stellar systems forms an ordered disk-like state, where the angular momentum vectors are aligned or anti-aligned.\footnote{For liquid crystals, alignment or antialignment is equally probable in the absense of an external magnetic field, and similarly for gravitating systems with zero net angular momentum. The aligned/antialigned fraction varies as a function of the external magnetic field for liquid crystals or for the net angular momentum for VRR \cite{Roupas_2017}.}

This paper is structured as follows. In Sec.~\ref{sec:theory}, we derive the mean field theory equilbrium of VRR by maximising the entropy for a two-component system by generalising Ref.~\cite{Roupas_2017}. We compare the limiting case of a heat bath to the exact two-component calculation and show analytically how the stellar orbital distribution depends on the mass and orbital radius. In Sec.~\ref{sec:results}, we present the results of the VRR mean field distribution under the heat bath approximation. We examine the conditions for the subdominant component to transit smoothly from an ordered to a disordered phase. We also determine how the distribution varies systematically with mass for different sets of total energy and angular momentum values. In Sec.~\ref{sec:phase_transition}, we explore the existence of phase tranisition in both the canonical and microcanonical ensemble. We also explore what conditions affect the existence of negative temperature equilibria in Sec.~\ref{sec:neg_temperature}. We summarize the results in Sec.~\ref{sec:conclu}.

\section{The two component mean field theory of VRR}\label{sec:theory}
We consider toy models of nuclear star clusters with two types of stellar components denoted by $\mathcal{C}_{1}$ and $\mathcal{C}_{2}$ of $N_1$ and $N_2$ number of stars, respectively, orbiting around the same central SMBH.  All stars in $\mathcal{C}_{1}$ have the same mass, semimajor axis, and angular momentum magnitude labelled $(m_1,a_1,l_1)$, and similarly for $\mathcal{C}_{2}$ with $(m_2,a_2,l_2)$. For simplicity, we assume that the distribution of the angular momentum vectors is axisymmetric and that the two components share the same axis of symmetry for their angular momentum vectors. The Hamiltonian of the system describing VRR in the leading order quadrupole approximation has the form of \cite{Roupas_2017}\footnote{Note that we have dropped the kinetic energy term in the Hamiltonian following Ref.~\cite{Kocsis_Tremaine2015}. Ref.~\cite{Roupas2020} confirmed that the kinetic energy term is indeed negligible if the mass of the SMBH dominates the potential.}:
\begin{align}\label{eq:HVRR}
    H_{\rm VRR} =& -\frac12\sum_{i,j\in \mathcal{C}_{1}} J_{1} P_{2}(\bm{n}_i\cdot \bm{n}_j) - \frac12\sum_{i,j\in \mathcal{C}_{2}} J_{2} P_{2}(\bm{n}_i\cdot \bm{n}_j) \nonumber\\
    &-  \sum_{i\in \mathcal{C}_{1}, j\in \mathcal{C}_{2}} J' P_{2}(\bm{n}_i\cdot \bm{n}_j) \nonumber\\
    =& -\sum_{i\in \mathcal{C}_{1}} \frac34\left[J_1 N_1 Q_1 +J'N_2Q_2 \right]q_i
    \nonumber\\
    &- \sum_{i\in \mathcal{C}_{2}} \frac34\left[J_2 N_2 Q_2 +J'N_1 Q_1 \right]q_i,
\end{align}
where $\bm{n}_i$ is the normalised angular momentum vector of star $i$, $P_{2}(x)=\frac32 x^2 - \frac12$ is the $2^{\rm nd}$ Legendre-polynomial, $q_i\equiv q(\bm{n_i}) = (s_i)^2 - \frac13$, $s_i$ is the $z$-component of $\bm{n}_i$ for star $i$ with respect to the symmetry axis of the cluster, $J_{1,2}$ are the coupling constants among stars within the same component $\mathcal{C}_{1,2}$, respectively, $J'$ is the inter-component coupling constant between $\mathcal{C}_{1}$ and $\mathcal{C}_{2}$, $Q_1=N_{1}^{-1}\sum_{i\in \mathcal{C}_{1}} q(\bm{n}_i) $, and similarly for $Q_2$. In Eq.~\eqref{eq:HVRR}, we have omitted constant terms which do not depend on $\bm{n}_i$. Note that $Q_{1,2}$ are ensemble averages, which we also write in the mean-field approximation as 
\begin{equation}\label{eq:Q_def}
Q_{1,2}= \langle q(\bm{n})\rangle_{\mathcal{C}_{1,2}} = \int_{-1}^{1} \left(s^2 - \frac13\right)f_{1,2}(s) ds,   
\end{equation}
where $f_{1,2}(s)$ is the distribution function of $s$ (i.e. the $z$ Cartesian component of $\bm{n}$)\footnote{i.e. $s=\cos \theta$ where $\theta$ is the inclination angle or the angular momentum vector's polar angle in spherical coordinates.} for stellar component $\mathcal{C}_{1,2}$, respectively, which are to be determined by maximising the Boltzmann entropy. Note that all other parameters (e.g. $m,a,l,N$) are constant during VRR. In the mean-field approximation, the total entropy of the system is a functional of $f_{1,2}(s)$. For circular orbits, the coupling constants are given as 
\begin{equation}
    \label{eq:J}
J_{1,2} = \frac{3Gm_{1,2}^2}{8a_{1,2}}, \quad J' = \frac{3Gm_1m_2\min(a_1,a_2)^2}{8\max(a_1,a_2)^3}\,.
\end{equation}
and we refer to Ref.~\cite{Kocsis_Tremaine2015} for the general eccentric case. 

Maximising the entropy subject to the constraints of fixed total energy and total angular momentum, the distribution function at equilibrium can be obtained by generalizing Ref. \cite{Roupas_2017}. For given $Q_1$ and $Q_2$, we get
\begin{align}
\label{eq:f1(s)}
    &f_{1}(s|Q_1,Q_2) = \frac{ e^{\frac{3}{2}\beta(J_1N_1Q_1+J^\prime N_2 Q_2)s^2+l_1\gamma s}}{\int_{-1}^{1}  e^{\frac{3}{2}\beta(J_1N_1Q_1+J^\prime N_2 Q_2)s^2+l_1\gamma s}d\textrm{s}},\\
    \label{eq:f2(s)}
&f_{2}(s|Q_1,Q_2) = \frac{ e^{\frac{3}{2}\beta(J_2N_2Q_2+J^\prime N_1 Q_1)s^2+l_2\gamma s}}{\int_{-1}^{1}  e^{\frac{3}{2}\beta(JNQ_2+J^\prime N_1 Q_1)s^2+l_2\gamma s}d\textrm{s}}\,.
\end{align}
Here $\beta$ and $\gamma$ are Lagrange multipliers corresponding to the constraints of total energy and total angular momentum, respectively, arising when maximising the Boltzmann entropy; where in terms of the thermodynamic temperature of the system $\beta =1/(kT)$ and $\gamma$ is related to the net rotation rate \cite{Roupas_2017,Roupas2020,Levin2022}. Note that $Q_{1,2}$ are the mean trace-removed quadruple moment of the angular momentum distribution for $\mathcal{C}_{1}$  and $\mathcal{C}_{2}$, respectively, as defined in Eq.~\eqref{eq:Q_def} above. Here $Q_{1,2}$ are also the order-parameters of the equilibrium distribution of the axisymmetric systems, where $Q = 0$ corresponds to an isotropic distribution and its maximum value $Q = 2/3$ represents a razor thin disk in physical space. Depending on $\gamma$, the stars may be orbiting both in the prograde and retrograde senses with respect to the total angular momentum.  Given Eqs.~\eqref{eq:f1(s)} and \eqref{eq:f2(s)}, $Q_{1,2}$ satisfy the self-consistency Eqs.~\eqref{eq:f1(s)} and \eqref{eq:f2(s)}, i.e. 
\begin{align}
  \label{eq:Q1}
 Q_1 &= \int_{-1}^{1} \left(s^2 -\frac{1}{3}\right) f_{1}(s|Q_1,Q_2)d\textrm{s},\\
  \label{eq:Q2}
 Q_2 &= \int_{-1}^{1} \left(s^2 -\frac{1}{3}\right) f_{2}(s|Q_1,Q_2)d\textrm{s}.
\end{align}

Given $\beta$ and $\gamma$, the total angular momentum and total VRR energy of the whole system can be evaluated as:
\begin{align}
    \label{eq:Ltot}
L &= {N_1l_1}\int_{-1}^{1} sf_1(s|Q_1,Q_2) d\textrm{s}
+{N_2l_2}\int_{-1}^{1} sf_2(s|Q_1,Q_2) d\textrm{s},\\
    \label{eq:Etot}
E &= -\frac{3}{4} J_1N_1^2Q_1^2 -\frac{3}{4} J_2 N_2^2Q_2^{2} -\frac{3}{2}J^\prime N_1 N_2 Q_1Q_2.
\end{align}
If the system is isolated, $L$ and $E$ are fixed and the system samples the microcanonical ensemble. Eqs.~\eqref{eq:Q1}--\eqref{eq:Etot} provide a closed system of equations to obtain the unknowns $(\beta,\gamma,Q_1,Q_2)$, and thereby the distribution function Eqs.~\eqref{eq:f1(s)}--\eqref{eq:f2(s)}.

\subsection{The heat bath approximation}\label{sec2}

An important limiting case is when one of the components, e.g. $\mathcal{C}_{1}$, dominates the total VRR energy and total angular momentum of the system. For the nuclear star cluster in the Milky Way, the dominant component may be a massive perturber, e.g. the galactic environment, and in particular the molecular gas torus (also known as the circumnuclear disk) of total mass $10^5-10^6\Msun$ at a distance of 2-7pc from the center \cite{nayakshin2005_torus,2009subr_torus,smith2014_torus,2017torus_Hsieh}, and the subdominant components are the stars around the SMBH in the nuclear star cluster. Similarly if there are IMBHs at a particular range of radii, they may represent the dominant component under which the less massive stellar components relax to find their statistical equilibrium distribution of orbital inclinations \cite{Girma2019_IMBH,arca_sedda_2019_IMBH,Deme2020_IMBH,szolgyen2021_IMBH}, or possibly the large population of nearly spherically distributed main sequence stars comprising the nuclear star cluster may represent the dominant component for the clockwise disk of massive stars in the Galactic centre \citep{Bartko_2009,yelda2014} and/or possible IMBHs.\footnote{As we will see, as long as the interaction among them is negligible this approximation leads to an analytic result even in cases where the subdominant components have not a single value but multiple mass, eccentricity, and semimajor axes.} 

From Eq.~\eqref{eq:Ltot} and Eq.~\eqref{eq:Etot}, this limiting case requires two conditions to hold:
\begin{align}\label{eq:ang_mtm_cond}
 N_1l_1 \langle s\rangle_{\mathcal{C}_1} &\gg N_2l_2 \langle s\rangle_{\mathcal{C}_2}\,,\\
 J_1N_1^2 Q_1^2 &\gg J_2 N_2^2 Q_2^2+2J' N_1N_2Q_1Q_2 \label{eq:energy_cond}
 \,.   
\end{align}
For circular orbits around an SMBH, this is equivelent to
\begin{align}\label{eq:conditions1}
 1 &\gg \bar{M} \bar{a}^{1/2} \bar{\langle s\rangle},\,\\
 1 &\gg \frac{\bar{M}^2}{\bar{a}} \bar{Q}^2 + 2\bar{M}\min(\bar{a}^{-3},\bar{a}^{2})\bar{Q},\, \label{eq:conditions3}
\end{align}
where we define the dimensionless quantities $\bar{X} = X_2/X_1$ for any quantity $X$ for the two components and $M=Nm$ is the total mass of each component. For instance, $\bar{M}$ is the total mass of the subdominant component relative to the dominant component.

As a result, the dominant component's distribution $f_1$ is approximately independent of $Q_2$ in Eq.~\eqref{eq:f1(s)}:
\begin{align}
\label{eq:fdom(s)}
    f_{1}(s|Q_1) &\approx \frac{ e^{\frac{3}{2}\beta J_1N_1Q_1 s^2+l_1\gamma s}}{\int_{-1}^{1}  e^{\frac{3}{2}\beta J_1N_1Q_1 s^2 +l_1\gamma s}d\textrm{s}},\\
   \label{eq:Q_dominant}
 Q_1 &\approx \int_{-1}^{1} \left(s^2 -\frac{1}{3}\right) f_{1}(s|Q_1)d\textrm{s}.    
\end{align}

Further, $L$ and $E$ are approximately determined solely by $f_1$ and $Q_1$ independently of $Q_2$ or $f_2$. Component  $\mathcal{C}_{1}$ thus also determines the corresponding values of $\beta$ and $\gamma$ independently of $Q_2$ or $f_2$. Component  $\mathcal{C}_{2}$  then settles in the background potential for a given fixed $(Q_1,\beta,\gamma)$ and obtains its equilibrium $Q_2$ through Eq.~\eqref{eq:Q2}. Thus component $\mathcal{C}_{1}$ may be regarded as a heat bath for component $\mathcal{C}_{2}$, which operates like a canonical ensemble. More generally, similar conclusions may hold for a multicomponent system with an arbitrary number of subdominant components in the background potential of the dominant component. 

In the following we label the dominant component with a `d' index for ``dominant'' and drop the label of the subdominant component. We assume that the angular momentum vector distribution is known and may be parameterized as
\begin{equation}
    \label{eq:f_d}
f_d(s) \propto e^{\kappa_d s^2 + c_d s },
\end{equation}
where $s=\cos \theta$ and $c_d$ and $\kappa_d$ are constant fitting coefficients. 
For $\kappa_d\ll 1$ and $c_d\ll 1$ this clearly represents a nearly isotropic distribution both in angular momentum space and in physical space, and for $\kappa_d\gg 1$ or $c_d\gg 1$ this represents a narrow cone in angular momentum space and a thin disk in physical space. 
The case of $c_d\gg 1$ may represent the massive gas torus observed around the Milky Way, irrespective of whether it is in a state of VRR equilibrium, as long as it is stationary. 
The values of $(\kappa_d,c_d)$ are directly measurable by fitting the observed distribution Eq.~\eqref{eq:f_d}. They are related to the system parameters $(Q_d, \beta, \gamma)$ via Eq.~\eqref{eq:Q_dominant} as 
\begin{equation}\label{eq:kappa_d}
\kappa_d = \frac32 \beta J_dN_d Q_d\,,{~\rm and~} c_d = \gamma l_d.    
\end{equation}
The value of $Q$ of the subdominant species, henceforth ``stars'', in this background is
\begin{equation}
    \label{eq:bath_Q}
 Q = \frac{\int_{-1}^{1} (s^2 -\frac{1}{3}) e^{\frac{3}{2}\beta(JNQ+J^\prime N_d Q_d)s^2+l\gamma s}d\textrm{s}}{\int_{-1}^{1}  e^{\frac{3}{2}\beta(JNQ+J^\prime N_d Q_d)s^2+l\gamma s}d\textrm{s}}, 
\end{equation}
where the quantities without the subscript (i.e. $J$, $N$, $Q$) describe the stars, and quantities with the `d' subscript are of the dominant component (i.e. the heat bath). 

A parametric solution to 
Eq.~\eqref{eq:bath_Q} can be found using the one-particle partition function \cite{Roupas_2017}\footnote{Note that we define $Z_0=(2\pi)^{-1}Z_0^{\rm RKT}$, where $Z_0^{\rm RKT}$ denotes the formula quoted in Ref.~\cite{Roupas_2017}}, 
\begin{align}
    \label{eq:bath_Z}
    Z_0(\kappa,c) =& \int_{-1}^{1} e^{\kappa(s^2-\frac{1}{3})+cs} ds
 = \frac{\pi^{1/2}}{2\sqrt{-\kappa}} \exp \left(-\frac{\kappa}{3}-\frac{c^2}{4\kappa}\right) \notag \\
&\times \left[\textrm{erf}\left(\frac{c-2\kappa}{2\sqrt{-\kappa}}\right)+\textrm{erf}\left(
-\frac{c+2\kappa}{2\sqrt{-\kappa}}\right)\right],
\end{align}
where
\begin{align}
\label{eq:kappa}
\kappa &= \frac{3}{2}\beta(JNQ+J^\prime N_d Q_d) = \frac{J'}{J_d} \kappa_{\rm d} [1+(J/J')\bar{N}\bar{Q}], \\
c &= l\gamma = \bar{l} c_d , \label{eq:c}
\end{align}
where $\bar{N} = N/N_d$, $\bar{Q} = Q/Q_d$. For circular orbits this becomes
\begin{align}\label{eq:kappacirc}
 \kappa &=   \bar{m} \min(\bar{a}^{2},\bar{a}^{-3})\left[1+ \bar{M} \max(\bar{a}^2,\bar{a}^{-3})
 \bar{Q}\right]  \kappa_d \nonumber\\
 &= \bar{m}  \left[\min(\bar{a}^{2},\bar{a}^{-3})+ \frac{\bar{M} \bar{Q}}{\bar{a}}
 \right]  \kappa_d, \\
 c &= \bar{m} \bar{a}^{1/2} c_d\label{eq:ccirc}.
\end{align}
Here $\bar{Q}$ depends implicitly on $\kappa$ and $c$. Given $Z_0(\kappa,c)$, $Q$ is given by\footnote{Similarly, the angular momentum simplifies analytically, which we henceforth denote with the dimensionless function $\bar{L}(\kappa,c)$ defined as
\begin{equation}\label{eq:bath_L_analy}
    \frac{L}{Nl} = \bar{L}(\kappa,c) \equiv \frac{\partial\ln Z_0(\kappa,c)}{\partial \kappa} = \frac{-c}{2\kappa} + \frac{ e^{\frac23 \kappa}\sinh c}{\kappa Z_0(\kappa,c)}\,.
\end{equation}} 
\begin{align}
    \label{eq:bath_Q_analy}
Q  &= \frac{\partial\ln Z_0(\kappa,c)}{\partial \kappa} \notag \\
&= \frac{c^2}{4\kappa^2} - \frac{1}{2\kappa} - \frac{1}{3}+\frac{e ^{\frac{2}{3}\kappa}}{\kappa \,Z_0(\kappa,c)}\left( \cosh c - \frac{c}{2\kappa} \sinh c\right).
\end{align}
A similar expression holds for $Q_d\equiv Q(\kappa_d,c_d)$ by replacing $(\kappa,c)\rightarrow(\kappa_d,c_d)$ in Eq.~\eqref{eq:bath_Q_analy}. 
Note that since $l_{i}$ is always positive for all components $\mathcal{C}_i$, hence $c_{i}=\gamma l_i>0$ must hold for all $i$ in order for the angular momentum to be in the positive direction (assumed by construction). Indeed in the axisymmetric case, two-component VRR, the parallel aligned configuration has higher entropy and lower free energy than the anti-aligned configuration. The numerical simulations of Ref.~\cite{Levin2022} and \cite{taras_2022_discs} confirm that an initially anti-aligned disc of young stars will align with the rotating spherical host star cluster in the Galactic Centre.

In these expressions $\kappa$ and $c$ are proportional to $\beta$ and $\gamma$ which represent dimensionless effective inverse temperatures conjugate to the VRR energy and total angular momentum, respectively (cf. Eqs.~\ref{eq:f1(s)} and \ref{eq:f2(s)}). 
We obtain a solution for $Q$ for the subdominant component in two ways:
\begin{enumerate}
\item[{\crtcrossreflabel{(HBp)}[{i:HBp}]}] without any further approximations beyond $(J'/J_d)\bar{N}\bar{Q} \ll 1$, i.e. obtain $Q_d$ for given $(\kappa_d,c_d)$ as mentioned below Eq.~\eqref{eq:bath_Q_analy}, then calculate $(\kappa,c,Q)$ by self-consistently solving Eqs.~\eqref{eq:kappacirc}--\eqref{eq:bath_Q_analy}.
    \item[{\crtcrossreflabel{(HB)}[i:HB]}] using the approximation that the self interaction between the stars is negligible compared to the coupling between the stars and the dominant component, $(J/J')\bar{N}\bar{Q} \ll 1$, such that Eq.~\eqref{eq:kappa} simplifies to 
\begin{equation}
    \label{eq:bath_approx_kappa}
    \kappa \approx \frac{J'}{J_d} \kappa_d = \bar{m} \min(\bar{a}^{2},\bar{a}^{-3})\kappa_d. 
\end{equation}
$Q$ is obtained by substituting $\kappa$ and $c$ in Eq.~\eqref{eq:bath_Q_analy}.
\end{enumerate}

Note that in the former case \ref{i:HBp} the self-interaction of the subdominant component is accounted for exactly on top of the effects caused by the dominant component, and in the latter \ref{i:HB} it is neglected. The equilibrium distribution function follows from Eq.~\eqref{eq:f2(s)}
\begin{equation}\label{eq:f(kappa,c)}
    \frac{f(s)}{f(1)} = e^{\kappa s^2 + c s}\,.
\end{equation}

  The meaning of the parameters $(\kappa,c)$ are as follows (see Appendix E in Ref.~\cite{Roupas_2017}). Generally $-\infty\leq \kappa\leq \infty$ and $0\leq c \leq \infty$ may be assumed without loss of generality for equilibria.   
  These parameters specify the order parameter $Q$ (Eq.~\eqref{eq:bath_Q_analy}) and $L$ (Eq.~\ref{eq:bath_L_analy}) such that $\kappa=c=0$ is the isotropic disordered state with $Q=0$.
  The $c$ parameter specifies the corotating vs. counterrotating angular momentum density at $s=1$ and $-1$ as $f(1)/f(-1)=e^{2c}$. Nonrotating clusters have zero net angular momentum for which $c=0$. The angular momentum increases monotonically between $\bar{L}=L/(Nl)=0$ and $1$ for fixed $\kappa$ as $c$ changes from 0 to $\infty$. The order parameter $Q$ increases monotonically with $\kappa$ between $-\frac13+(L/Nl)^2$ and $\frac23$, where a razor-thin disk (maximally ordered state) has $\kappa\rightarrow\infty$ and/or $c\rightarrow\infty$, and $\kappa=0$ state has the maximum disorder with $Q\geq 0$ among states with a given $L$. For small $\kappa$ and $c$,
 \begin{align}\label{eq:Qisotropic}
     Q&\approx\frac{4}{45}\kappa + \frac{8}{945}\kappa^2 + \frac{2}{45}c^2 
     &\;(|\kappa|\ll 1, c\ll 1),\\
    \label{eq:L_small}
\frac{L}{Nl} &\approx \frac13c+\frac{4}{45}\kappa c &\;(|\kappa|\ll 1, c\ll 1).
\end{align}
Further, $Q=0$ is obtained at finite negative $\kappa$ for $c\neq 0$. The stable equilibrium distribution function has local maxima at $s=1$ and $-1$ ($\theta=0$ and $\pi$) for $\kappa\geq 0$. Equilibria also exist with $\kappa<0$, where $f(s)$ possibly peaks at $0\leq s\neq 1$. These states with small to intermediate negative $\kappa$ and $c\neq 0$ may correspond to negative absolute temperatures for which $Q>0$, discussed further in Section~\ref{sec:neg_temperature}. States with large negative $\kappa$ have $Q<0$, which are unstable in the one-component case, but in some cases stable in the two component case (see Sec.~\ref{sec:neg_temperature} and Appendix \ref{app:max_E}).

Ref.~\cite{Levin2022} finds that $c\sim 0.3$ for the observed spherical distribution of old stars in the nuclear star cluster in the Galactic center.

\subsection{Region of validity of the heat bath approximation}
\begin{figure}
    \centering
    \includegraphics[scale=0.5]{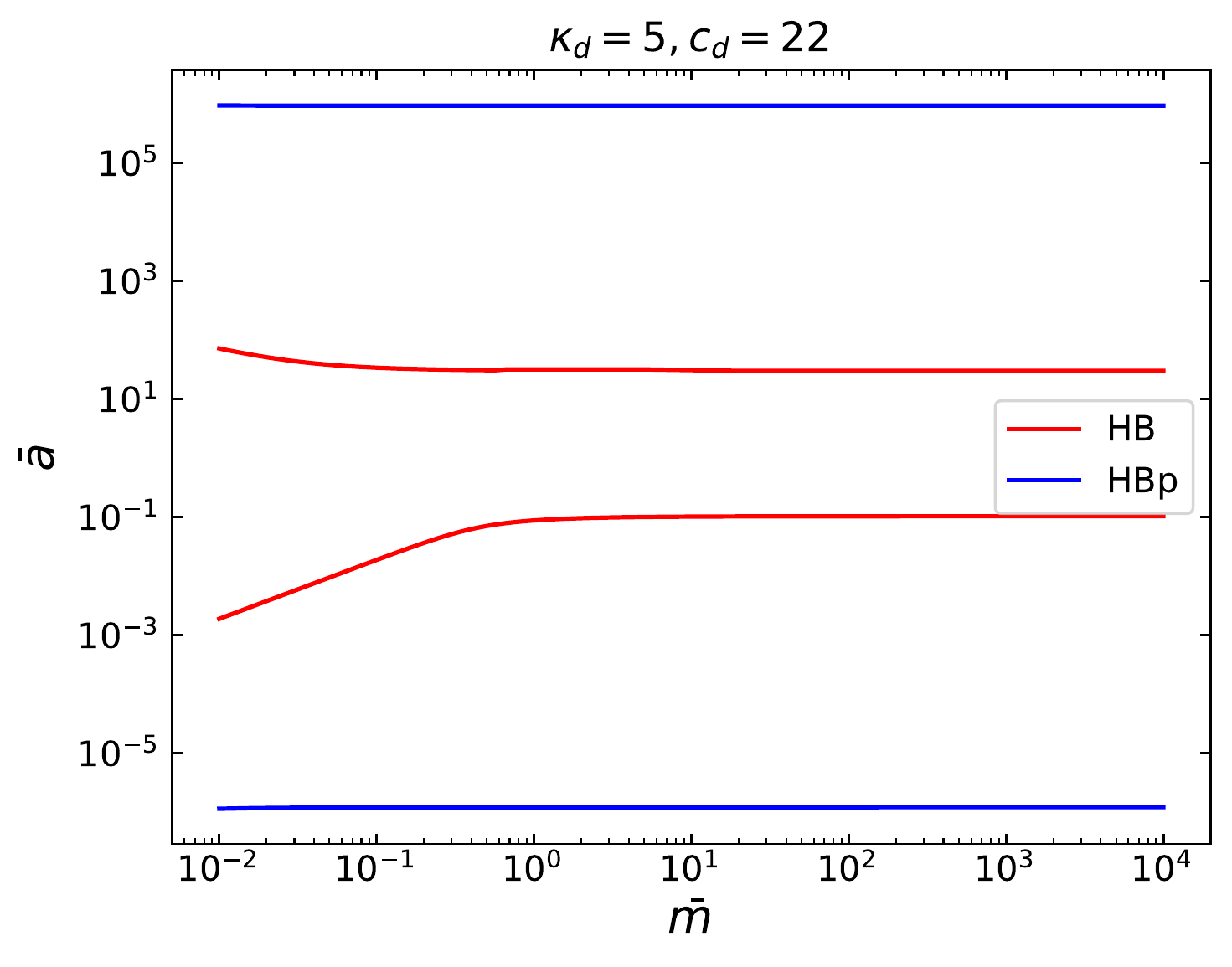}
    \includegraphics[scale=0.5]{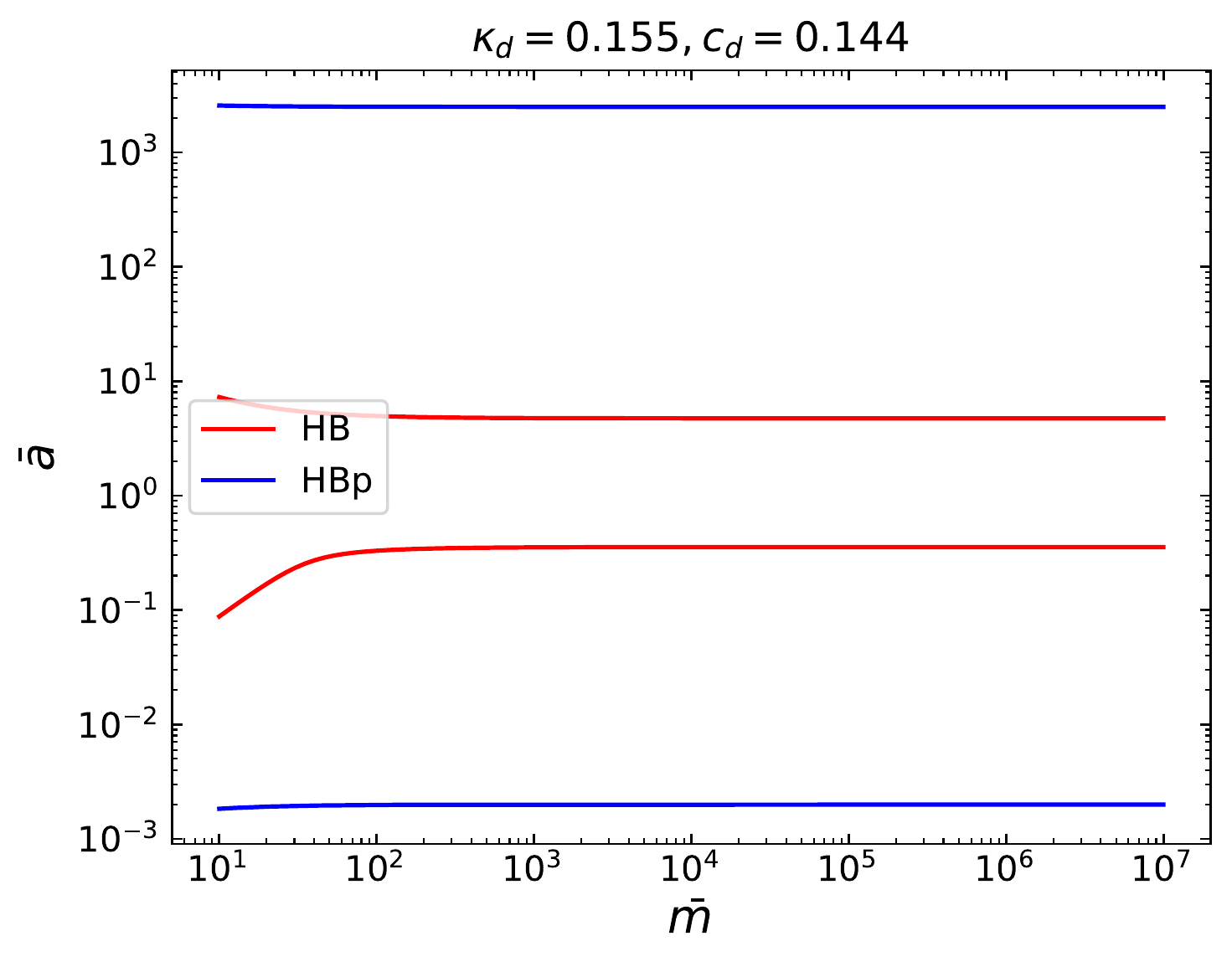}
    \caption{The region of validity of the heat bath approximation as a function of stellar mass ($\bar{m}=m/m_d$) and semimajor axis  and ($\bar{a}=a/a_d$) relative to the dominant component. In the region between the blue curves,
    the self-energy of the dominant component is larger than the interaction energy (i.e \ref{i:HBp} conditions stated in Eqs.~\eqref{eq:conditions1} and \eqref{eq:conditions3}). In the region between the red curves, the interaction energy dominates over the self-energy of the subdominant component (i.e. \ref{i:HB} approximation). The upper and lower panels correspond to a disk-like and a spherical dominant component, respectively, Eq.~\eqref{eq:f_d} with $(\kappa_d,c_d)$ as labelled with relative total mass $\bar{M} = 10^{-3}$ for both panels. }
    \label{fig:parameter_spcae}
\end{figure}

Figure~\ref{fig:parameter_spcae} shows the region of the parameter space of semimajor axis and mass $(\bar{a},\bar{m})$ where the heat bath approximation \ref{i:HBp} is valid for two representative cases $(\bar{M},\kappa_d, c_d)=(10^{-3},5,22)$ and $(10^{-3},0.155,0.144)$.  Eqs.~\eqref{eq:conditions1} and \eqref{eq:conditions3} marginally hold using the exact two-component calculation.

The red boundary curve in Figure~\ref{fig:parameter_spcae} shows the region of validity for the \ref{i:HB} approximation which neglects the self interaction of the subdominant component. This is derived by setting $(J/J')\bar{N}\bar{Q} = 1$ in the \ref{i:HBp} calculation.
The region of validity is strongly limited in semimajor axis for \ref{i:HB}, but it is not limited in the allowed individual stellar mass. The energy and the angular momentum of the subdominant component becomes nonnegligible compared to the interaction energy respectively beyond the lower and upper red boundaries in $\bar{a}$. The region within the blue boundaries is where the \ref{i:HBp} approximation is valid, where the interaction energy and angular momentum of the subdominant component is smaller than the self-energy and angular momentum of the dominant component.

\subsection{Exact calculation} \label{exact_calculation}

We compare the axisymmeric equilibrium distribution function of the subdominant component with the \ref{i:HB} and \ref{i:HBp} approximations to the exact calculation for the same fixed values of $(\kappa_d,c_d)$ of the dominant component. The value of $Q_d$ is found from $Q_d(\kappa_d,c_d)$ by replacing $(\kappa,c)\rightarrow(\kappa_d,c_d)$ in Eq.~\eqref{eq:bath_Q_analy}. When the interaction energy with the subdominant component is not neglected then Eq.~\eqref{eq:kappa_d} is replaced by
\begin{equation}
    \label{eq:kappa_d_exact}
    \kappa_d =  \frac32 \beta (J_dN_d Q_d + J' N Q) = \frac32 \beta J_dN_d Q_d \left(1 + \frac{J'}{J_d} \bar{N} \bar{Q}\right) .
\end{equation}
Here Eqs.~\eqref{eq:kappa} and \eqref{eq:kappacirc} are modified as

\begin{align}
\kappa &= \frac{3}{2}\beta(JNQ+J^\prime N_d Q_d) 
= \kappa_d \frac{J^\prime}{J_d} \left(\frac{1+(J/J')\bar{N}\bar{Q}}{1+(J^\prime/J_d)\bar{N}\bar{Q}} \right)\nonumber\\
&=\bar{m} \min(\bar{a}^{2},\bar{a}^{-3})\frac{[1+ \bar{M} \max(\bar{a}^2,\bar{a}^{-3}) \bar{Q}]}
{[1+ \bar{M} \min(\bar{a}^2,\bar{a}^{-3}) \bar{Q}]}\kappa_d 
\nonumber\\
&=
\bar{m} \frac{[\min(\bar{a}^{2},\bar{a}^{-3})+ \bar{M} \bar{a}^{-1} \bar{Q}]}
{[1+ \bar{M} \min(\bar{a}^2,\bar{a}^{-3}) \bar{Q}]}\kappa_d 
.\label{eq:kappaexact}
\end{align}

For the exact two-component calculation, we solve Eq.~\eqref{eq:kappaexact} simultaneously with  Eqs.~\eqref{eq:ccirc} and \eqref{eq:bath_Q_analy} as discussed in Appendix~\ref{app:analytic}.

\section{Equilibrium disks of massive objects}\label{sec:results}

 Here we present the results of the VRR mean field model and examine the conditions for the component comprised of heavier objects to form a disk as a function of mass and semimajor axis.

We present the results of the \ref{i:HB} and \ref{i:HBp} approximations and the exact calculation for fixed ratio of total mass $\bar{M} = 10^{-3}$ between the two components, and for two cases where the dominant component is disk-like with $(\kappa_d,c_d)=(5,22)$ and when it is nearly spherical $(\kappa_d, c_d)=(0.155, 0.144)$ with order parameter $Q_d=0.60$ and 0.014, respectively, and mean angular momentum $L_d/N_dl_d =0.97$ and $0.054$, respectively (Eqs.~\ref{eq:Qisotropic}--\ref{eq:L_small}). Note for reference that if $N_{\rm d}$ bodies are drawn from an isotropic distribution their order parameter and mean angular momentum satisfy $\langle Q_d\rangle= 0$, $\langle Q_d^2\rangle^{1/2} = 1/[(\sqrt{5})N_d]$, and $L_d/N_dl_d = 1/\sqrt{N_d}$
, respectively, implying that $N_d= 32$ and $340$ are required to yield $Q_d=0.014$ and $L_d/N_dl_d = 0.014$, respectively. Thus, more abundant isotropic shells of stars with higher $N$ may be in principle even closer to being isotropic than in our spherical example. 
\footnote{If the stars are drawn independently from an isotropic distribution, $\langle s
\rangle^2 = \left< \Vec{n}\right>\cdot\left<\Vec{n}\right> = N^{-2}\sum_i\sum_j\left<\bm{n}_i\cdot \bm{n}_j\right>= N^{-2}\sum_i\sum_j\delta_{ij}=N^{-1} $. Hence $\left<s\right> = 1/\sqrt{N}$. }

\subsection{Anisotropic mass segregation}\label{sec:mass_ratio}

\begin{figure}
    \centering
    \includegraphics[scale = 0.5]{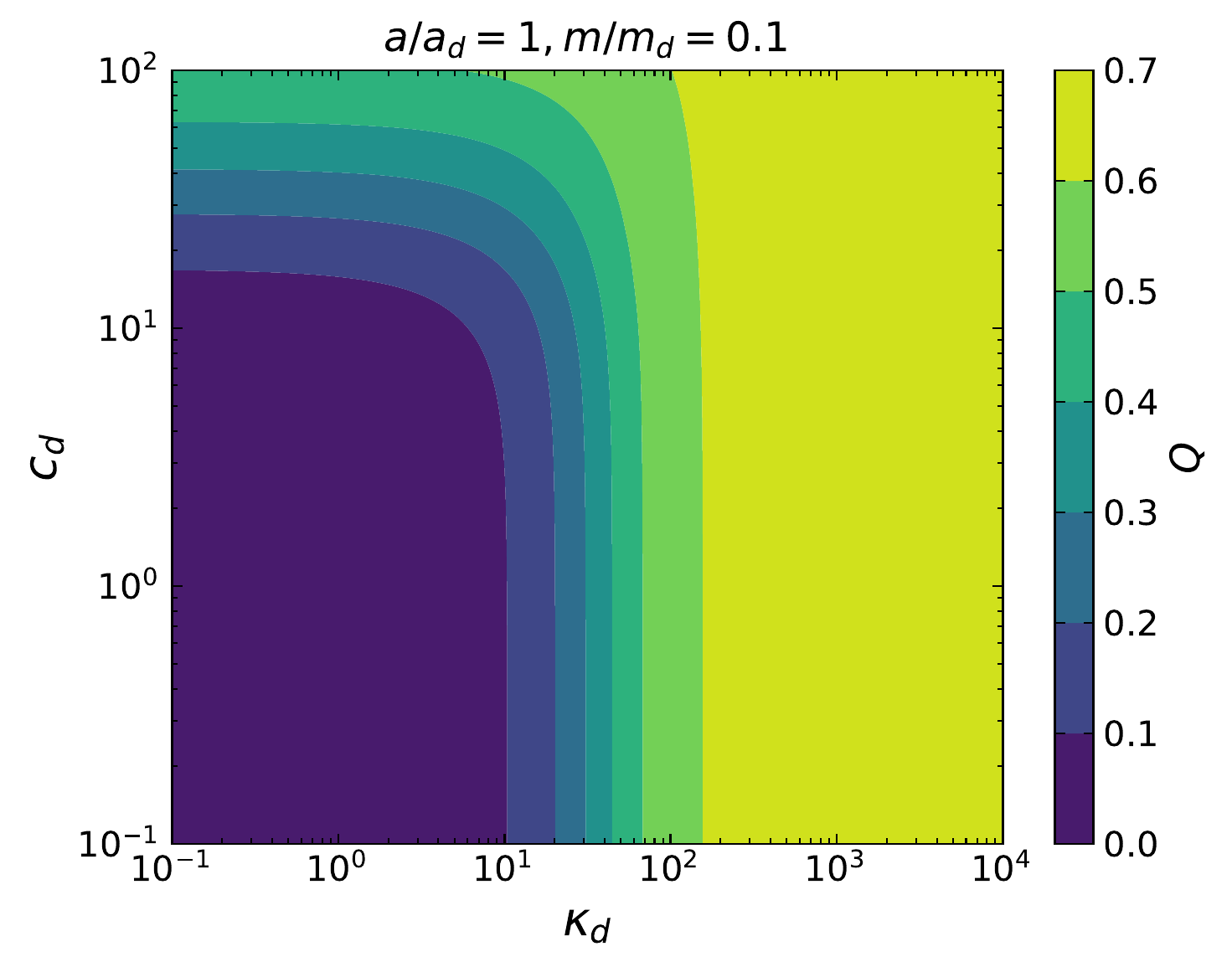}
      \includegraphics[scale = 0.5]{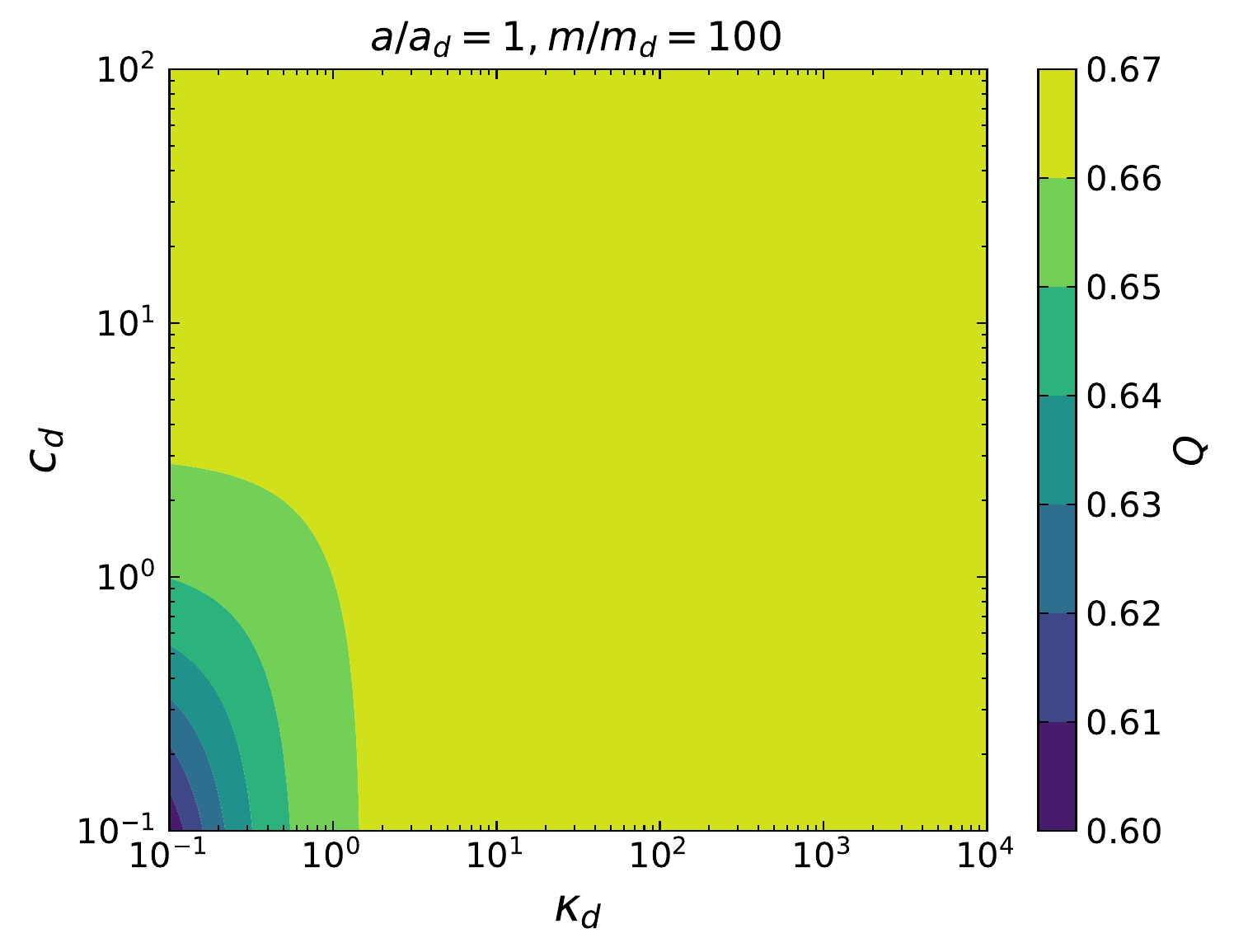}
    \caption{The order parameter $Q$ of the subdominant component for different distribution functions of the dominant component parameterized by $\kappa_d$ and $c_d$ in Eq.~\eqref{eq:f_d}. The ratio of total mass is $\bar{M} = 10^{-3}$ for both panels, while top and bottom panels have different individual stellar mass $m$ as labelled. The distribution approaches the isotropic distribution ($Q\sim 0$) for small $c_d$ and small $\kappa_d$ for small mass $m/m_d=0.1$ and it is a thin disk ($Q\sim\frac23$) for $m/m_d=100$.}
    \label{fig:Q_k_c_contour}
\end{figure}

Figure~\ref{fig:Q_k_c_contour} shows how the angular momentum vector distribution of the subdominant component depends on the dominant component's parameters $(\kappa_d,c_d)$ for a system where all objects have the same semimajor axis and $\bar{M}=10^{-3}$. The figure shows the value of the order parameter $Q$ of the subdominant component determined via \ref{i:HBp} for two different values of $\bar{m}=0.1$ and 100, respectively as labelled. Note that larger $\kappa_d$ and $c_d$ values correspond to a dominant component that is more flattened while smaller $\kappa_d$ and $c_d$ values correspond to a dominant component that is more spherical (see Eq.~\ref{eq:FWHM} below). 
Recalling that $Q$ changes between 0 and $\frac23$ between the isotropic (disordered) and razor thin (ordered) cases. Figure~\ref{fig:Q_k_c_contour} shows that if the subdominant component has a much larger $m$, the distribution is very much flattened for a wide range of parameters including cases where the dominant component is close to spherical $(\kappa_d,c_d)=(0.1,0.1)$. If the subdominant component has a smaller individual stellar mass it is typically nearly spherical unless the dominant component is very much flattened with $c_d\gg 1$ or $\kappa_d\ggg 1$. We determine an analytical criterion for the subdominant component to form a disk next.

\subsubsection{The case of a dominant disk}
When the dominant component is disk-like, for small angles near the axis of symmetry (here $\theta$ is the polar angle in spherical coordinates of the angular momentum vector direction), Eq.~\eqref{eq:f_d} simplifies to
\begin{align}
    \label{eq:f_d_smalltheta}
f_d(\cos \theta) \propto e^{\kappa_d +c_d - (\kappa_d+\frac{1}{2}c_d)\theta^2} \propto e^{- (\kappa_d+\frac{1}{2}c_d)\theta^2} \quad (\theta\approx 0).
\end{align}
Large $\kappa_d$ implies an order parameter of \cite{Roupas_2017}
\begin{equation}\label{eq:Qdisk}
Q_d=\frac{2}{3} - \frac{1}{\kappa_d} + \frac{c_d \tanh c_d - 1 }{2\kappa_d^2} + \mathcal{O}(\kappa_d^{-3}),
\end{equation}
which represents a thin disk with a full-width half maximum (FWHM) thickness of 
\begin{align}\label{eq:FWHM}
    \Delta \theta_d \approx \frac{(\ln 2)^{1/2}}{(\kappa_d+\frac{1}{2}c_d)^{1/2}}.
\end{align}

\begin{figure}
    \centering
    \includegraphics[scale = 0.5]{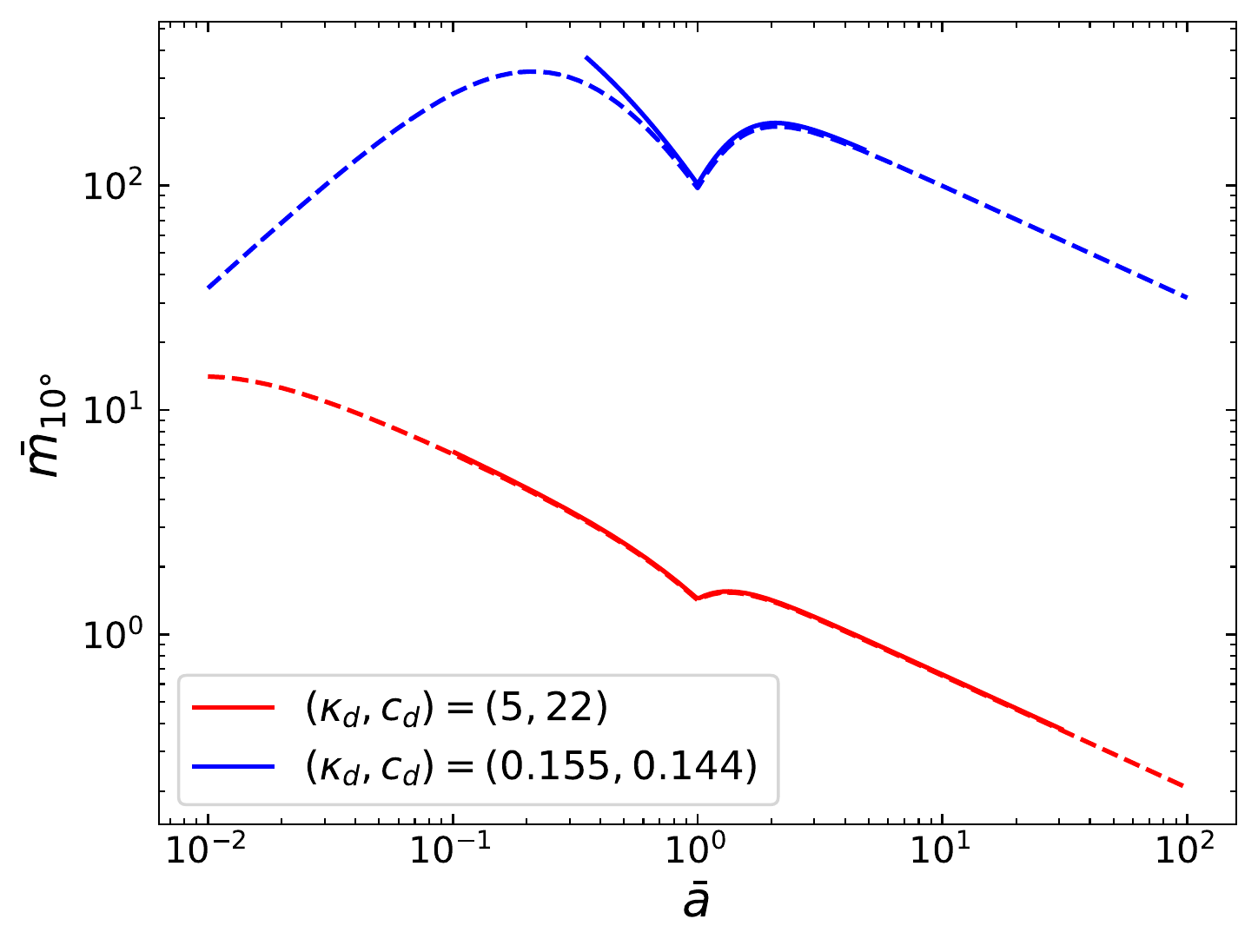}
    \caption{
    The critical semimajor axis and mass $(\bar{a},\bar{m})=(a/a_d,m/m_d)$ at which the objects of the subdominant component gradually transit to a disk-like phase with angular momentum vector distribution FWHM thickness of $10^\circ$. Two cases are presented where the heat bath is disk like $(\kappa_d,c_d)=(5,22)$ (red curve) and when it is nearly spherical $(0.155,0.144)$ (blue curve). The solid curves assume no self-interaction within the subdominant component (Eq.~\ref{eq:alpha_m_at_disk}, \ref{i:HB} approximation) valid only for the plotted domain (see Figure~\ref{fig:parameter_spcae}). The dashed curves are evaluated by also accounting for the self-interaction with the \ref{i:HBp} calculation by solving Eqs.~\eqref{eq:kappacirc}--\eqref{eq:bath_Q_analy}.}
    \label{fig:aa_am_curve}
\end{figure}

Let us examine the critical $m/m_d$ where the stars transit from an isotropic state to a disk-like configuration. In the \ref{i:HB} approximation, Eqs.~\eqref{eq:ccirc} and \eqref{eq:bath_approx_kappa} can be used to derive how the thickness of the disk depends on mass and semimajor axis. The FWHM is analogous to Eq.~\eqref{eq:FWHM}, we get\footnote{Here $(\kappa_d,c_d)$ may be arbitrary, the dominant component needs not be disk-like}
\begin{equation}
    \label{eq:FWHM_stars}
    \Delta \theta \approx \frac{(\ln 2)^{1/2}}{(\kappa+\frac{1}{2}c)^{1/2}}\approx\frac{(\ln 2)^{1/2}}{\sqrt{\bar{m} \left[ \min(\bar{a}^2,\bar{a}^{-3})\kappa_d + \frac{1}{2}\bar{a}^{1/2}c_d  \right]}} .
\end{equation}
The FWHM thickness of the subdominant component reaches $\Delta \theta=10\degree$ when $\bar{m} $ is larger than approximately
\begin{equation}
    \label{eq:alpha_m_at_disk}
\bar{m}_{10^{\circ}} \approx \frac{23}{ \min(\bar{a}^2,\bar{a}^{-3})\kappa_d + \frac{1}{2}\bar{a}^{1/2}c_d  }.
\end{equation}
The solid curves in Figure~\ref{fig:aa_am_curve} shows $\bar{m}_{10^{\circ}}$ for cases with a disk-like and spherical dominant components, respectively. Generally $\bar{m}_{10^{\circ}} \leq 45.5/(c_d \,\bar{a}^{1/2})$ where equality holds asymptotically for $\bar{a}\ll [c_d/(2\kappa_d)]^{2/3}$ or $\bar{a}\gg [c_d/(2\kappa_d)]^{-2/7}$.
The critical mass $\bar{m}_{10^{\circ}}$ is generally much larger if the dominant component is nearly spherical, but even so the subdominant component transitions to a disk-like state at large $\bar{a}$.\footnote{This conclusion is limited by the range of validity of the \ref{i:HB} approximation shown in Figure~\ref{fig:parameter_spcae}.} Note however that these results assume \ref{i:HB} that a dominant component drives the evolution which is only valid in the restricted domain of the solid lines in Figure~\ref{fig:aa_am_curve}. The dashed curves shows the numerical values of $\bar{m}_{10^{\circ}}$ evaluated using the more general \ref{i:HBp} calculation which is valid throughout the plotted range. The two calculatons clearly agree in the overlapping region of validity. In fact the extrapolation of the \ref{i:HB} curve (Eq. \ref{eq:alpha_m_at_disk}) to larger semimajor axis matches the \ref{i:HBp} calculation. However, at very small $\bar{a}$ values, the critical mass $\bar{m}_{10^{\circ}}$ evaluated with the general \ref{i:HBp} solution becomes approximately independent of the perturbing component $(\kappa_d,c_d)$ and the two curves converge.

We first present the equilibria as a function of mass in the heat bath approximation for a disk-like dominant component with $(\kappa_d,c_d) = (5,22)$ and order parameter $Q_d=0.6$. The distribution peaks at $\theta=\pi$, and the fraction of counterrotating objects for this choice of $\kappa_d$ and $c_d$ on axis is practically zero: $f_d(\theta=\pi)/f_d(\theta=0)=e^{-2c_d}=e^{-44}=10^{-11}$. The FWHM angular thickness of the disk is approximately $12\degree$ from Eq.~\eqref{eq:FWHM}. The angular momentum of the dominant component $L_d/N_dl_d = 0.97$ is close to the maximum value.

\begin{figure*}
	\centering 
\includegraphics[scale = 0.6]{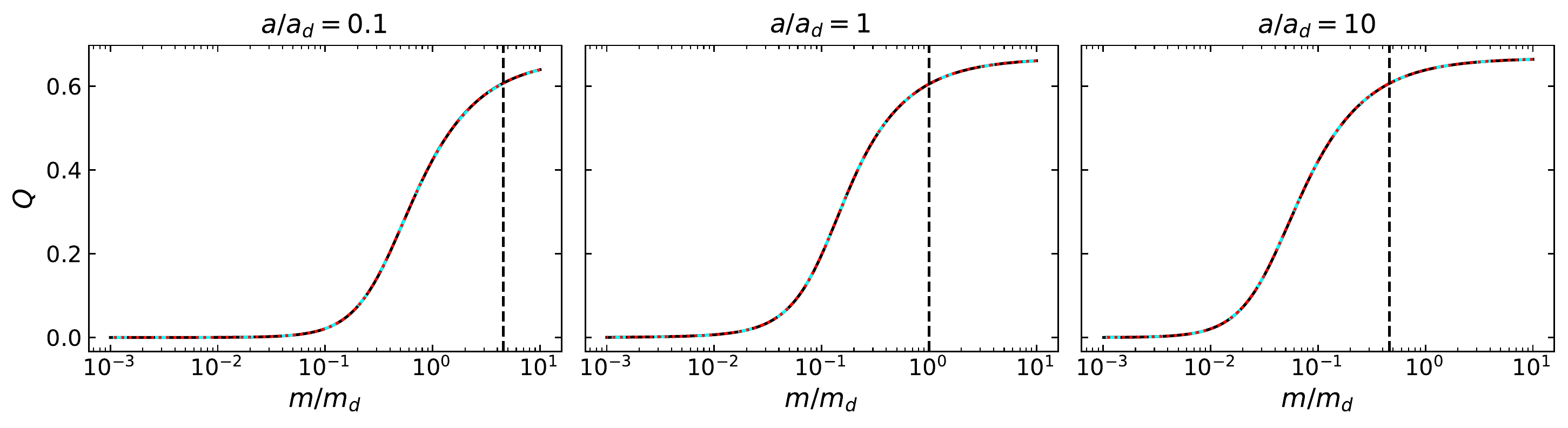}
	\caption{The order parameter $Q$ of the orbital angular momentum vector direction distribution (Eq.~\ref{eq:Q_def}) of the subdominant component (e.g. stars or BHs) as a function of mass ratio with respect to a flattened dominant component (e.g. a gaseous circumnuclear disk or a population of IMBHs). The dominant component is assumed to represent a thin disk (Eq.~\ref{eq:f_d} with $\kappa_d = 5$ and $c_d =22$), and the total mass of the subdominant component is  $M/M_d =10^{-3}$. The dashed cyan curve shows the values of $Q$ in the heat bath \ref{i:HB} approximation neglecting the self interaction of the subdominant component (Eq.~\ref{eq:bath_approx_kappa}). The solid black curve shows the values of $Q$ obtained from solving the  self-consistency equation~\eqref{eq:kappacirc}--\eqref{eq:bath_Q_analy} in the \ref{i:HBp} approximation with accounting for the self-interaction of the subdominant component as a perturbation. The red dotted curve shows the solution of the exact two-component VRR calculation (Eq.~\ref{eq:kappaexact}).  The three panels have different semimajor axes $a/a_d = 0.1$ (left), $a/a_d =1$ (middle), and $a/a_d = 10$ (right). In all cases, the distribution exhibits vertical mass segregation as the distribution changes from isotropic ($Q=0$) for low mass stars to nearly disk like ($Q\sim \frac23$) for high mass stars. The black vertical dashed line shows the value of $m/m_d$ at which the stars transit to a disk-like state of FWHM width of $10\degree$ as predicted by \eqref{eq:alpha_m_at_disk}.
	} 
	 \label{fig:Q_m_bath}
\end{figure*}
\begin{figure*}
	\begin{centering}
    \includegraphics[scale = 0.40]{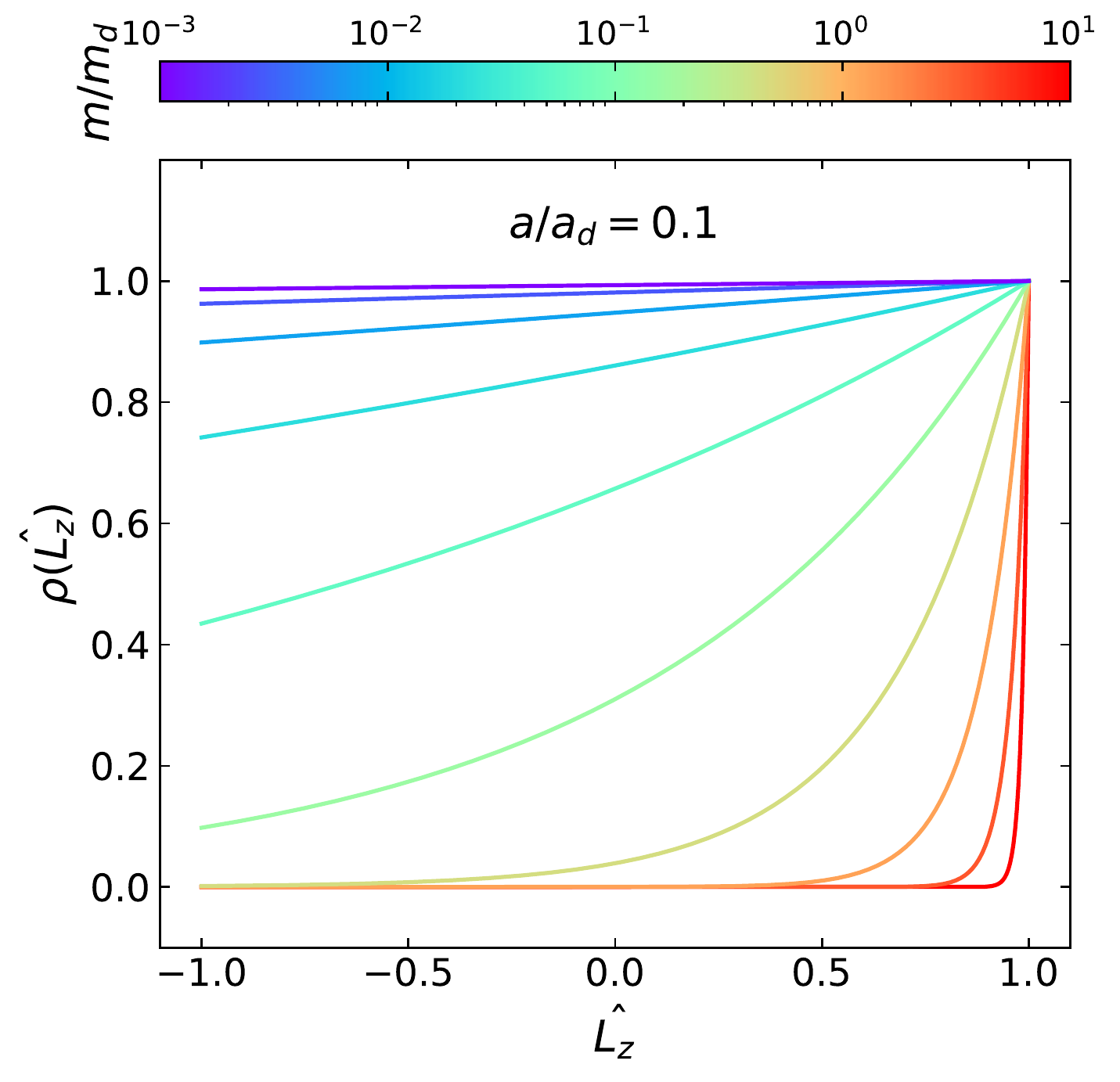}
    \includegraphics[scale = 0.40]{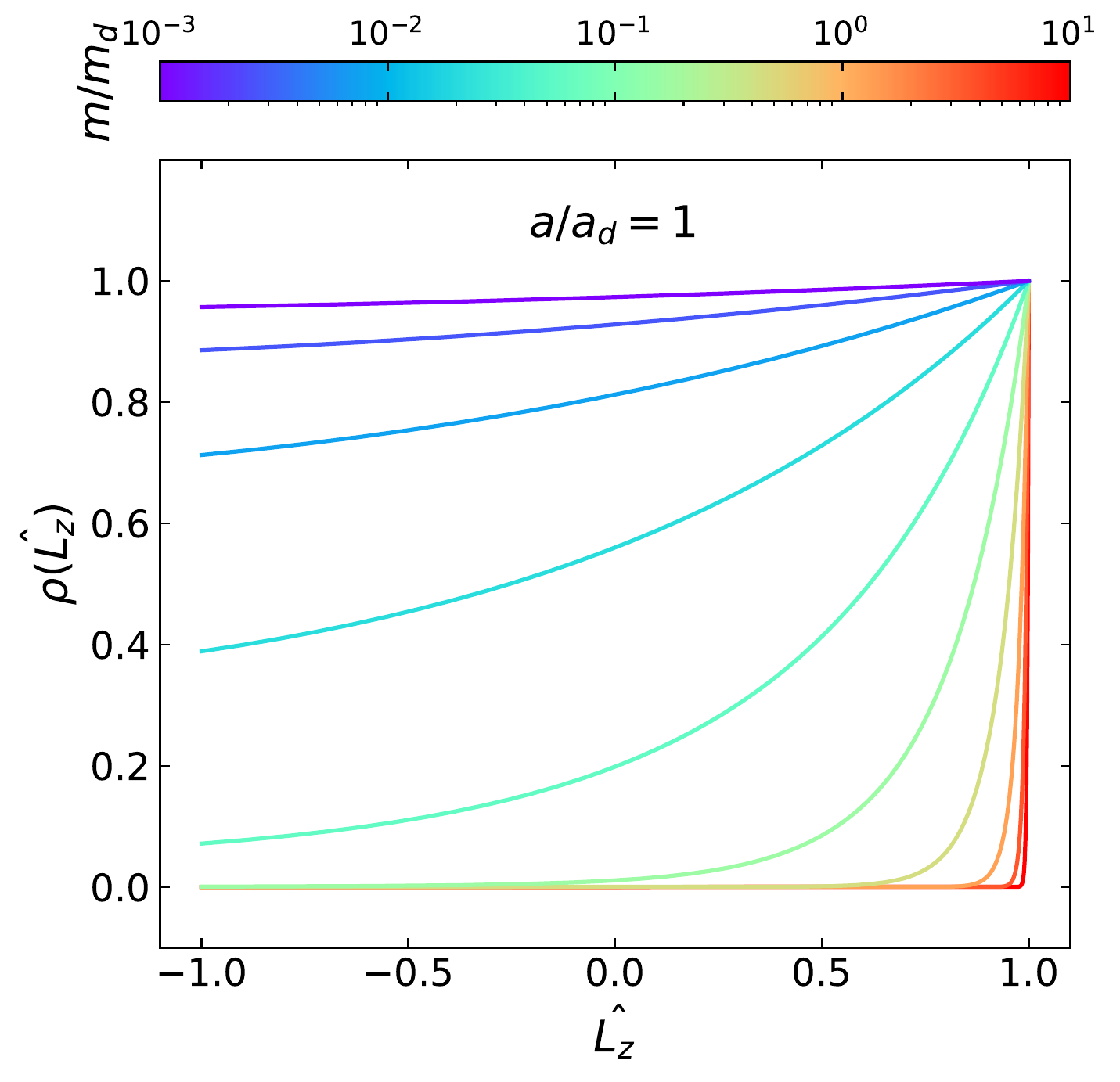}
	\includegraphics[scale=0.40]{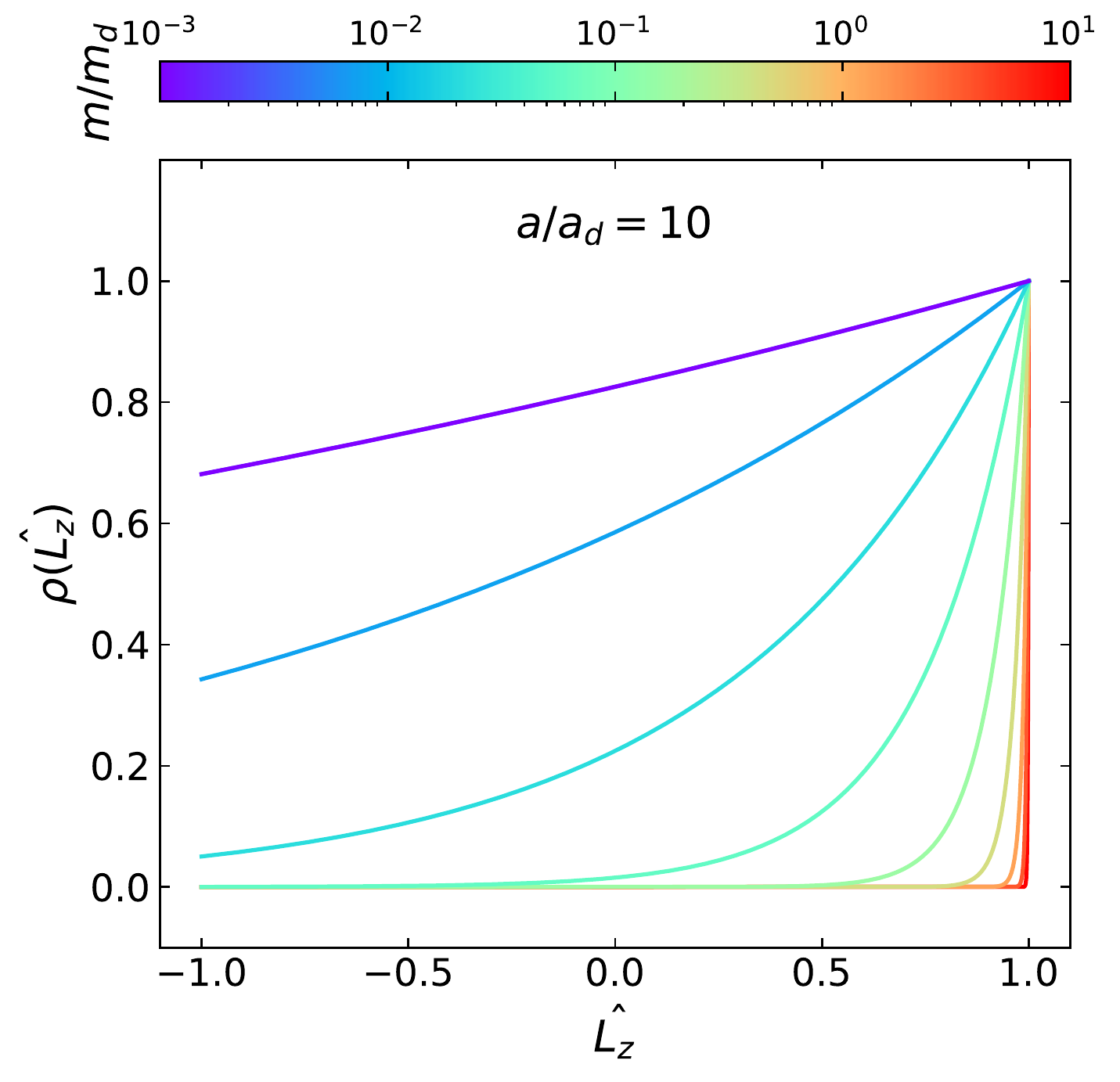}
	\end{centering}
	\caption{ The distribution function of the normalised orbital angular momentum vector direction $\cos\theta$ for the subdominant component, i.e. along the axis of symmetry, $\hat{L}_z$ in the presence of a flattened dominant component as defined in Figure~\ref{fig:Q_m_bath}. The distribution function is normalised such that $\rho(\hat{L}_z=1) = 1$. Individual curves have fixed $m/m_d$ mass ratios between $10^{-3}$ (violet) to 10 (red) as shown in the colorbar, the mass values are selected uniformly on a log scale. The three panels have different semimajor axes $a/a_d = 0.1$ (left), $a/a_d =1$ (middle), and $a/a_d = 10$ (right). The distribution shows vertical mass segregation; more massive objects are distributed in a disk while the distribution of low mass objects is nearly isotropic.} \label{fig:rho(L)_m_bath}
\end{figure*}
\begin{figure*}
	\begin{centering}
 \includegraphics[scale = 0.58]{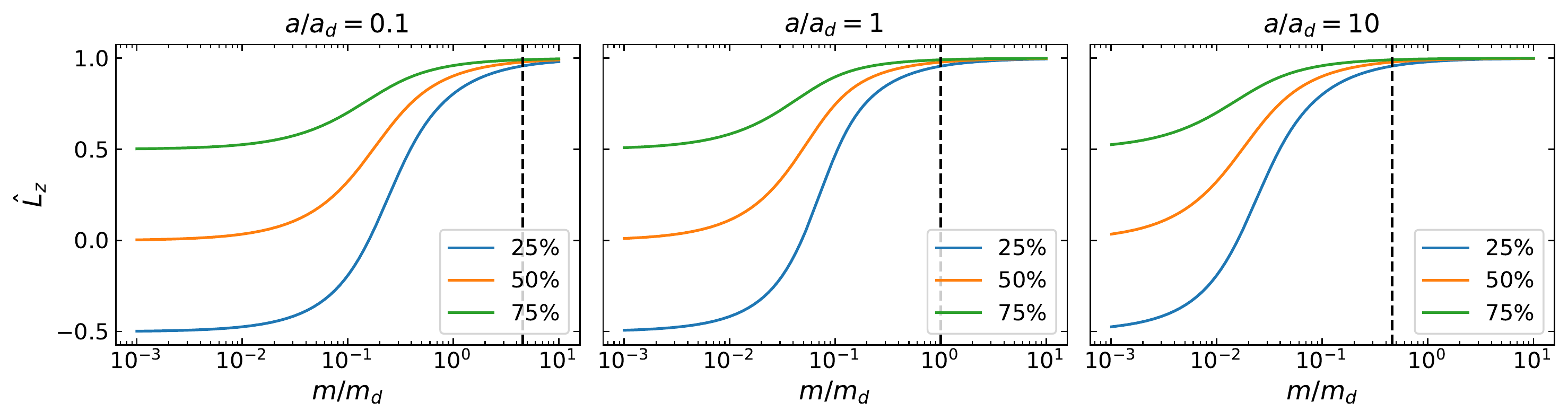}   
 \end{centering}
	\caption{ The cumulative distribution levels of $\hat{L}_z$ as a function of mass for the subdominant component in the presence of a flattened dominant component as in Figures~\ref{fig:Q_m_bath} and \ref{fig:rho(L)_m_bath} worked from the \ref{i:HBp} calculation. The plots show the value of $\hat{L}_z$ at which the cumulative distribution function reaches $25\%,50\%,75\%$ for three different semimajor axes in different panels as labelled. The distribution exhibits vertical mass segregation as the distribution changes from isotropic ($m/m_d\ll 0.1$) to narrowly peaked around the axis of symmetry $\hat{L}_z=1$ for $m/m_d\gtrsim 1$. The black vertical dashed line shows the value of $m/m_d$ at which the stars transit from an isotropic state to a disk-like state of FWHM width about $10\degree$ as predicted by \eqref{eq:alpha_m_at_disk}.
	} \label{fig:lvl_m_bath}
\end{figure*}

We investigate the distribution function of the subdominant component as a function of mass and number ratio $(\bar{m}, \bar{N})$ while fixing the total mass ratio $\bar{M}=10^{-3}$ to ensure that the dominant total energy and total angular momentum condition are satisfied. We explore the cases when the subdominant components are radially either inside ($\bar{a} = 0.1$), outside ($\bar{a} = 10$), or they are radially overlapping ($\bar{a} = 1$) with the dominant component to examine how
the distribution function transitions from a more isotropic state to a disk-like state  \cite{Szolgyen_Kocsis2018,magnan2021orbital,Mathe:2022azz}.

In all three cases of orbital radii, the stellar distribution changes smoothly between a sphere for small $m/m_d$ to a disk at large $m/m_d$, where the transition radius depends on the orbital radius $a/a_d$ and $N/N_d$ (discontinuous transitions are possible for smaller $a/a_d$, see Sec.~\ref{sec:phase_transition} below). The analytical estimate of the transition point from the isotropic state to the disk-like state, $\bar{m}_{10^\circ}$ (Eq.~\ref{eq:alpha_m_at_disk}) shown with a vertical dashed line is clearly consistent with Figures~\ref{fig:Q_m_bath} and \ref{fig:lvl_m_bath}.

Figure~\ref{fig:rho(L)_m_bath} shows the distribution of the angular momentum vectors of the stars for different mass ratios and $a/a_d$ for the \ref{i:HBp} model. Given the assumption of axisymmetry we plot the distribution as a function of $s=\cos\theta$, which is also the z-component of the normalised angular momentum $\hat{L}_z$.
The figure shows the distribution function $\rho(\hat{L}_z)=f(s)/f(1)$, normalised for clarity such that $\rho(1) = 1$. A flatter distribution corresponds to a more isotropic distribution. Figure~\ref{fig:lvl_m_bath} shows the values of $\hat{L}_z$ at which the cumulative distribution reaches 25\%, 50\% and 75\% respectively. For an isotropic distribution, these three angles lie close to $\hat{L}_z =-0.5$, 0 and 0.5.

Figures~\ref{fig:Q_m_bath}, \ref{fig:rho(L)_m_bath}, and \ref{fig:lvl_m_bath} show that in all three radial regions with different $a/a_d$, the distribution becomes more flattened for larger stellar masses, as expected for vertical mass segregation. For higher $a/a_d$, the disk-isotropic transition is shifted towards lower $m/m_d$ and conversely for smaller $a/a_d$ as explained by Eq.~\eqref{eq:alpha_m_at_disk}.
The models demonstrate not only the mass dependence of a two-component stellar distribution, but the figures may also be interpreted as showing the distribution of a multimass stellar cluster under the influence of a massive component provided that the coupling between the stellar components is neglected, i.e. for model \ref{i:HB}. 

\subsubsection{The case of a dominant sphere}
Consider now a dominant component such that $\bar{M} = 10^{-3}$ in a nearly isotropic state: $\kappa_d = 0.155$ and $c_d = 0.144$, such that $Q_d\approx 0.015$ 
and $L_d/(N_dl_d)\approx 0.05$ from Eqs~\eqref{eq:Qisotropic} and \eqref{eq:L_small}.

\begin{figure}
    \centering
    \includegraphics[scale = 0.5]{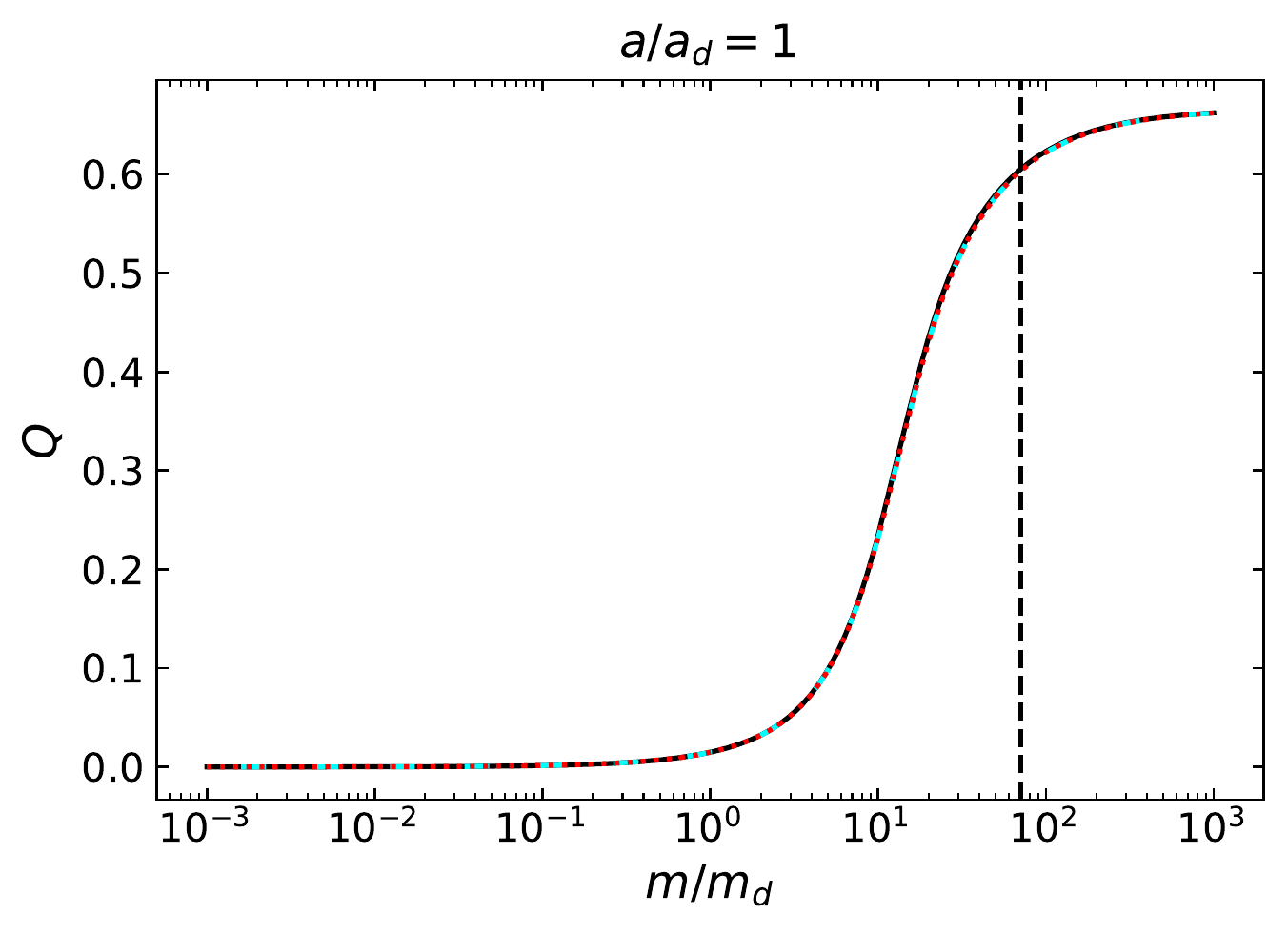}
    
    \caption{Similar to Figure~\ref{fig:Q_m_bath} showing the order parameter $Q$ of the orbital angular momentum vector direction distribution (Eq.~\ref{eq:Q_def}) of the subdominant component as a function of mass ratio driven by a spherical dominant component ($\kappa_d = 0.155, c_d =0.144$). $\bar{M} = 10^{-3}$ and $a/a_d = 1$.}
    \label{fig:Q_m_bath_sphere}
\end{figure}
\begin{figure}
    \centering
    \includegraphics[scale= 0.5]{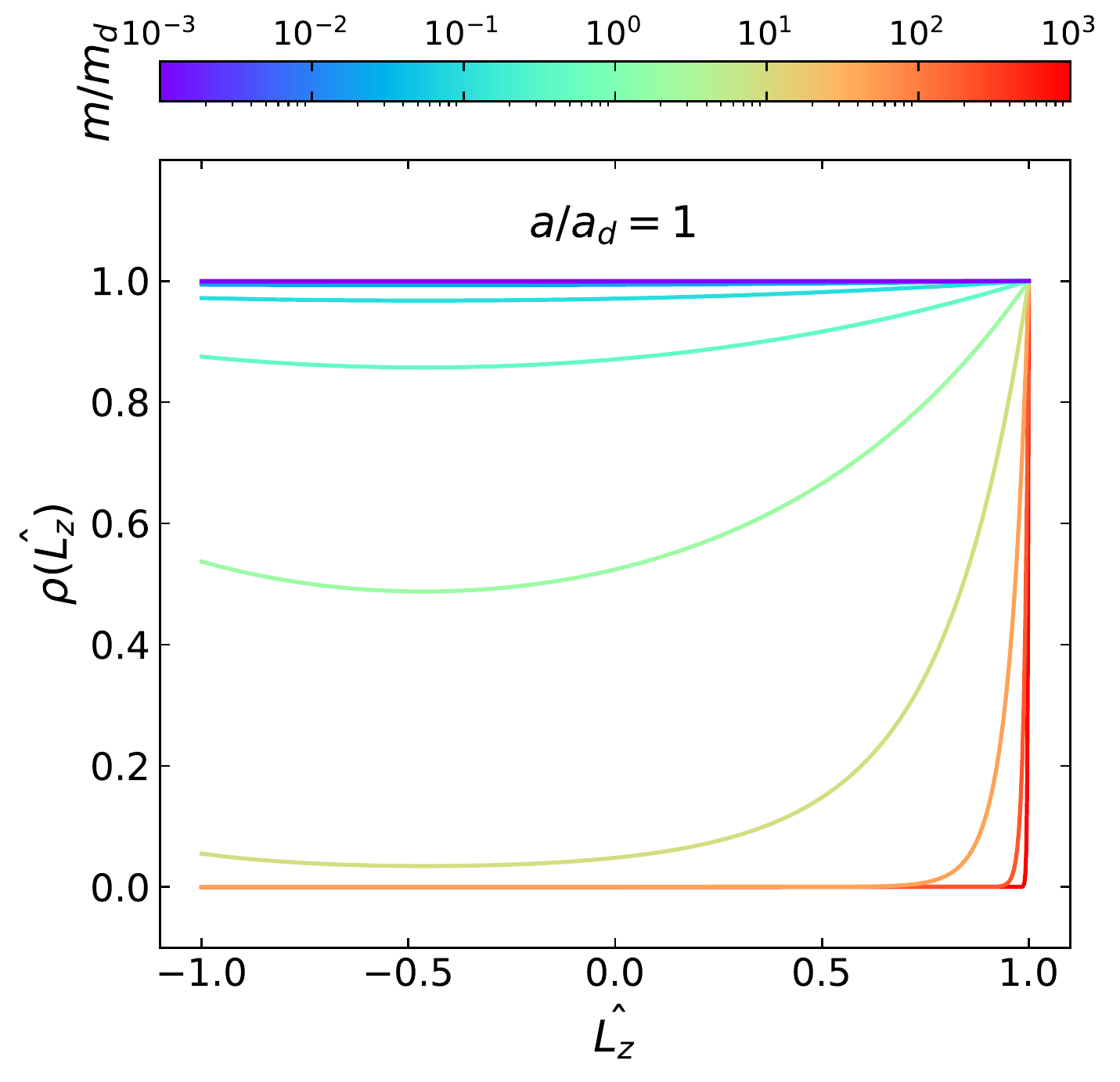}
    \caption{Similar to Figure~\ref{fig:rho(L)_m_bath} showing the distribution function of the normalised orbital angular momentum vector direction $\cos\theta$ for the subdominant component along the axis of symmetry, $\hat{L}_z$ in the presence of a spherical dominant component ($\kappa_d = 0.155, c_d =0.144$). The distribution function is normalised such that $\rho(\hat{L}_z=1) = 1$. Individual curves have fixed $m/m_d$ mass ratios between $10^{-3}$ (violet) to $10^3$ (red) as shown in the colorbar, the mass values are selected uniformly on a log scale.  $\bar{M} = 10^{-3}$ and $a/a_d = 1$.}
    \label{fig:selected_bath_sphere}
\end{figure}
\begin{figure}
    \centering
    \includegraphics[scale = 0.5]{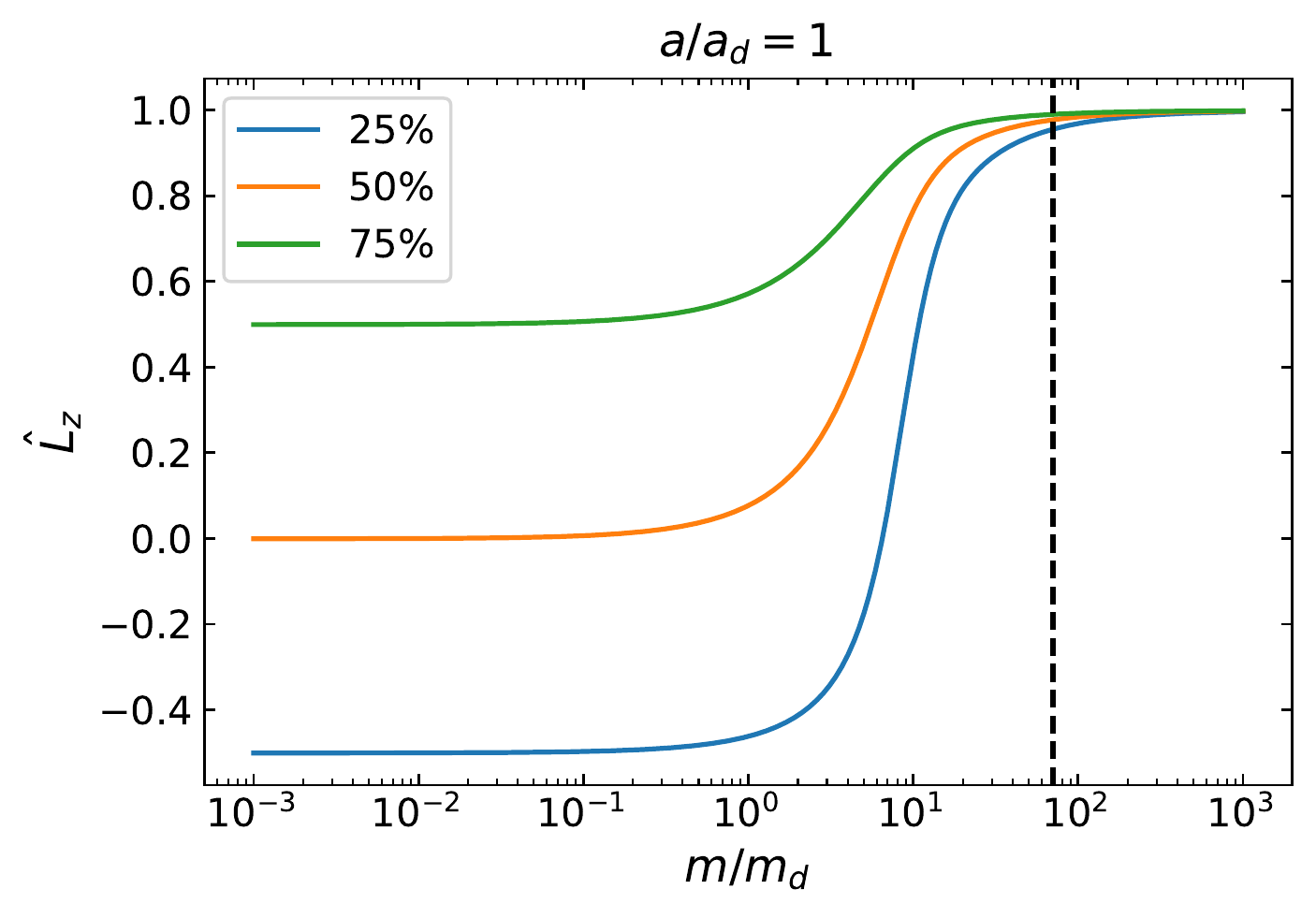}
    \caption{Similar to Figure~\ref{fig:lvl_m_bath} but showing the cumulative distribution levels of $\hat{L}_z$ as a function of mass for the subdominant component in the presence of a spherical dominant component $(\kappa_d, c_d, M/M_d,a/a_d)= (0.155, 0.144, 10^{-3},1)$}
    \label{fig:lvl_bath_spherical}
\end{figure}
Figures~\ref{fig:Q_m_bath_sphere}, \ref{fig:selected_bath_sphere}, and \ref{fig:lvl_bath_spherical} show the parameter dependence of vertical mass segregation as in Figures~\ref{fig:Q_m_bath}, \ref{fig:rho(L)_m_bath}, and \ref{fig:lvl_m_bath}. For this choice of $(\kappa_d,c_d)$ vertical mass segregation may still take place but only for $m/m_d$ values much larger than for a flattened dominant component. The smooth transition to a disk-like phase occurs at much larger $m/m_d$ values, consistent with the predictions of Eq.~\eqref{eq:alpha_m_at_disk}. This result could potentially explain the lack of vertical mass segregation signature in the direct N body simulations by Ref. \cite{taras_2022_discs} where the initial condition is nearly isotropic with $(E_{\textrm{norm}},L_{\textrm{norm}}) \sim (10^{-4}, 10^{-2})$ and a relatively narrow mass range ($10^{-2}\leq m/m_d \leq 1$)\footnote{For $(\kappa_d,c_d) = (0.155, 0.144)$,  $(E_{\textrm{norm}},L_{\textrm{norm}})\approx (5\times 10^{-4},5\times 10^{-2})$
using the definition in Eqs.~\eqref{eq:E_norm}--\eqref{eq:L_norm}. }.

We emphasize that Figures~\ref{fig:Q_m_bath}--\ref{fig:lvl_bath_spherical} are expected to be valid for a general multicomponent model with an arbitrary spectrum of masses for the subdominant component for the given values of $\bar{a}$ and $\bar{M}$ since these systems are in the \ref{i:HB} heat bath regime where the self-interaction is negligible.

\subsection{Dependence on orbital radius} \label{bath_radii}
We examine the systematic variation of the anisotropy with orbital radius $\bar{a}$ under the influence of a flattened dominant component comprised of massive bodies such that $(\kappa_d,c_d,\bar{M},\bar{m},\bar{N})=(5,22,0.001,0.01,0.1)$. The assumptions of a dominant total energy and total angular momentum are satisfied if $0.1\leq \bar{a}\leq 10$, where the self-interaction within the subdominant component is negligible (Figure~\ref{fig:parameter_spcae}). These models may be used to explore how a disk of massive perturbers (e.g. the circumnuclear disk or a population of IMBHs) at a particular radius affects the distribution of lower mass objects (e.g. main sequence stars or black holes) as a function of distance. 

\begin{figure}
	\centering 

\includegraphics[scale = 0.5]{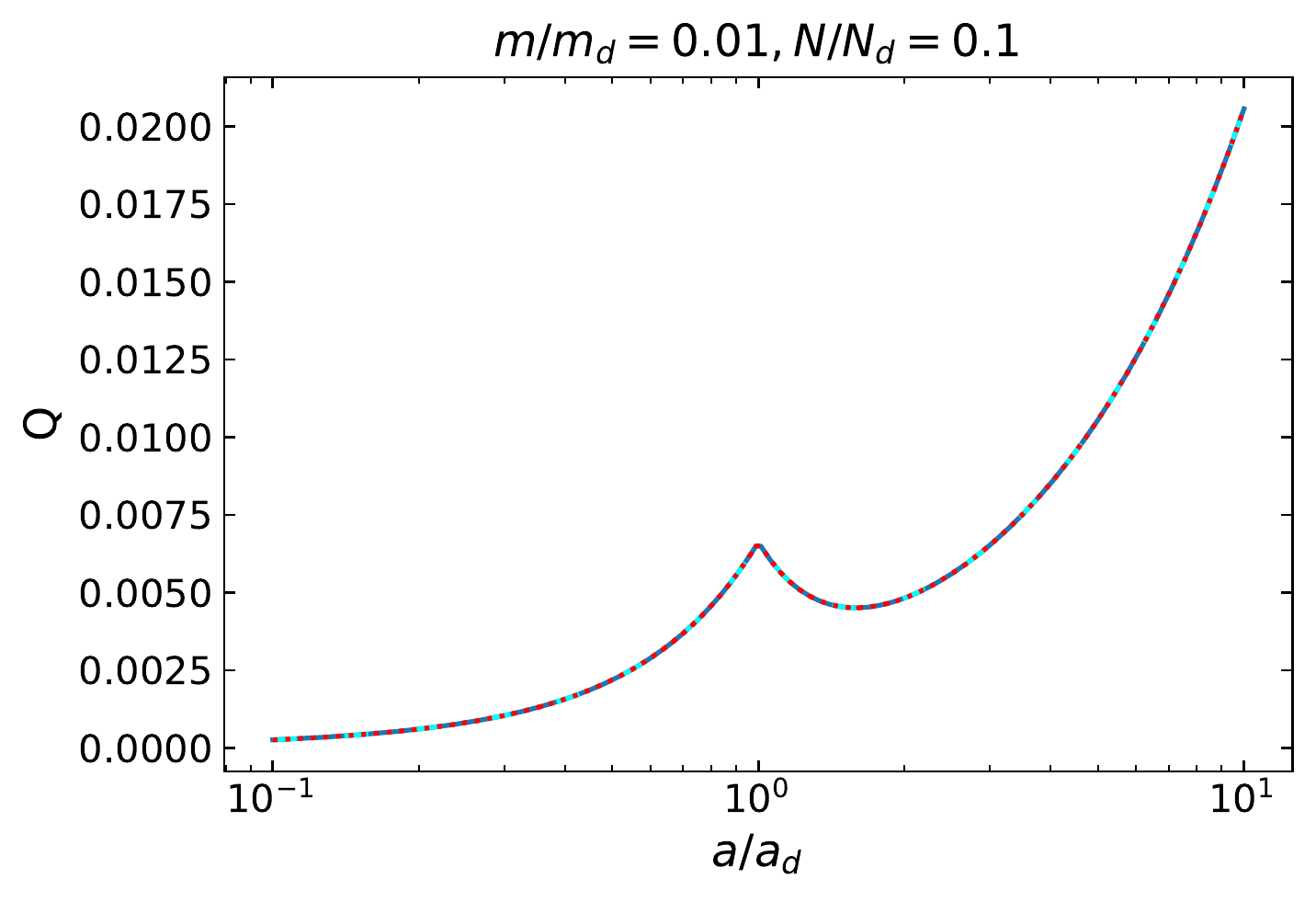}
	\caption{
	Similar to Figure~\ref{fig:Q_m_bath} but showing the order parameter $Q$ for stars under the influence of a disk of massive perturbers with parameters $(\kappa_d,c_d, M/M_d,m/m_d) =(5,22, 10^{-3},0.01)$ as a function of stellar orbital radii for the two heat bath models \ref{i:HB}, \ref{i:HBp}, and the exact two-component calculation, all of which approximately overlap. The stellar distribution is approximately spherical but slightly more anisotropic at higher $a/a_d$. } \label{fig:Q_a_bath}
\end{figure}
\begin{figure}
	\centering 

\includegraphics[scale = 0.5]{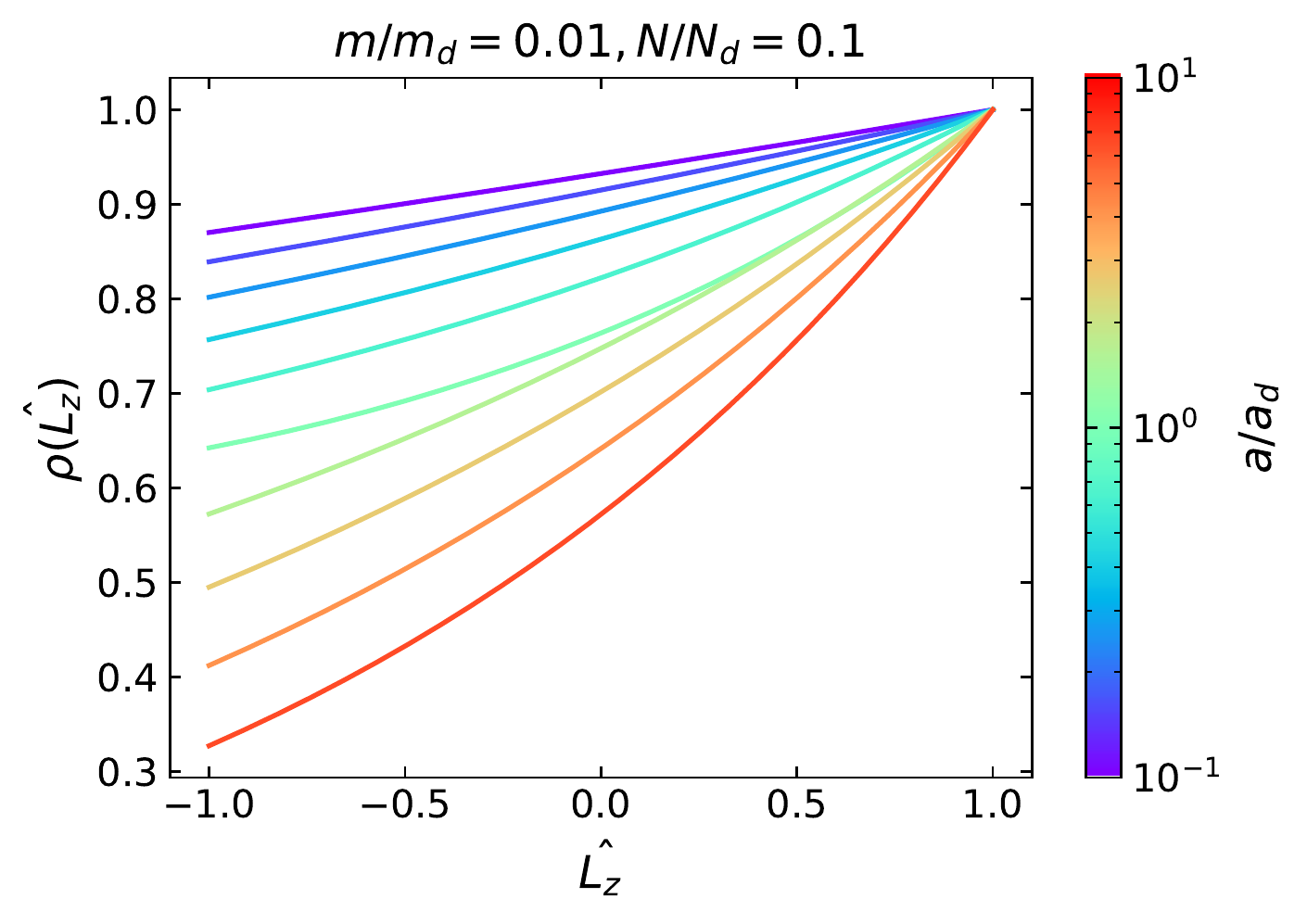}
	\caption{Similar to Figure~\ref{fig:rho(L)_m_bath} but showing the distribution function of $\hat{L}_z$ for the subdominant component for 10 selected radius ratios $a/a_d$ from 0.1 to 1 in log scale under the influence of a disk of massive perturbers with the same parameters as in Figure~\ref{fig:Q_a_bath} $(\kappa_d,c_d, M/M_d,m/m_d) =(5,22,10^{-3},0.01)$.} \label{fig:rho(L)_a_bath}
\end{figure}
\begin{figure}
	\centering 
\includegraphics[scale = 0.5]{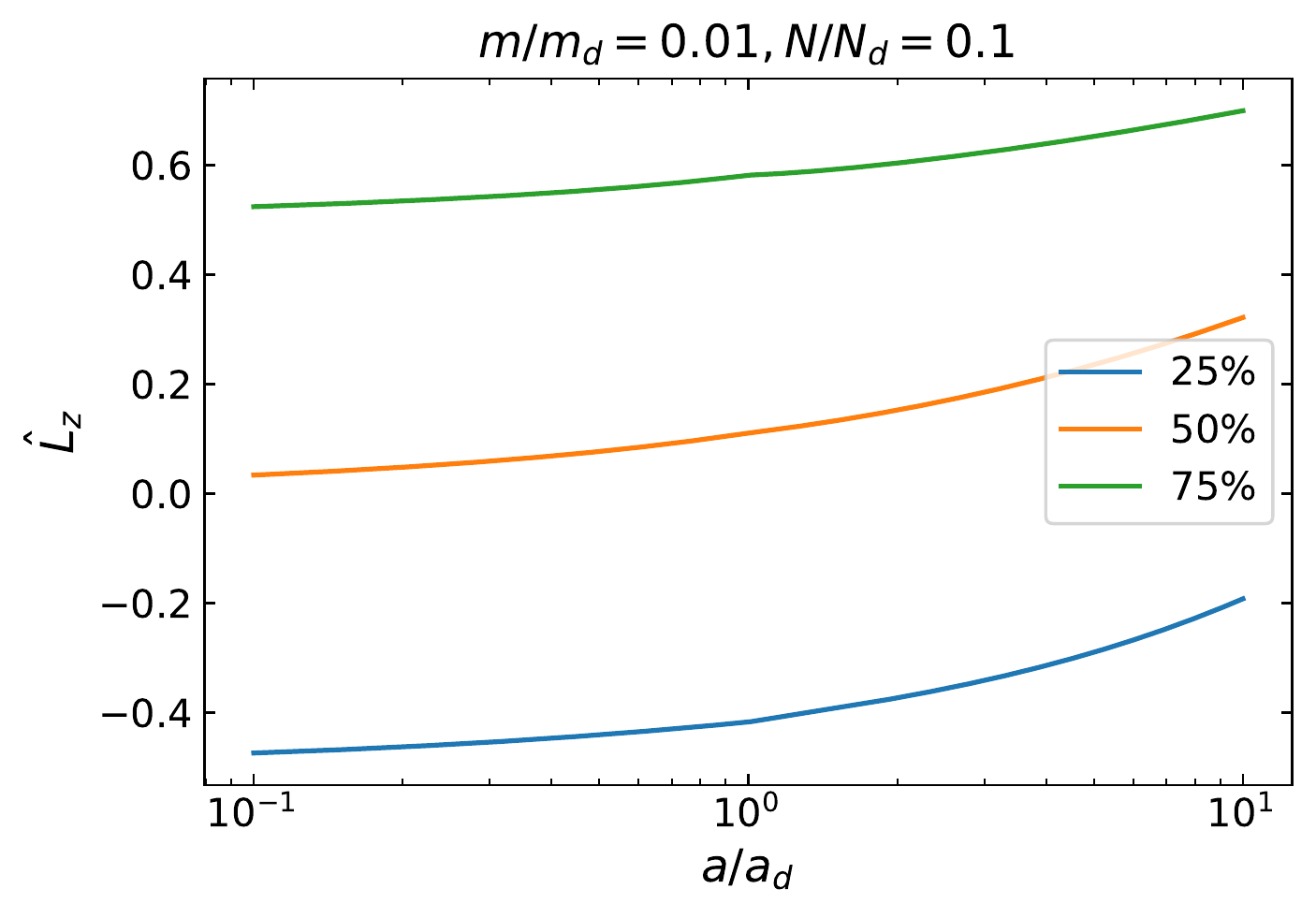}
	\caption{Similar to Figure~\ref{fig:lvl_m_bath} but showing the cumulative distribution levels of $\hat{L}_z$ for different orbital radii for the subdominant component. The plots show the value of $\hat{L}_z$ at which the cumulative distribution function reaches $25\%,50\%,75\%$, respectively. $(\kappa_d,c_d, M/M_d,m/m_d) =(5,22,10^{-3},0.01)$.} \label{fig:lvl_a_bath}
\end{figure}

Figures~\ref{fig:Q_a_bath}, \ref{fig:rho(L)_a_bath}, \ref{fig:lvl_a_bath} show the value of the order parameter $Q$, the distribution of angular momentum vector directions, and the cumulative distribution levels for different radius ratios. Again, the \ref{i:HB}, \ref{i:HBp} models and the exact calculation give consistent results in this parameter range. The figures show that the spacial distribution is highly disordered, but the distribution is slightly more flattened at larger radii and the distribution is more isotropic at moderately smaller radii. 
In Sec.~\ref{sec:phase_transition}, we study the behaviour at much smaller $\bar{a}$ where the heat bath approximation breaks and find evidence for a discontinuous change in the order parameter there to higher values (Figure~\ref{fig:region_phase_bath}).

\subsection{Mass segregation without a dominant component; exploring the $(E_{\rm norm},L_{\rm norm})$ landscape}\label{mass_seg_general_two}
Let us now examine mass segregation in a two component model without a dominant component. We relax the heat bath approximations and solve Eqs.~\eqref{eq:Q1}--\eqref{eq:Q2} exactly. An analytical solution exist when $a_1 = a_2$, in other cases we resort to a numerical solution, see Appendix~\ref{app:analytic}.
Unlike previously in Sec.~\ref{exact_calculation} where we used $(\kappa_d,c_d)$ to characterise the system, here, we characterise the system with the conserved quantities: the normalized total 
VRR energy and total angular momentum in Eqs.~\eqref{eq:Ltot}--\eqref{eq:Etot} defined as \cite{Mathe:2022azz}:
\begin{align}
{E_{\textrm{norm}}} &= \frac{3E}{J_1N_1^2+J_2N_2^2+2J'N_1N_2},   \label{eq:E_norm} \\
{L_{\textrm{norm}}} &= \frac{L}{N_1l_1+N_2l_2}.     \label{eq:L_norm}
\end{align}
These quantities are bounded by $0\le L_{\textrm{norm}}\le 1$ and $-1 \le E_{\textrm{norm}} \le  0$ (see Appendix \ref{app:max_E}). We explore four representative combinations in the parameter space  $(L_{\textrm{norm}},E_{\textrm{norm}}) = (0.15,-0.66)$, $(0.82,-0.81)$, $(0.16,-0.09)$, $(0.38,-0.09)$ following Ref. \cite{Mathe:2022azz}. This investigation aims to map out the possible behaviour of two component systems in the qualitatively different regions of parameter space. However, given that the results are qualitatively similar to the previously stated conclusions, we defer the plots of this section to the Appendix~\ref{app:plots}.

We explore systems with $a_1 = a_2$ and $N_1 =N_2$, and vary the mass ratio $m_2/m_1$ from 0.02 to 1 with cluster $\mathcal{C}_2$ being the lighter component. Figure~\ref{fig:Q_m_two} shows the value of $Q_1$ and $Q_2$ at different mass ratios. Figure~\ref{fig:L_m_two} shows the angular momentum of each component as a fraction of the total angular momentum. Figure~\ref{fig:rho_m_two} shows the distribution of angular momentum vectors at different angles of both components for different mass ratios. Figure~\ref{fig:lvl_m_two} shows the cumulative distribution levels of the lighter component $\mathcal{C}_2$ for different mass ratios. Clearly for all four combinations of total energy and total angular momentum, the less massive component distribution becomes more anisotropic as the mass ratio $m_2/m_1$ becomes larger, similar to the the multi-component systems examined in Ref. \cite{Mathe:2022azz}. Figures~\ref{fig:Q_m_two} and \ref{fig:rho_m_two} show that the order parameter of the more massive component is insensitive to the mass ratio.

We also explore the behaviour of systems with nearly isotropic configurations with $(E_{\rm norm},L_{\rm norm})=(5\times 10^{-4}, 0.05)$, as in the last part of Sec.~\ref{sec:mass_ratio}. Here we examine the case of a dominant component with $M_2/M_1 = 10^{-3}$, and $a_2/a_1 =1$. Similarly to Figures~\ref{fig:Q_m_two}, \ref{fig:rho_m_two} and \ref{fig:lvl_m_two}, Figures~\ref{fig:two_sphere_QQ}, \ref{fig:two_sphere_selected} and \ref{fig:two_sphere_lvl} shows the distribution of the two components using the general two component calculation for different orbit mass ratios $m_2/m_1$. Results are consistent with the heat bath approximation of Figures~\ref{fig:Q_m_bath_sphere}, \ref{fig:selected_bath_sphere} and \ref{fig:lvl_bath_spherical}. The heavier component $\mathcal{C}_1$ maintains an isotropic distribution for all values of $m_2/m_1$ while the lighter component $\mathcal{C}_2$ exhibits vertical mass segregation but in this case transits to a disk-like state only at very large values of $m_2/m_1$. 

\section{Phase transition in two component systems}\label{sec:phase_transition}

\subsection{Phase transition of an inner low-mass component influenced by an outer massive perturber} \label{heat_phase_bath}

In the previous section we have examined the equilibria of the subdominant component under the influence of the dominant component with given $(\kappa_d,c_d)$ and explored cases without a dominant component with given ($E_{\rm norm},L_{\rm norm}$). Let us now study how the equilibria change with the total energy and angular momentum of the system and identify possible discontinuities.
Since the interaction with a heat bath generates the canonical ensemble for the subdominant component, the angular momentum distribution of the subdominant component may be expected to undergo a first order phase transition from an ordered disk phase to a disordered spherical phase when varying the parameters of the heat bath as found previously for one component systems \cite{Roupas_2017,Roupas2020}. The phase transition is characterised by a discontinuous change in the order parameter $Q$ between an ordered $Q_{\rm ord}$ and a disordered $Q_{\rm dis}$ state, where $Q_{\rm ord}=0.286014$ and $Q_{\rm dis}=0$ for a one component model with no rotation \cite{Roupas_2017}.  Equilibria with $Q_{\rm dis}<Q<Q_{\rm ord}$ are inaccessible to the system as they are either metastable or unstable.

However, Eqs.~\eqref{eq:c} and \eqref{eq:bath_approx_kappa} show that a system strongly driven by a dominant component responds continuously to changes in the dominant component, indicating that a discontinuous phase transition is \textit{not} possible when the \ref{i:HB} model applies. Thus, a phase transition is prohibited in the region bounded by the red lines in Figure~\ref{fig:parameter_spcae}). The lack of a phase transition is due to the intercomponent coupling, which is strongly nonnegligible here. This is in stark contrast with additive short-range interacting systems where the intercomponent coupling is absent/negligible and where a first order phase transition \textit{is} possible, such that the system exhibits phase separation during the transition. Nevertheless, here we demonstrate that a phase transition is also possible for VRR for an isolated two component systems where the self-interaction is non-negligible in the energy equation (i.e. this leads to the $\bar{Q}$ terms in Eqs.~\ref{eq:kappacirc} and \ref{eq:kappaexact}). This happens when $(J/J')\bar{N}\bar{Q} \ll 1$ is violated, i.e. in the region below the bottom red line in Figure~\ref{fig:parameter_spcae}, corresponding to a system influenced by an outer massive perturber.

\begin{figure}
    \centering
    \includegraphics[scale = 0.5]{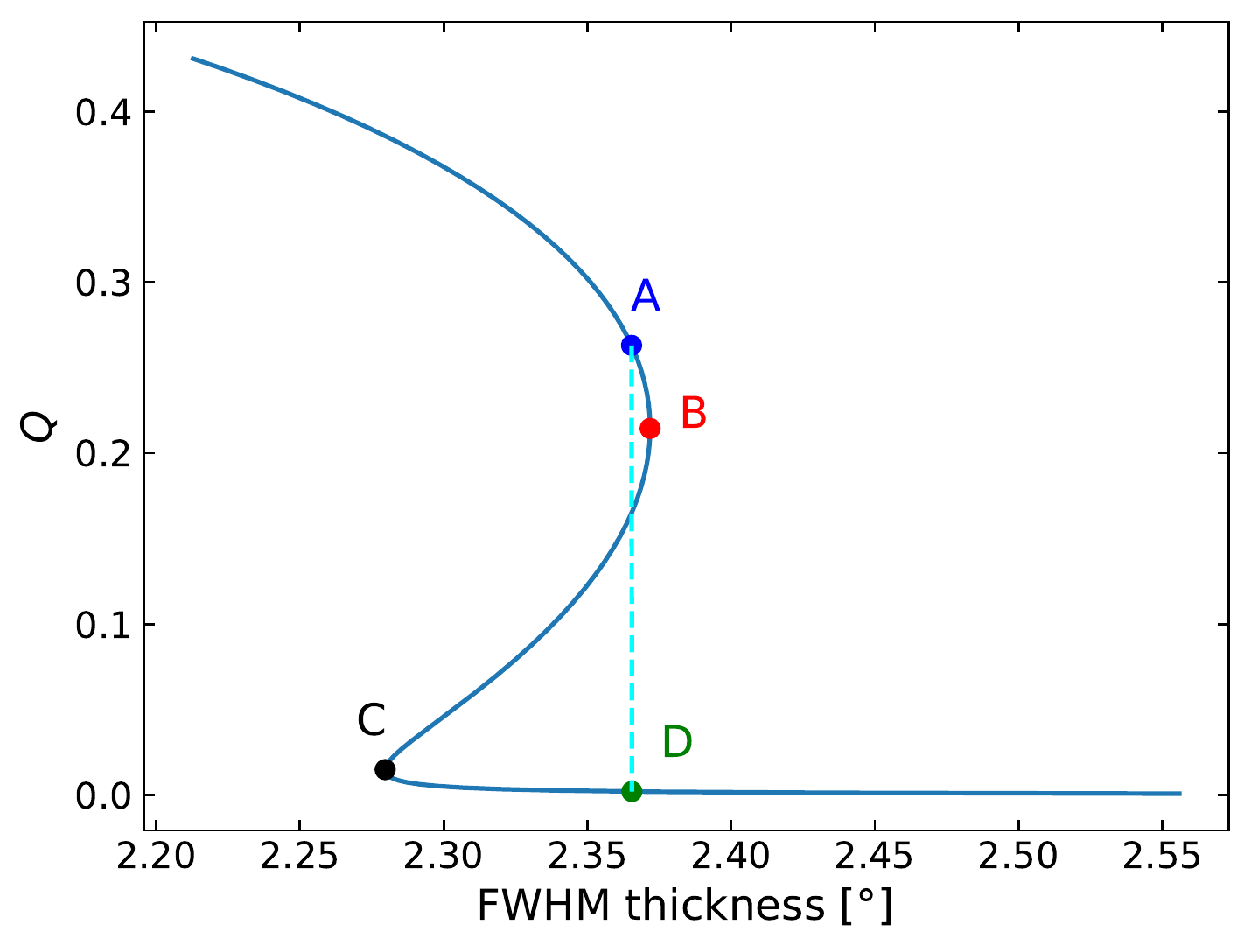}
    \includegraphics[scale=0.5]{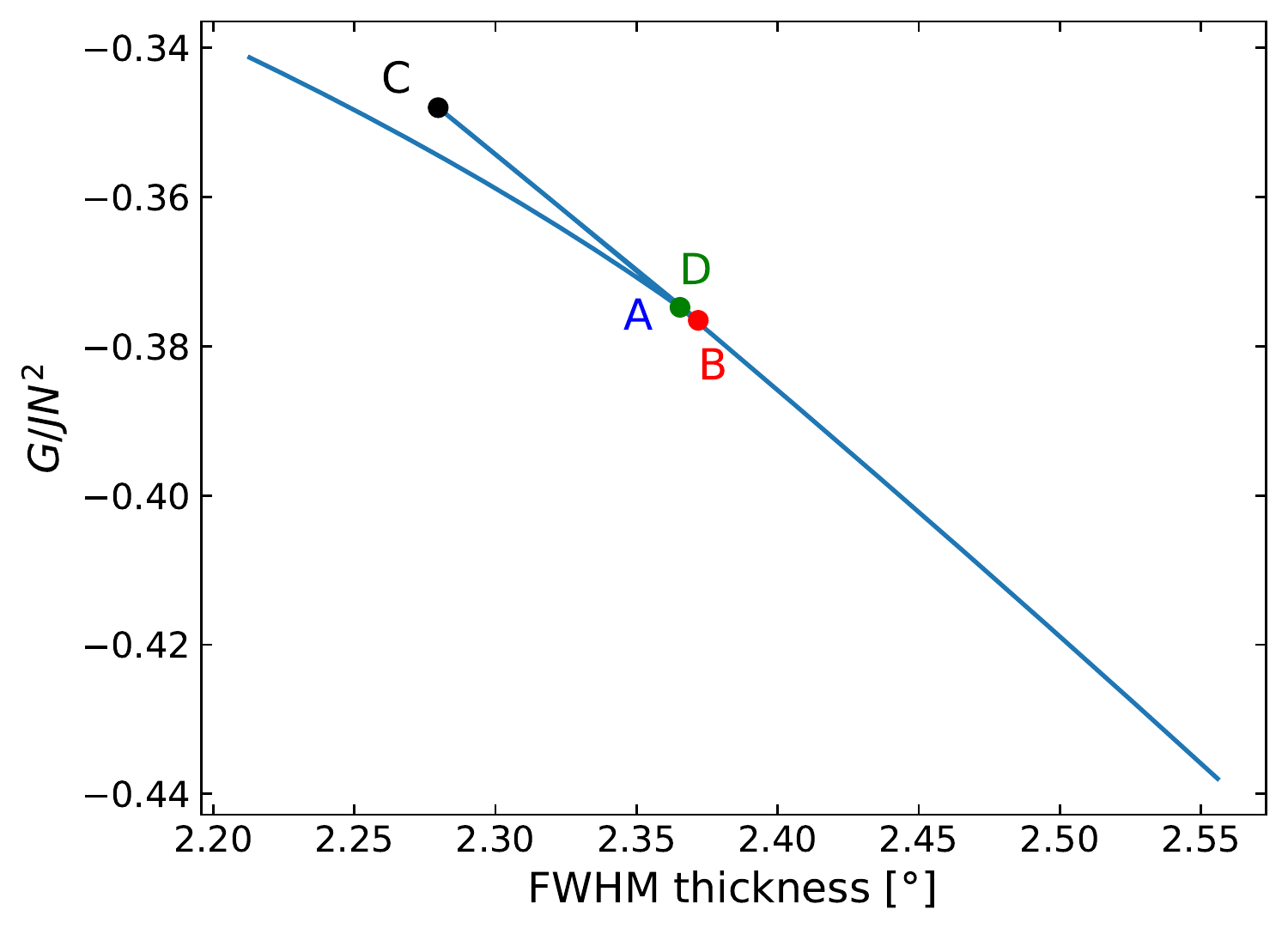}
    \caption{The order parameter $Q$ of the inner low-mass component (top) and the Gibbs free energy (bottom) as a function of the FWHM thickness of the outer massive component $\Delta \theta_d$ in degrees assuming $c_d = 5$ and $(m/m_d,N/N_d,a/a_d)= (0.1, 0.01, 0.006)$. The thickness of the inner component changes discontinuously when the FWHM of the outer component is $2.37\degree$ 
    which corresponds to $Q_d = 0.664$ and $\kappa_d = 404$. A phase transition takes place from point $A$ to point $D$, which have the same Gibbs free energy. }
    \label{fig:Q_F_phase_bath}
\end{figure}

For a proof-of-concept, we present an example of such a phase transition in a two component model with an outer massive perturber such that $(\bar{M},\bar{m},\bar{a})=  (10^{-3},0.1,0.006)$. 
We vary the FWHM of the angular momentum distribution of the outer massive component between 
$\Delta \theta_d = 1.3^\circ$ and $4.7^\circ$ with a fixed negligible fraction of counterrotating objects using the distribution function of Eq.~\eqref{eq:f_d} with parameters $100<\kappa_d<1500$ and $c_d=5$ (see Eq.~\ref{eq:FWHM}).\footnote{$Q_d$ for these parameters (Eq.~\ref{eq:Qdisk}) stays close to $2/3$, i.e. 0.66.} This is equivalent to varying the temperature (Eq.~\ref{eq:kappa_d}) which induces a change in the value of the $\kappa$ parameter of the inner low-mass component (i.e. its dimensionless effective inverse temperature, Eq.~\ref{eq:kappacirc}) and hence the order parameter $Q$ of the subdominant component (Eqs.~\ref{eq:kappacirc}--\ref{eq:bath_Q_analy}). The top panel of Figure~\ref{fig:Q_F_phase_bath} shows the equilibrium value of $Q$ of the subdominant component at different $\Delta \theta_d$. Clearly, there are three possible equilibria for the subdominant component if the FWHM of the dominant component is between $\Delta\theta_{d,C}$ and $\Delta\theta_{d,B}$.

If a subsystem exchanges both energy and angular momentum with a heat bath, it will search for the global minimum value of the Gibbs-like free energy \cite{Roupas_2017,Roupas2020}:
\begin{equation}
    \label{eq:F}
    G = E - \omega L - TS.
\end{equation}
where $\omega$ is defined by $\gamma = \beta\omega$. Here the energy $E$ and angular momentum $L$ denotes that of the subdominant component.
The entropy $S$ is given by  
\begin{equation}
    \label{eq:S}
S = -k_{\rm B}\int f(\bm{n})\ln f(\bm{n}) d\Omega .
\end{equation}

\begin{figure}
    \centering
    \includegraphics[scale = 0.5]{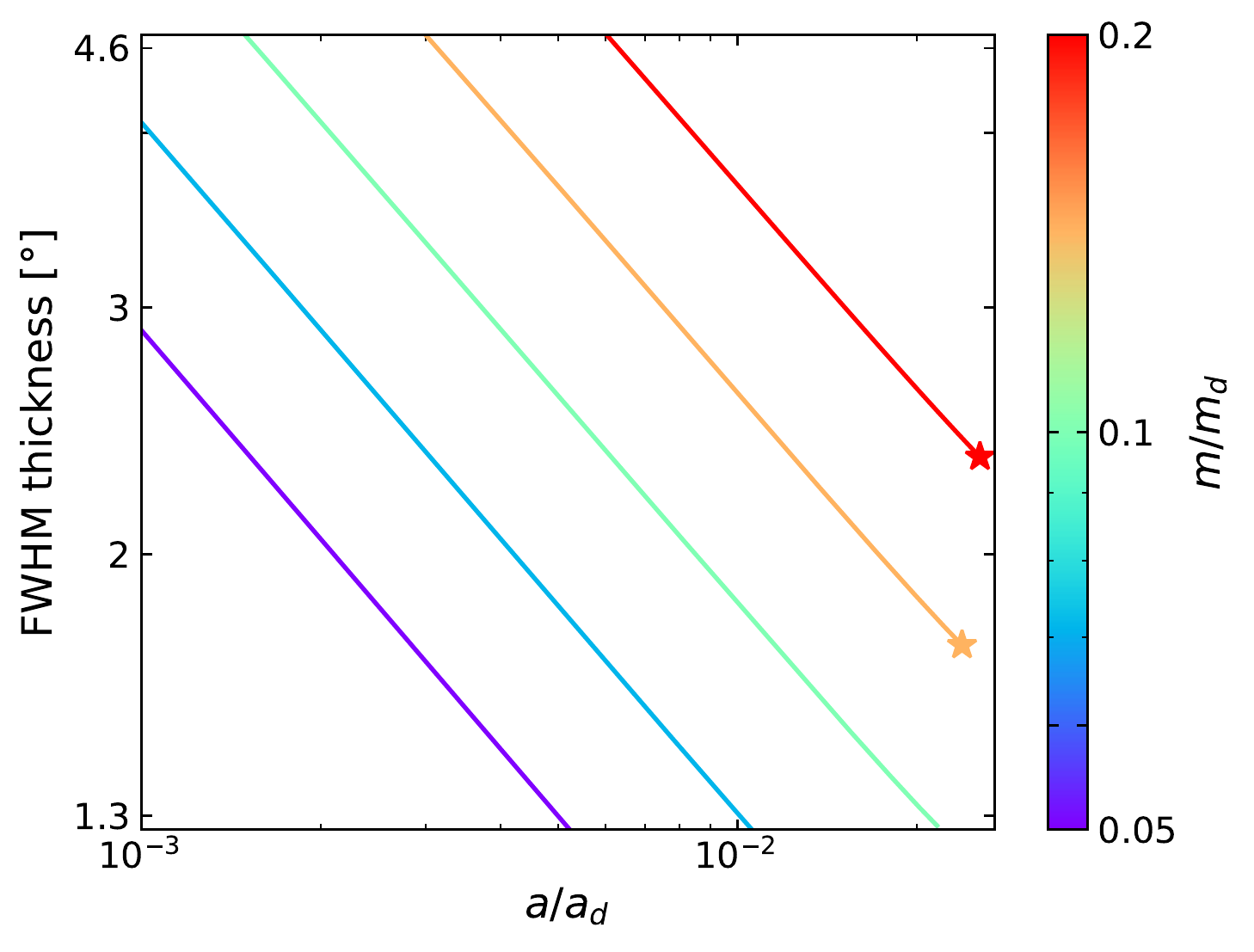}
  \includegraphics[scale = 0.5]{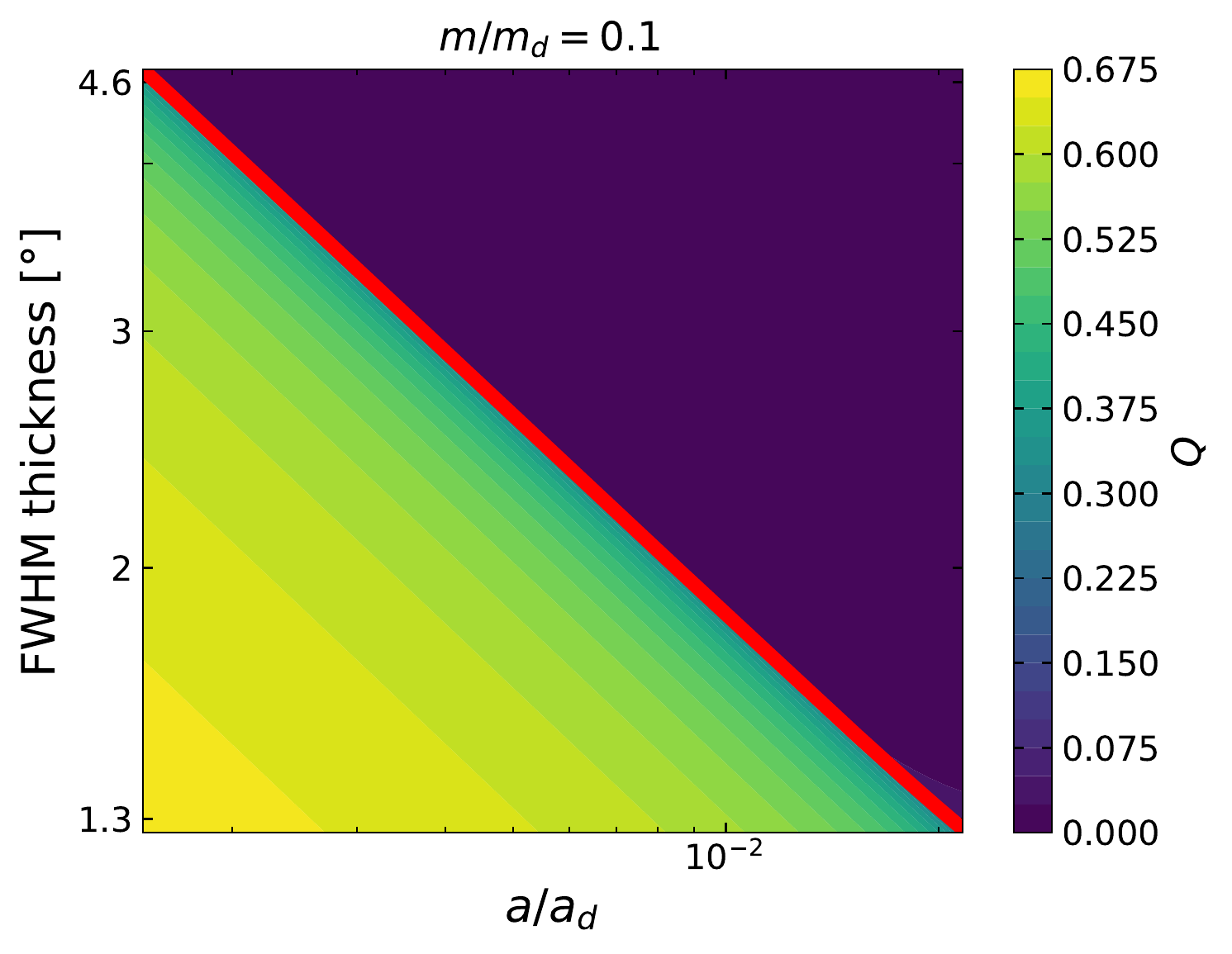}
    \caption{
    \textit{Top panel:} the phase diagram of the inner component for different $a/a_d$, $m/m_d$, and outer component's FWHM  $\theta_{d}$ (Eq.~\ref{eq:FWHM}) and fixed $c_d=5$ and $N/N_d=0.01$. A first-order phase transition takes place 
    for different stellar masses selected uniformly on a log scale (see colorbar). The star symbols mark the critical points where a second-order phase transition takes place. \textit{Bottom panel:} The order parameter for $m/m_d = 0.1$. The color contours show the order parameter $Q$ of the subdominant component for different $a/a_d$ or $\theta_{d}$. The thick red curve highlights the discontinuity in $Q$, which represents the same phase boundary curve as in the top panel for $m/m_d = 0.1$ (i.e. the middle green curve). }
    \label{fig:region_phase_bath}
\end{figure}

The bottom panel in Figure~\ref{fig:Q_F_phase_bath} shows the free energy at different $\Delta \theta_d$. While changing the order parameter of the dominant component the free energy changes in a nonmonotonic way, the free energy along the intervals $A$-$B$, $B$-$C$ and $C$-$D$ is higher than along the equilibria with the same $\Delta \theta_d$ outside of states $A$ and $D$. Consequently, when the dominant component undergoes an increase in $\Delta \theta_d$ near point $A$ in Figure~\ref{fig:Q_F_phase_bath}, instead of moving along the $A$-$B$-$C$-$D$ smooth curve, the system jumps from point $A$ to point $D$ which are at the same temperature and free energy. The first derivative of the free energy with temperature is discontinuous at point $A$, implying that a first-order the phase transition takes place here.
Both the $Q$ curve and $G$ curve have similar features as the canonical ensemble of a one-component system Ref. \cite{Roupas_2017}. 

For a first order phase transition to occur, the equilibria in the $Q-T$ plane must be multi-valued in some temperature range. This can be ensured if $T$ has two extrema with respect to $Q$. Since $T\equiv T(\kappa,\bar{Q})$ for fixed $\{J,J_d,J',\bar{N}\}$ (Eqs.~\ref{eq:kappa_d} and \ref{eq:kappa}), $Q_d$ is approximately constant (Eq.~\ref{eq:Qdisk}) while changing $\kappa_d$ across the phase transition, and $Q=Q(\kappa,c)$ (Eq.~\ref{eq:bath_Q_analy}), 

\begin{equation}
    \label{eq:dT_dQ}
    \left.\frac{\partial T}{\partial Q}\right|_{\kappa,Q_d} = \left.\frac{\partial T}{\partial \kappa}\right|_{Q,Q_d} \left.\frac{\partial \kappa}{\partial Q}\right|_c,
\end{equation}
and $Q$ is a monotonic function of $\kappa$ at fixed $c$, we identify the extrema with $\partial T/\partial \kappa|_{\bar{Q}} = 0$ using Eq.~\eqref{eq:kappa}.
We fix $c_d =5$, $N/N_d =0.01$ and vary the opening angle of the dominant component from $1.3\degree$ to $4.7\degree$ as described earlier. We arbitrarily restrict the range of semimajor axis to $a/a_d \geq 10^{-3}$, where the \ref{i:HBp} conditions are very well satisfied (Figure~\ref{fig:parameter_spcae})\footnote{The blue boundary curves in Figure~\ref{fig:parameter_spcae} for $(\kappa,c_d)=(5,22)$ lie close to the boundary curves for the systems considered here.}. We find that a phase transition takes place only if $a/a_d \ll 1$.

The top panel of Figure~\ref{fig:region_phase_bath} shows the phase diagram with respect to $\bar{a}$ and the FWHM angular thickness of the dominant component, $\Delta \theta_d$ (Eq.~\ref{eq:FWHM}), for 5 fixed values of mass ratios from $m/m_d = 0.2$ to $m/m_d = 0.05$ separated uniformly on a log scale.  
The bodies of the subdominant component condense into a disk phase when $\Delta \theta_{d}$ or $\bar{a}$ are smaller than the phase curves in Figure~\ref{fig:region_phase_bath} and become spherical above the curves for each fixed mass ratio. When the system crosses the phase curves, it undergoes a first-order phase transition with a discontinuous jump in the order parameter $Q$. The critical points in the figure are labelled with star symbols at which the system undergoes a second-order phase transition, with a continuous change in the order parameter $Q$ but a discontinuity in $\partial Q/\partial T|_c$ and in the second derivative of the free energy $\partial ^2G/\partial T^2|_c$. At larger $\bar{a}$ than the critical points of the phase curves, there is no phase transition but a smooth crossover to a disk-like state as seen in Sec.~\ref{sec:mass_ratio}. The critical points for the green, blue and purple curves fall outside of the studied range $\Delta \theta_d\in [1.3\degree, 4.7\degree]$ and $\bar{a} \geq 10^{-3}$ and do not appear in Figure~\ref{fig:region_phase_bath} for this reason. We leave a detailed exploration of the parameter space allowing phase transitions to a follow-up study.

The bottom panel of Figure~\ref{fig:region_phase_bath} shows the order parameter $Q$ (see color-bar) as a function of $(\Delta \theta_d, \bar{a})$ for a fixed mass ratio of $m/m_d = 0.1$. Clearly the order parameter exhibits a discontinuity highlighted by a thick red curve at the same place as in the phase diagram of the top panel. The order parameter $Q$ changes smoothly in regions outside of this line,  it is in the ordered phase $(Q \gtrsim 0.3)$ and in the disordered phase ($Q \lesssim 0.075$) below and above the red line, respectively. When the system crosses the red curve, $Q$ decreases discontinuously between the ordered and disordered phases. This demonstrates a first-order phase transition.
Clearly, the phase transition takes place at a larger thickness hence larger temperature at smaller $a/a_d$ or larger $m/m_d$. This is because $J$ increases with decreasing $a$ or increasing $m$. 

\begin{figure}
    \centering
    \includegraphics[scale = 0.5]{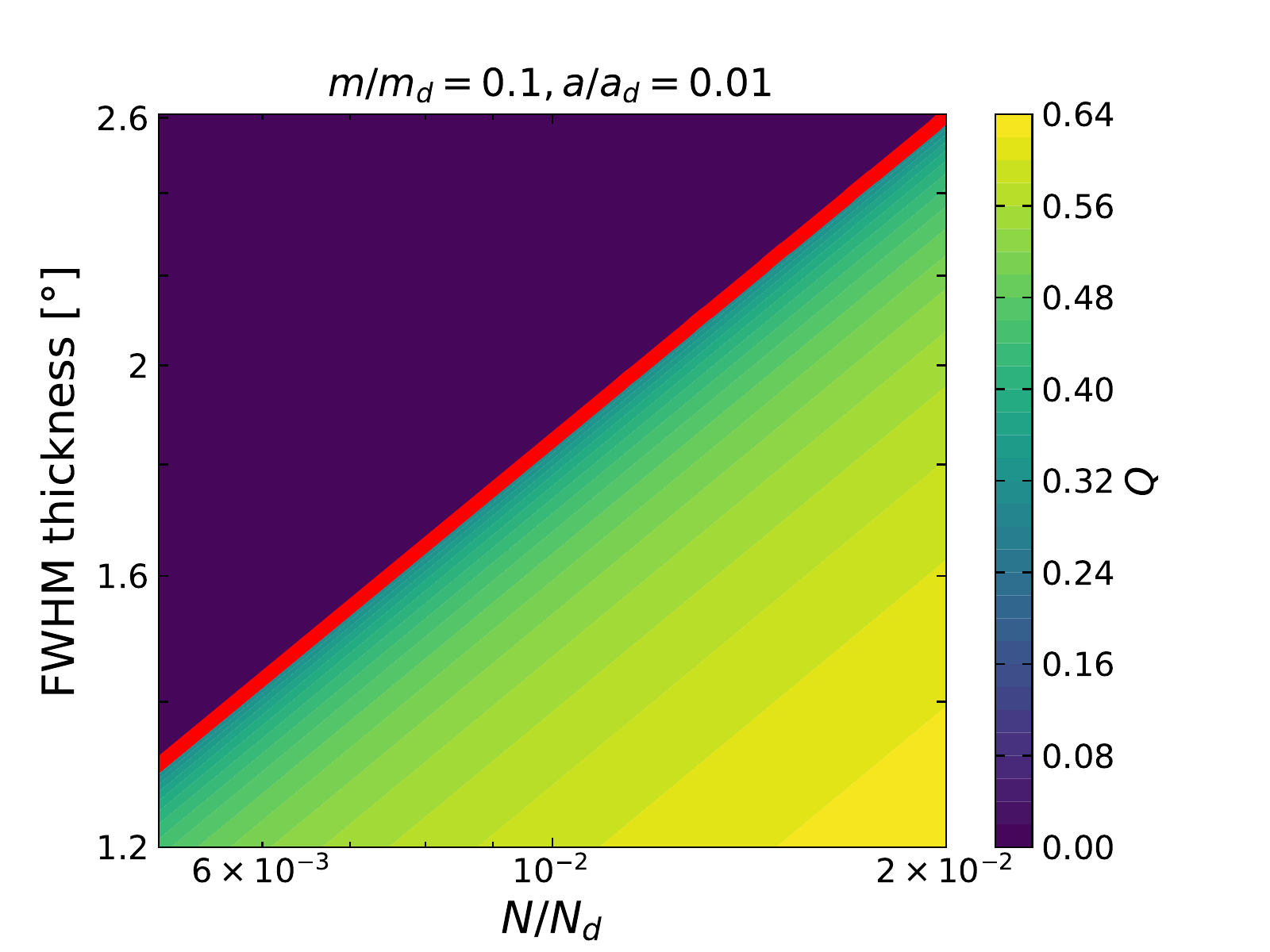}
    \caption{The phase diagram similar to the bottom panel of Figure~\ref{fig:region_phase_bath}, but with fixed $a/a_d = 0.01$, $c_d =5$ and $m/m_d = 0.1$, and varying $N/N_d$ hence $M/M_d$. }
    \label{fig:phase_changeN}
\end{figure}

We find that the phase transition occurs at similar values of the dimensionless temperature 
\begin{align}
\tau &= \frac{kT}{JN} 
= \frac{3}{2} \frac{\bar{a} Q_d(\kappa_d,c_d)}{ \bar{N}\bar{m}^2\kappa_d} \nonumber\\
&= \frac{\bar{a}}{ \bar{N}\bar{m}^2\kappa_d}
\left(1- \frac{3}{2\kappa_d} + \frac{3c_d \tanh c_d - 3 }{4\kappa_d^2}\right)  + \mathcal{O}(\kappa_d^{-4}) \nonumber\\
&\approx \frac{\Delta \theta_d^2\bar{a} }{ (\ln 2)\bar{N}\bar{m}^2} \left(1 + \frac{(c_d-3)\Delta \theta_d^2}{(2\ln 2)}\right) + \mathcal{O}(\Delta\theta_d^{6}),
\end{align}
with $\tau=\tau_{PT} \in [0.148,0.155]$
universally along all phase transition curves with different $\bar{m}$ in the top panel of Figure~\ref{fig:region_phase_bath}. In the second and third lines we used Eqs.~\eqref{eq:Qdisk} and \eqref{eq:FWHM}, respectively.
Thus at the phase transition
\begin{align}\label{eq:a_N_relation_transition}
\left.\frac{\bar{a}}{\bar{N}}\right|_{\rm PT} &= \frac{2}{3}\frac{\bar{m}^2\tau_{\rm PT}\kappa_d}{Q_d(\kappa_d,c_d)}
\approx 
\tau_{\rm PT}\bar{m}^2 \left(\frac{ \ln 2}{\Delta \theta_d^2} - \frac{(c_d-3)}{2}\right). 
\end{align}
The first term dominates as $\Delta \theta_d^2$ approaches zero, explaining the phase curves in Figures~\ref{fig:region_phase_bath} and \ref{fig:phase_changeN}.
The distribution function collapses to a disk as $\bar{a}$ decreases below this critical value or if $\bar{N}$, $\bar{m}$, or $\bar{M}$ increase above the corresponding critical value or if $c_d$ is decreased below a critical value. 

Indeed, Figure~\ref{fig:phase_changeN} shows the phase diagram at $a/a_d =0.01$, $c_d =5$ and $m/m_d =0.1$ while $N/N_d$ is changed from 0.005 to 0.02 for fixed values of the $\Delta \theta_d$ FWHM of the dominant component. In this case $\tau_{PT} \in [0.1496,0.1505]$. Note however that we restricted attention to the region where $\bar{a}$ is in the regime where the self-energy of the outer massive perturber dominates over the interaction energy between the components. We leave a detailed exploration of the full parameter space allowing a phase transition to a future study. 
\subsection{Phase transition in the microcanonical ensemble} \label{sec:two_comp_phase}

In short-range interacting additive systems, the canonical and microcanonical ensembles are asymptotically equivalent for large $N$ implying that phase transitions are possible in both ensembles. A first order phase transition exhibits phase separation, i.e. a mixture of ordered and disordered subsystems whose fraction depends on the energy of the system between that of the ordered and disordered states. However, in many non-additive systems phase separation is prohibited by the large interaction energy between subsystems and phase transitions do not exist in the microcanonical ensemble but they do in the canonical ensemble, which are manifestly different for long-range interacting systems. Indeed, this was confirmed previously for a one-component VRR systems \cite{Roupas_2017,Roupas2020}. 
In a two-component VRR system, the two components can be well-separated radially such that the intercomponent coupling $J'$ becomes arbitrarily small, indicating that a phase transition may be possible even in the microcanonical ensemble. We solve the two-component isolated system with a fixed total angular momentum and examine the order parameters $Q_1$ and $Q_2$ at different values of the total energy. Hence, we examine the components using the microcanonical ensemble as opposed to the canonical ensemble treatment in the heat bath approximation in Sec.~\ref{heat_phase_bath}. 

\begin{figure*}
    \centering
    \includegraphics[scale = 0.5]{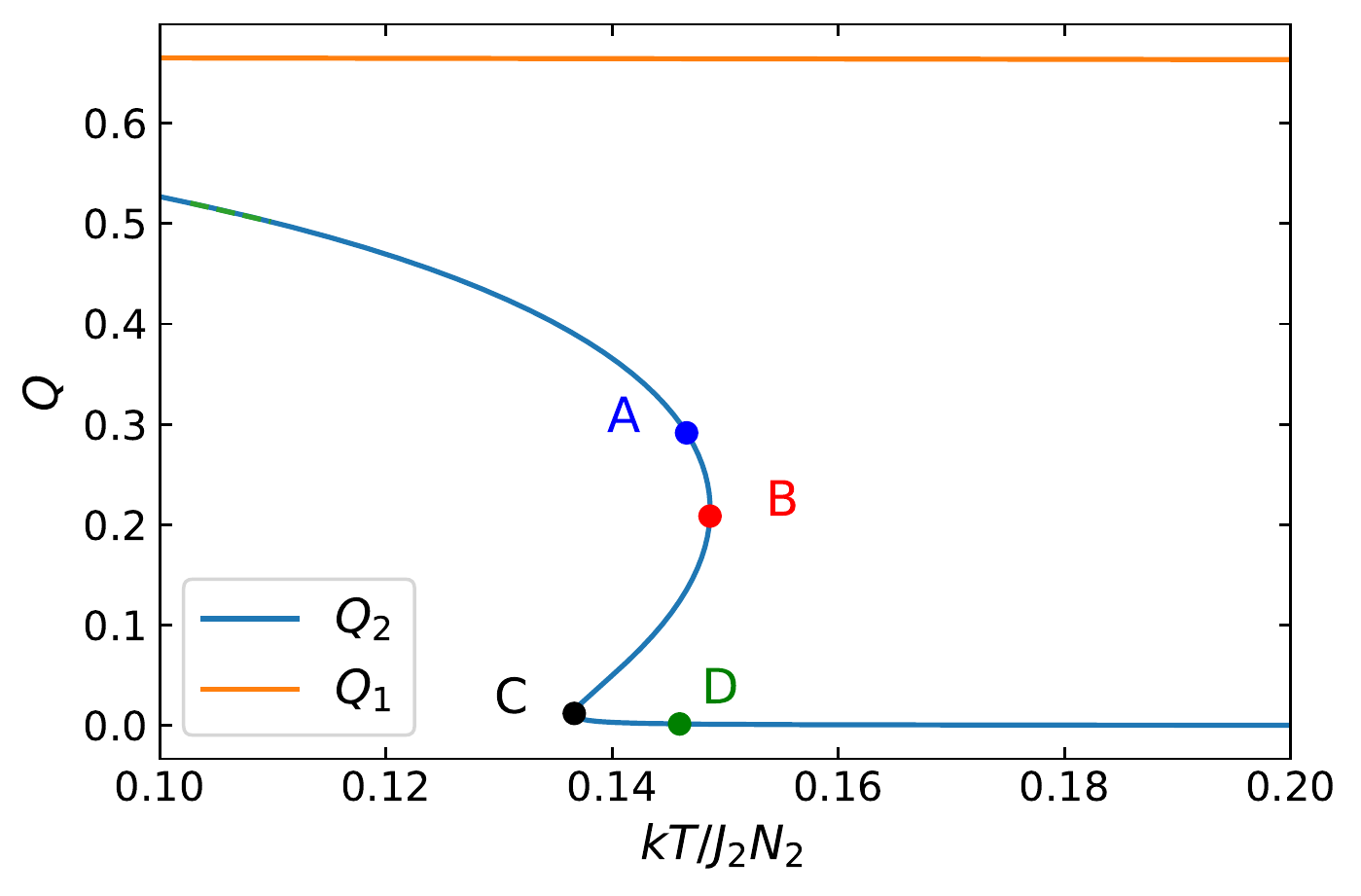}
        \includegraphics[scale=0.5]{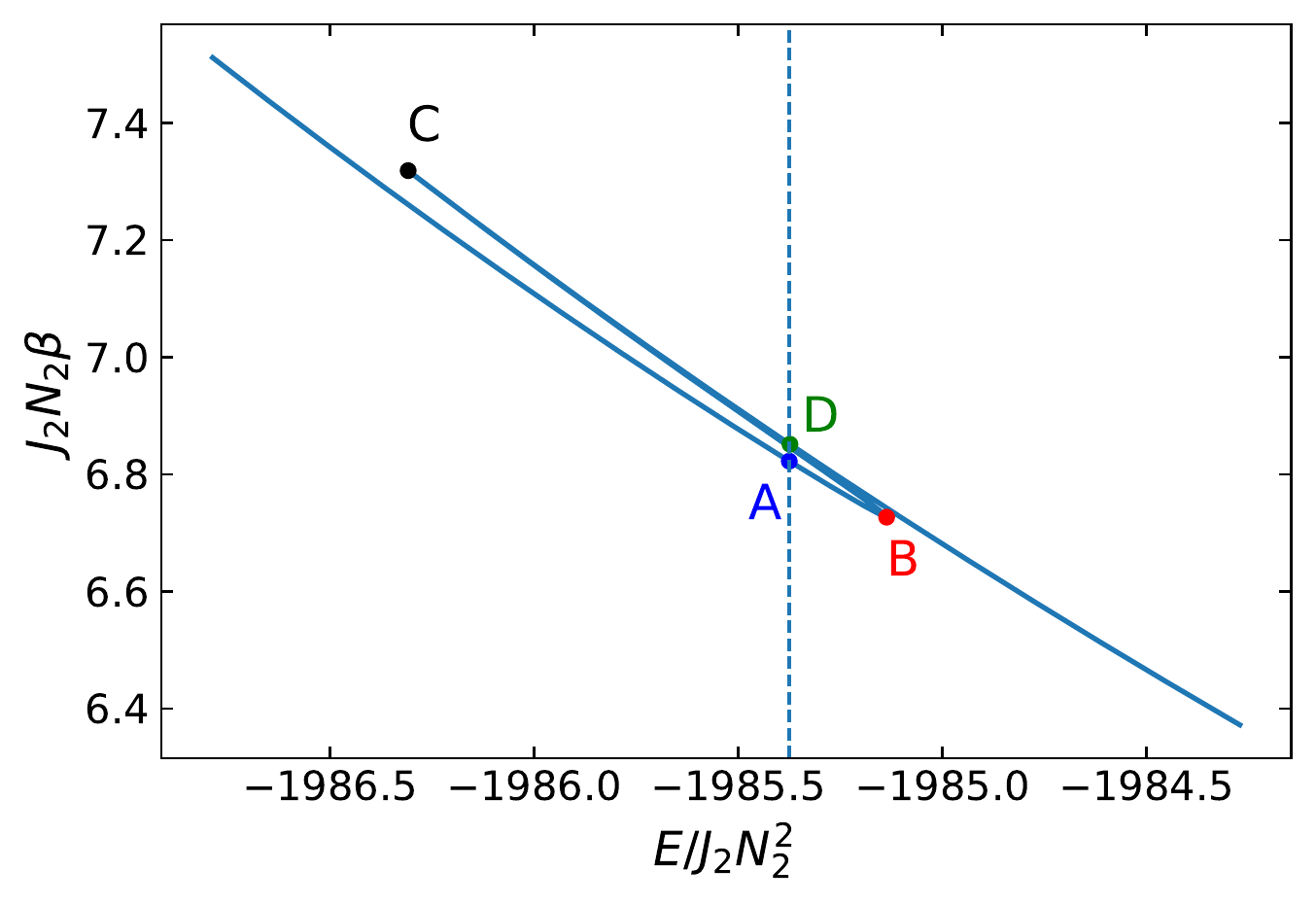}
        \includegraphics[scale=0.5]{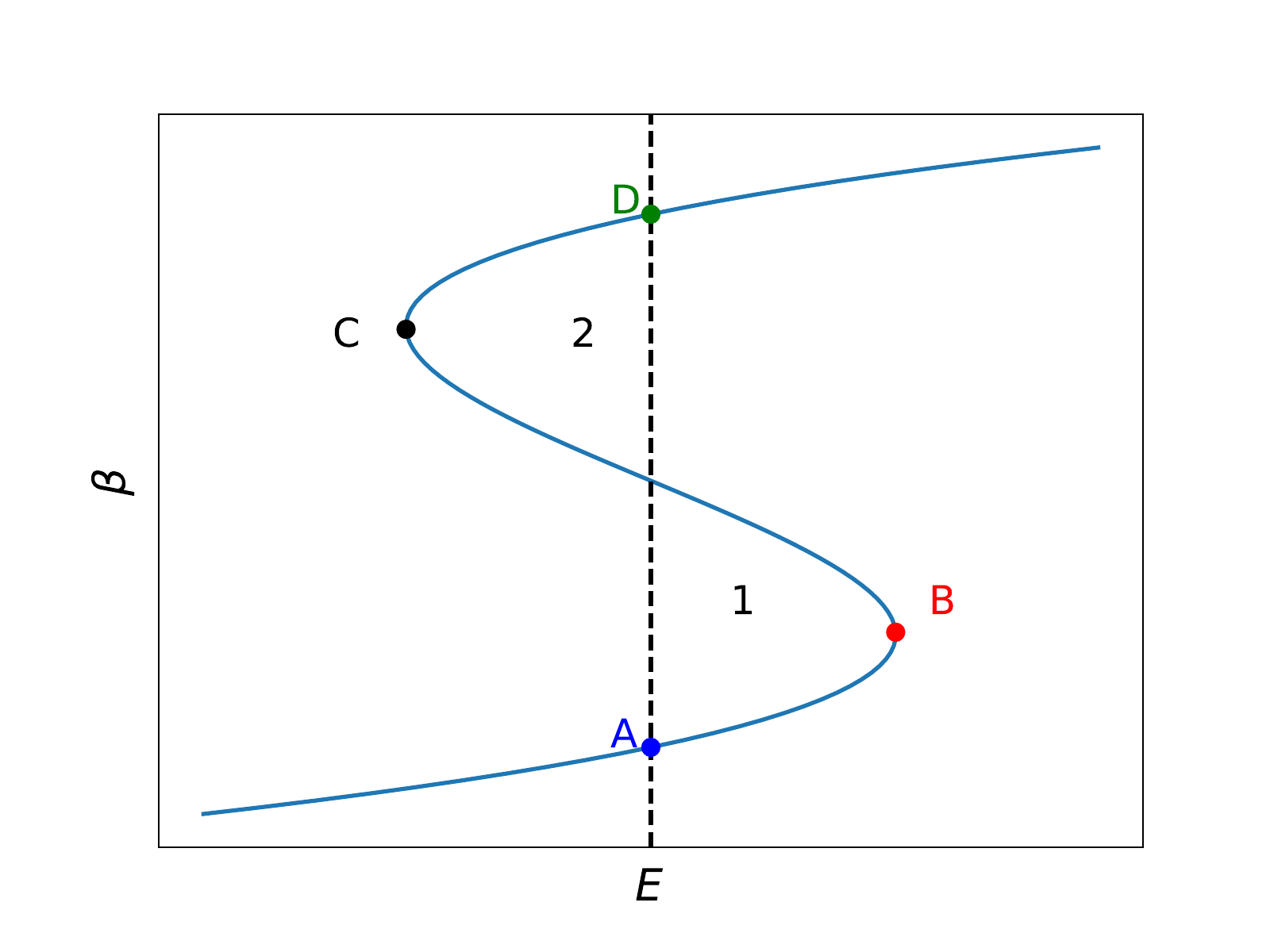}
    \caption{$m_2/m_1 = 0.1$, $N_2/N_1 =0.01$, $a_2/a_1 =0.006$ and $L/N_2l_2 = 760.11$. The top left panel shows the value of $Q_1$ and $Q_2$ at different temperatures in units of $J_2N_2$ of the less massive component, near the phase transition region. The top right panel is the caloric curve around the region of phase transition which shows the value of $\beta$ in units of $(J_2N_2)^{-1}$ at different values of energy in units of $J_2N_2^2$. The bottom panel shows a schematic diagram of the $\beta -E$ graph not drawn to scale  and skewed (the A-B branch is lowered and the C-D branch is raised) for clarity. Phase transition occurs when areas labeled 1 and 2 are equal.} 
    \label{fig:phase_two}
\end{figure*}

As a proof-of-concept we demonstrate the existence of a phase transition for an isolated two-component system with $(m_2/m_1, N_2/N_1, a_2/a_1) = (0.1, 0.01, 0.006)$ and $L/N_2l_2 = 760.11$. This set of parameters produces similar initial conditions as the phase transition example for the canonical ensemble in Figure~\ref{fig:Q_F_phase_bath}. 
The top left panel of Figure~\ref{fig:phase_two} shows the value of $Q_1$ and $Q_2$ at different temperatures. Clearly, the $Q_2$ curve has a similar shape and scale as the top panel of Figure~\ref{fig:Q_F_phase_bath} in Sec.~\ref{sec:phase_transition}, justifying the validity of the heat bath approximation there. 

In the microcanonical esemble, the relevant thermodynamic potential is the total entropy. Phase transition occurs when two distinct states have the same energy and entropy. This is most easily seen from the $\beta-E$ caloric curve in the top right panel of Figure~\ref{fig:phase_two} since $\beta = dS/dE$. Using the Maxwell construction in the microcanonical ensemble described in Refs. \cite{Campa2014} \cite{chavanis_review2006}, the area bounded by $A$, $B$, and the vertical dashed line connecting $A$ and $D$ is equal to the area bounded by $C$, $D$, and the vertical dashed line. Hence, the entropy of point $A$ is equal to that at point $D$. The bottom panel of Figure~\ref{fig:phase_two} gives a schematic plot of the $\beta-E$ curve not drawn to scale. Phase transition occurs when area 1 equals area 2 as labelled. The system undergoes a phase transition from $A$ to $D$ without following the $A$-$B$-$C$-$D$ path, which corresponds to a discontinuity in the order parameter $Q_2$ of the less massive component. There is a temperature jump from $A$ to $D$, as opposed to an energy jump from $A$ to $D$ in the heat bath approximation calculation in Figure~\ref{fig:Q_F_phase_bath}. This is possible due to the ensemble inequivalence for VRR. However, the phase transition takes place at a similar dimensionless temperature $kT/J_2N_2$ (from 0.1466 at A to 0.1460 at $D$) as the canonical ensemble case in Figure~\ref{fig:Q_F_phase_bath} ($kT/JN = 0.1467$ at phase transition). 
The entropy-energy curve 
has a similar shape to the self-gravitating fermions in Figure 9 of Ref. \cite{Chavanis_2002}.

\section{Negative temperature equilibria}\label{sec:neg_temperature}
Negative temperature equilibria are possible if a stable equilibrium state decreases its entropy with increasing energy. Such equilibria are found in other long-range systems such as the 2-dimensional vortices which have deep analogy with self-gravitating systems in their statistical mechanics \cite{chavanis_vortices}. Ref. \cite{Roupas_2017} showed that for a one-component VRR system, negative temperature equilibria are always stable and a larger total angular momentum allows for a larger range of states with negative temperatures. The angular momentum distribution is highly disordered for these states. If $Q>0$ then in this case $\kappa<0$ and $\ln f(s)$ is a concave function of $s$, which may have a maximum at $s<1$, as opposed to positive temperature equilibria for which $\ln f(s)$ is convex in $s$ and has a maximum always at $s=1$.  

\begin{figure}
	\centering 
	\includegraphics[scale=0.5]{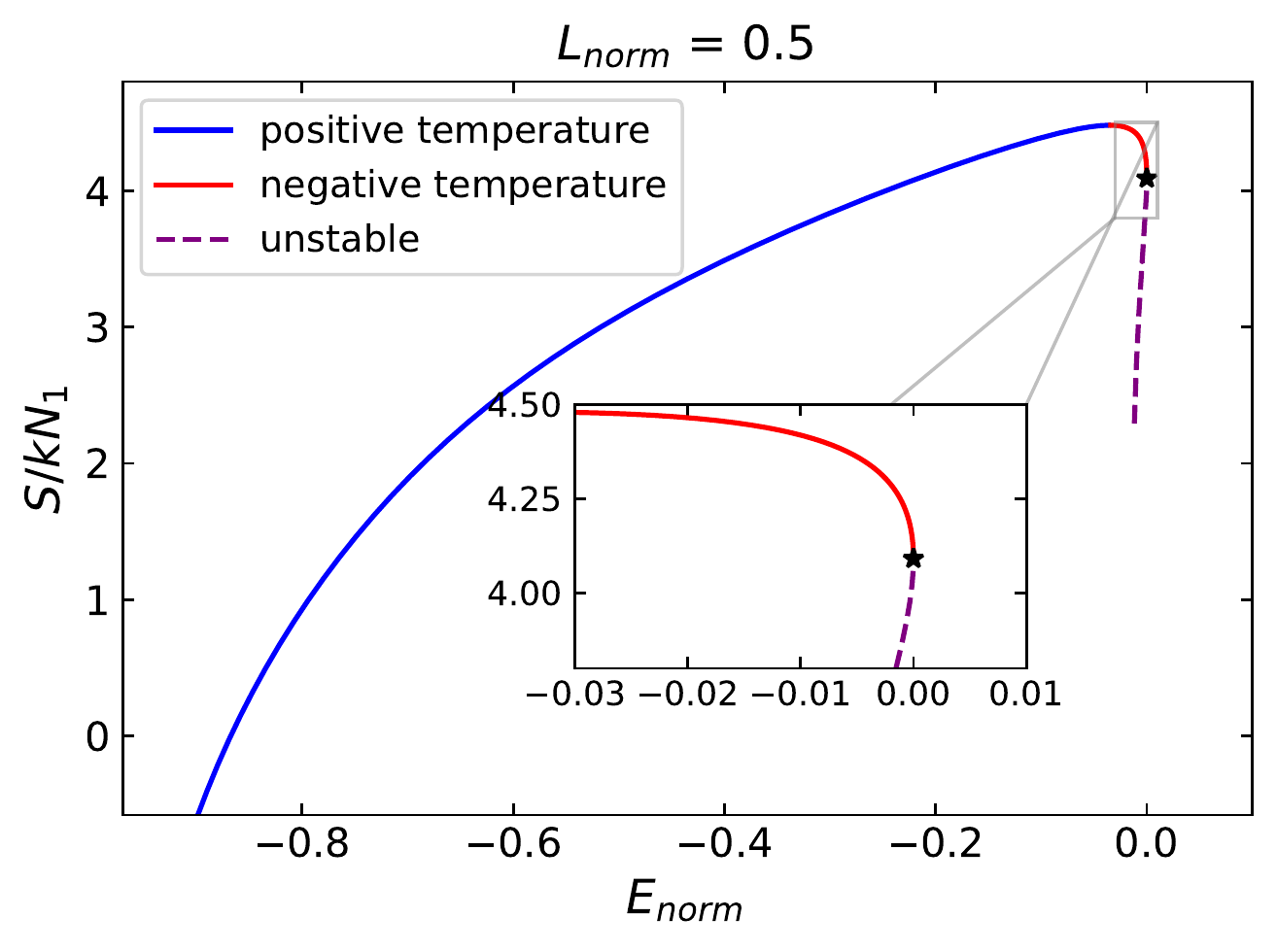}
	\includegraphics[scale=0.5]{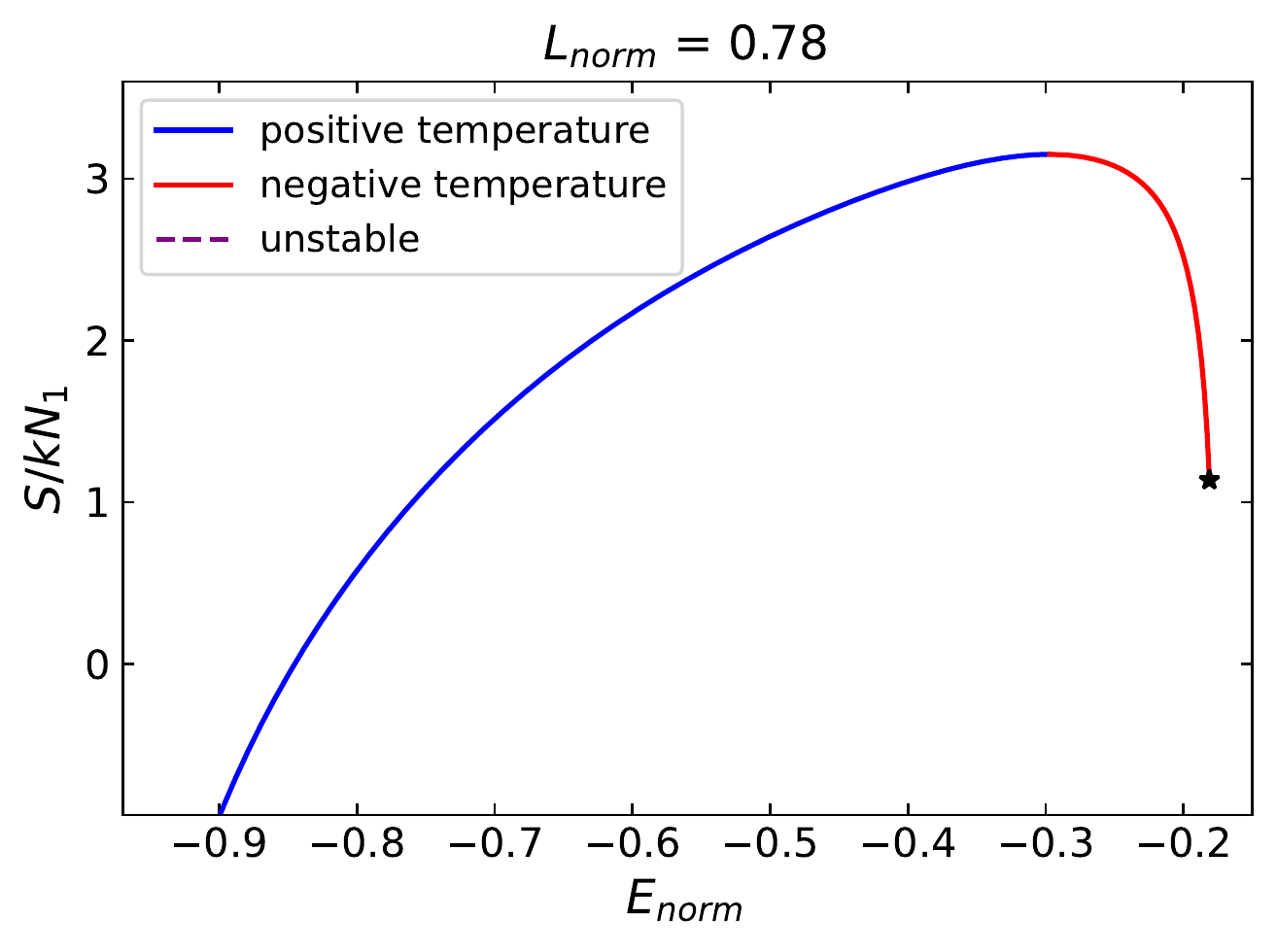}
	\caption{
 The total entropy against the normalised total energy (as defined by Eq.~\ref{eq:E_norm}) of the two-component system for $(m_2/m_1,N_2/N_1,a_2/a_1) = (0.2, 1, 1)$ and $L_{\textrm{norm}} = 0.5$ (top panel) or $L_{\textrm{norm}} = 0.78$ (bottom panel).  The gradient of the curve gives the inverse temperature of the system. The equilibria shown with the blue curves have positive temperatures, while the red curve corresponds to negative temperature equilibria. The purple dotted line represents the unstable equilibria  (cf. $O_2-Q_3$ in Ref. \cite{Roupas_2017} for the one-component case). These states do not exist for $L_{\textrm{norm}}\geq 1/\sqrt{3}$ as in the one component case \cite{Roupas_2017}. The black stars mark the maximum energy state. 
 } \label{fig:neg_T}
\end{figure}

Here we show that two-component VRR systems also exhibit negative temperature equilibria. 
The top panel of Figure~\ref{fig:neg_T} shows the entropy-energy plot for $L_{\textrm{norm}} = 0.5$ and the bottom panel for $L_{\textrm{norm}} =0.78$. The other parameters have values of $m_2/m_1 = 0.2$, $N_2/N_1 = 1$, $a_2/a_1 = 1$. The blue curve has a positive gradient hence corresponds to the positive temperature equilibria. The equilibria with the purple dashed curve also have positive temperatures but with a smaller entropy at the same energy than the states shown with a red curve, purple states are unstable. The red curve has $dS/dE<0$, these are the negative temperature equilibria. 
The particular equilibria presented in this paper have a positive $Q_1$ and either positive or negative $Q_2$. The negative temperature equilibria have $\kappa_{1,2}<0$.  The $L_{\textrm{norm}} = 0.78$ higher angular momentum case in the bottom panel leads to a larger range of energy with negative temperatures (red curve), i.e. for $E_{\textrm{norm}} \in [-0.30,-0.17]$ (as defined by Eq.~\ref{eq:E_norm}). 
The lower total angular momentum case ($L_{\textrm{norm}} = 0.5$) has negative temperature equilibria at $E_{\textrm{norm}} \in [-0.034,0]$. 

The maximum energy state is marked with a black star in Figure~\ref{fig:neg_T}. For the low angular momentum case with $L_{\rm norm} < 1/\sqrt{3}$, the system has attains its maximum energy at $E_{\rm norm}=0$ with $Q_1+\bar{m}Q_2=0$ (see detailed explanations in Appendix~\ref{app:max_E}). The maximum energy state also corresponds to the zero temperature state approaching from $T<0$ \cite{Roupas_2017}. This is possible as $Q_1+\bar{m}Q_2=0$ ensures that $\kappa_{1,2}$ remains finite at infinite $\beta$, see Eqs.~\eqref{eq:kappa_two_comp}--\eqref{eq:kappa_two_comp_2}. At this point the red curve reaches infinite gradient, $kT= 1/\beta = 0^-$ and the system reaches negative zero temperature.\footnote{Note that similar to the one-component system there are possibly three zero temperature states for $L_{\rm norm}\leq 1/\sqrt{3}$. One for the positive temperature stable equilibria, one for the positive temperature unstable states and one for negative temperature equilibria. The first two can be found by extending the blue and purple curves to the minimum energy configuration $dS/dE=\infty$.} For the high angular momentum case with $L_{\rm norm} > 1/\sqrt{3}$, it is not possible to have $Q_1+\bar{m}Q_2=0$ due to Eq.~\eqref{eq:Q_bound}. The maximum energy state has $E<0$ as given by Eq.~\eqref{eq:max_E_highL}, consistent with the numerical solution shown in Figure~\ref{fig:neg_T}. In this case the black star indicating the maximum energy state in the bottom panel of Figure~\ref{fig:neg_T} has a finite nonzero negative temperature, less negative temperature equilibria (closer to zero) do not exist.

\section{Conclusion}\label{sec:conclu}
We examined the statistical mechanics of VRR, the dominant gravitational mechanism that determines the direction of angular momentum vectors of objects orbiting in a spherical background potential. This is a long-range interacting system where the subsystems' energies are not additive due to the significant interaction energy between the subsystems. We extended the mean-field theory of a one-component system in Ref. \cite{Roupas_2017} to an isolated two-component system of bodies orbiting a supermassive black hole. We used the principle of maximising Boltzmann entropy at fixed total energy and total angular momentum. The system admits an implicit analytical solution using the one-component partition function in case where one of the two components dominates the energy and angular momentum budget and serves as an effective heat bath for the subdominant component. However we have shown that the canonical ensemble of the subdominant component that arises due to the energy exchange with the heat bath is different from the canonical ensemble of an additive one-component system due to the interaction energy with the heat bath. 

We found that the distribution exhibits the so-called vertical mass segregation effect (i.e. in the direction perpendicular to a disk, as opposed to the more commonly studied radial mass segregation) in both the heat bath approximation case and the general two-component calculation unless the component comprised of the heavy objects is spherically distributed and dominates the total mass of the cluster, consistently with previous numerical studies of multi-component system  \cite{Szolgyen_Kocsis2018,magnan2021orbital,Mathe:2022azz,taras_2022_discs}. We have shown that for comparable semimajor axis the interaction among the subdominant component may be negligible if their total mass is much smaller than that of the dominant component. These subsystems relax independently, and the results of these two component models may be superposed to find the equilibria of multicomponent systems.

We explored the parameter space of energy and angular momentum and found evidence of vertical mass segregation in all regions of the parameter space. When one component strongly drives the evolution of the subdominant components, the transition from a spherical disordered state to a flattened ordered state is continuous as a function of stellar mass, semimajor axis, eccentricity, and net angular momentum. We determined the mass beyond which the objects settle into a disk (Eq.~\ref{eq:alpha_m_at_disk}).
Asymptotically for large semimajor axis, disk formation depends only on the relative angular momentum of the two components and the angular momentum of the dominant component but it is otherwise independent of the thickness of the spatial distribution of the dominant component. In the case of very nearly isotropic initial conditions, this critical mass may be very large. This helps to explain the lack of vertical mass segregation signatures in spherically dominated systems found in recent direct N-body simulations Ref. \cite{taras_2022_discs} while Ref. \cite{Mathe:2022azz} found vertical mass segregation for a wider range of initial conditions. We have shown that the distribution may become flattened even for highly spherically dominanted systems, albeit only for very large individual object masses or for much larger/smaller semimajor axes relative to that of the dominant component  (Figures~\ref{fig:Q_m_bath_sphere} and \ref{fig:region_phase_bath}). If the mass distribution does not extend to such high masses or the radial distribution does not extend to sufficiently low or high orbital radii, the system may not exhibit the disk phase within the spherical phase. 

The analytical heat bath models give insight into the response of the stellar system to a massive perturber such as a gaseous circumnuclear disk or a population of IMBHs which may represent the heat bath. The mass and semimajor axis dependence of the stellar angular momentum vector distribution can help to determine the features of these massive perturbers from the observations of the stellar orbits. 

Another interesting aspect of this work is the study of phase transitions of isolated long-range interacting systems or systems influenced by an outer massive perturber. During a phase transition, the angular momentum vector distribution undergoes a discontinuous change when the system is subject to a small change in its system parameters, leading to a change in the temperature or total energy.
We find that a phase transition is not possible if a dominant stellar component drives the evolution of a subdominant component in the test particle (no self-gravity) approximation to VRR. In this case the subdominant component responds continuously to changes in the properties of the dominant component. However an abrupt first order phase transition is observed when the self-gravity of the subdominant component is non-negligible, i.e. when it is perturbed by a distant outer massive perturber (below red line in Figure~\ref{fig:parameter_spcae}). In this case, changes in the thickness of the dominant component or the relative semimajor axis, mass, and number of objects induce a discontinuous change in the statistical equilibrium distribution of angular momentum vectors 
(Figures~\ref{fig:Q_F_phase_bath}, \ref{fig:region_phase_bath}, \ref{fig:phase_changeN}, \ref{fig:phase_two})
similar to the nematic-isotropic phase transition of liquid crystals.

We also examined the microcanonical ensemble for the general two-component calculation. While Refs. \cite{Roupas_2017} and \cite{magnan2021orbital} did not observe phase transitions in the micro-canonical ensemble for the studied one-component and multi-component systems, we find evidence for the possibility of phase transitions when the two components have a very different total mass (i.e. $\bar{M} \sim 10^{-3}$) and there is a large radial gap between the inner and outer stellar components. The distribution of the less massive component around the phase transition is similar to the result evaluated with the heat bath approximation, but the exact details of the transition such as the change in energy are slightly different due to ensemble inequivalence.

We have also shown that two component systems exhibit negative absolute temperature equilibria as found previously for one-component \cite{Roupas_2017} systems and multi-component simulations \cite{magnan2021orbital} at highly isotropic energies. These negative temperature equilibria are disordered and spherical, similar to the high positive temperature equilibria. However, at negative temperature equilibria, the system has population inversion: the more energetic microstates are relatively more populated and the angular momentum distribution function's logarithm is concave \cite{Roupas_2017}. 

Having presented cases of astrophysical systems with vertical mass segregation and phase transition for circular orbits, we plan to extend the model to less idealized assumptions in the future. Importantly we resticted attention to axisymmetric configurations, which are incompatible with bending waves that are expected to be prominent in the thin disk limit \cite{kocsis2011,Batygin18}. Further, we assumed bodies on circular orbits around the central point mass. Eccentric orbits satisfy the same self-consistency equations in equilibrium as in Eqs.~\eqref{eq:Q1}-\eqref{eq:Q2} but the coupling constants $J,J_d,J'$ have algebraically slightly more complicated forms that depend on both the semi-major axes and eccentricities, and whether the orbits are radially overlapping or non-overlapping \cite{Kocsis_Tremaine2015}. The VRR equilibria and possibility of a phase transition can be straightforwardly obtained for eccentric orbits by generalising the circular case. Indeed the statistical equilibria depend only on the underlying coupling constants $\{J,J_d,J'\}$ and angular momenta $\{l,l_d\}$, in particular $\{ J/J_d, J'/J_d, l/l_d\}$. We will also explore the possibility of phase co-existence or phase separation in two component VRR systems similar to ice and water in the orbits of stars and black holes in the galactic center.

\acknowledgments
We thank Jean-Baptiste Fouvry for useful discussions. This work received funding from the European Research Council (ERC) under the European Union’s Horizon 2020 Programme for Research
and Innovation ERC-2014-STG under grant agreement No. 638435 (GalNUC). This work was supported by the Science and Technology Facilities Council Grant Number ST/W000903/1. This work received funding from Department of Physics, Astrophysics, University of Oxford.

\appendix
\section{Analytical solution of the general two-component system}\label{app:analytic}
\subsection{The case of $\bar{a}=1$}
For $\bar{a}=1$, the parameters defining the distributions of the two components (Eqs.~\ref{eq:ccirc} and \ref{eq:kappaexact}) simplify to
\begin{align}
\kappa_1 &= \frac{3}{2}\beta J_1N_1(Q_1+\bar{M} Q_2)\,,\quad
\label{eq:kappa_two_comp}
c_1 = l_1\gamma\\
    \kappa_2 &= \bar{m} \kappa_1, \quad
    c_2 = \bar{m} c_1\label{eq:kappa_two_comp_2},
\end{align}
where $\bar{m}=m_2/m_1$ and $\bar{M}=M_2/M_1 = N_2m_2/(N_1m_1)$. 
The one-particle generating functions of the two components are defined as in Eq.~\eqref{eq:bath_Z}.
Eqs.~\eqref{eq:Q1}--\eqref{eq:Ltot} may be written as 
\begin{align}
 Q_1&\equiv Q(\kappa_1,c_1) = \left.\frac{\partial \ln Z_0(\kappa_1,c_1)}{\partial \kappa_1}\right|_{c_1},\label{eq:Q_two_comp}\\
 Q_2 &\equiv Q(\kappa_2,c_2) = \left.\frac{\partial \ln Z_0(\kappa_2,c_2)}{\partial \kappa_2}\right|_{c_2},\label{eq:Q2_two_comp} \\
\frac{L}{N_1l_1} &= \frac{\partial\ln  Z_0(\kappa_1,c_1)}{\partial c_1} 
+\bar{M} \left.\frac{\partial\ln Z_0(\kappa_2,c_2)}{\partial c_2}\right|_{\shortstack{$\kappa_2=\bar{m}\kappa_1$\\$c_2=\bar{m}c_1$}}.
\label{eq:L_two_comp}
\end{align}
The derivatives appearing on the right hand side simplify analytically using Eqs.~\ref{eq:bath_L_analy} and \ref{eq:bath_Q_analy}.
From Eq.~\eqref{eq:kappa_two_comp} 
\begin{equation}
    \label{eq:T_two_comp}
\frac{kT}{J_1N_1} = \frac{3}{2\kappa_1}\left[Q(\kappa_1,c_1)+\bar{M}Q(\kappa_2,c_2)\right].
\end{equation}
The entropy is given by 
\begin{align}
    \label{eq:S_two_comp}
\frac{S}{kN_1} =& - \kappa_1 (Q_1+\bar{M}Q_2) - \frac{L c_1}{N_1l_1}+\ln Z_1 + \bar{N}\ln Z_2.
\end{align}
We select a value of $\kappa_1$ and use Eq.~\eqref{eq:L_two_comp}  to numerically solve for $c_{1}$ for given $L$ total angular momentum, given Eq.~\eqref{eq:kappa_two_comp_2} which specifies $\kappa_2$ and $c_2$ for any $(\kappa_1,c_1)$. Here $L\equiv L(\kappa_1,\kappa_2,c_1,c_2)$ is a strictly monotonically increasing function of $c_1$ for any given $\kappa_1$ if setting $\kappa_2=\bar{m}\kappa_1$ and $c_2=\bar{l}c_1$.
Thus for fixed $L$, we have the functions $\{\kappa_2(\kappa_1),c_1(\kappa_1),c_2(\kappa_1)\}$ at our disposal which we can tabulate by letting $\kappa_1$ span all possible values between $\pm\infty$. The corresponding values of $Q_1$, $Q_2$, $T$, $S$, and $E$ then follow immediately by substituting into Eq.~\eqref{eq:Q_two_comp}, \eqref{eq:Q2_two_comp}, \eqref{eq:T_two_comp}, \eqref{eq:S_two_comp}, and \eqref{eq:Etot}. Thus, we obtain a parametric solution for the equilibria in the planes of $(T,Q_1)$, $(T,Q_2)$, and $(E,S)$ in this way parameterised by $\kappa_1$.

\subsection{The case of arbitrary $\bar{a}$}
The nonlinear mean-field selfconsistency equations (Eqs.~\ref{eq:ccirc} and \ref{eq:kappaexact}) may only be partially decoupled in the most general two-component case with different semimajor axes, where
\begin{align}\label{eq:kappa1_ex}
\kappa_1 &= \bar{\beta}(Q_1+\bar{J}' \bar{N} Q_2)\,,\quad
c_1 = l_1\gamma\\
    \kappa_2 &= \bar{\beta}(\bar{J} \bar{N} Q_2+ \bar{J}' Q_1), \quad
    c_2 = \bar{l} c_1,\label{eq:kappa2_ex}\\
    L_{\rm norm} &= \frac{\bar{L}_1+\bar{N}\bar{l}\bar{L}_2}{1+\bar{N}\bar{l}} 
    \label{eq:L_ex}\\
    E_{\rm norm} &= -\frac{9}{4}\frac{Q_1^2 + \bar{J}\bar{N}^2Q_2^2 +2\bar{J}'\bar{N}Q_1 Q_2 }{1+\bar{J}\bar{N}^2+2\bar{J}'\bar{N}}
    \label{eq:E_ex}    
\end{align}
where $\bar{\beta} = \frac{3}{2}\beta J_1 N_1$, $\bar{J}'=J'/J_1$, $\bar{J}=J_2/J_1$, $Q_1\equiv Q(\kappa_1,c_1)$, $Q_2\equiv Q(\kappa_2,c_2)$, $\bar{L}_1\equiv \bar{L}(\kappa_1,c_1)$, $\bar{L}_2\equiv \bar{L}(\kappa_2,c_2)$, where the dimensionless functions $Q(\kappa,c)$ and $\bar{L}(\kappa,c)$ are given explicitly by Eqs.~\eqref{eq:bath_Q_analy} and \eqref{eq:bath_L_analy}. Given that these functions are nonlinear, the system of equations may have several solutions, and it is useful to decouple the equations in terms of $\kappa_1$, $\kappa_2$, $c_1$, and $c_2$ analytically as follows.

\begin{enumerate}
    \item Start with some given $\beta$.  Solve Eq.~\eqref{eq:kappa1_ex} for $Q_2$, express it as a linear combination of $\kappa_1$ and $Q_1$ for any given $\beta$,
    \begin{equation}
            Q_2 = \frac{1}{\bar{J}'\bar{N}}\left(\frac{\kappa_1}{\bar{\beta}}-Q_1\right).     \label{eq:Q2_in_Q1}
    \end{equation}
    \item Substitute the result $Q_2(\kappa_1,Q_1,\beta)$ in Eq.~\eqref{eq:kappa2_ex} to obtain an expression for $\kappa_2$ as a linear combination of $\kappa_1$ and $Q_1=Q(\kappa_1,c_1)$, 
    \begin{equation}
            \kappa_2 = \frac{\bar{J}^{1/2}}{j}\kappa_1
    - \bar{\beta}\bar{J}^{1/2}(j^{-1}-j)Q(\kappa_1,c_1).
    \label{eq:k2_in_k1Q1beta} 
    \end{equation}
    Here we have introduced the dimensionless parameter
$j=J'/\sqrt{J_1J_2}=\bar{J}'/\bar{J}^{1/2} = \min(\bar{a}^{2.5},\bar{a}^{-2.5})$, which satisfies $0< j\leq 1$. 
    Furthermore $c_2=\bar{l} c_1$ and  $\kappa_2(\kappa_1,c_1,\beta)$ may be substituted back into Eq.~\eqref{eq:kappa1_ex} to obtain an expression between $\kappa_1$ and $c_1$ completely independent of the other component.
\begin{align}
     \frac{\kappa_1}{\bar{\beta}} =&  j\bar{J}^{1/2}\bar{N}
     Q\left(\frac{\bar{J}^{1/2}}{j}\kappa_1
    - \bar{\beta}\bar{J}^{1/2}(j^{-1}-j)Q(\kappa_1,c_1), \bar{l}c_1\right)
     \notag \\
    &+ Q(\kappa_1,c_1). \label{eq:k1_c1_alone}
\end{align}
\item For each value of $\bar{\beta}$, we create a table of the whole range of the $(\kappa_1,c_1)$ values, and compute the convergence of Eqs.~\eqref{eq:L_ex} and \eqref{eq:k1_c1_alone}:
\begin{align}
&G_1(\kappa_1,c_1) =   L_{\rm norm} -\frac{\bar{L}(\kappa_1,c_1)+\bar{N}\bar{l}\bar{L}(\kappa_2(\kappa_1,c_1,\beta),\bar{l}c_1)}{1+\bar{N}\bar{l}}, \label{eq:analy_L_converg} \\
    &G_2(\kappa_1,c_1) = \frac{\kappa_1}{\bar{\beta}}- Q(\kappa_1,c_1) \notag \\
    &-j\bar{J}^{1/2}\bar{N}
     Q\left(\frac{\bar{J}^{1/2}}{j}\kappa_1
    - \bar{\beta}\bar{J}^{1/2}(j^{-1}-j)Q(\kappa_1,c_1), \bar{l}c_1\right).
\end{align}
We plot the contours of $G_1$ and $G_2$ with respect to $(\kappa_1,c_1)$ and find the intersection points of $G_1(\kappa_1,c_1)=G_2(\kappa_1,c_1)=0$. We find one or three intersection points depending on the value of $\bar{\beta}$, hence solutions of $(\kappa_1,c_1)$.
\item We substitute the solution of $(\kappa_1,c_1)$ into Eq.~\eqref{eq:k2_in_k1Q1beta} to obtain $\kappa_2$ and $c_2 = \bar{l} c_1$. We then compute the value of $Q_1,Q_2$ and the total energy using Eq.~\eqref{eq:E_ex}. By calculating the solutions of $(\kappa_1,\kappa_2,c_1,c_2)$ over the allowed domain of $-\infty<\bar{\beta}<\infty$, we obtain the equilibrium solutions for the complete range of energy values.    
\end{enumerate}

The equilibria are globally stable if the entropy is maximized at given $(E_{\rm norm},L_{\rm norm})$, where
\begin{align}
\frac{S}{kN_1} =& -Q_1\kappa_1-\bar{N}Q_2\kappa_2 - (1+\bar{N}\bar{l})L_{\rm norm} c_1
\notag\\&+\ln Z_0(\kappa_1,c_1) + \bar{N}\ln Z_0(\kappa_2,c_2).
\end{align}

\subsection{Numerical method}
When the semimajor axes of the two components are different, Eqs.~\eqref{eq:Q1}, \eqref{eq:Q2} and \eqref{eq:Ltot} can also be solved numerically using Newton's method. This is done by looking for the zeros of the three functions $F_i(\bm{X})$ of the unknowns $\bm{X}=(Q_1,Q_2,\gamma)$ iteratively for a given $\beta$:
\begin{align}
F_1 &= Q_1 -\frac{\int_{-1}^{1} (s^2 -\frac{1}{3}) e^{\frac{3}{2}\beta(J_1N_1Q_1+J^\prime N_2 Q_2)s^2+l_1\gamma s}d\textrm{s}}{\int_{-1}^{1}  e^{\frac{3}{2}\beta(J_1N_1Q_1+J^\prime N_2 Q_2)s^2+l_1\gamma s}d\textrm{s}}, \label{eq:consistent_1} \\
F_2 &= Q_2 - \frac{\int_{-1}^{1} (s^2 -\frac{1}{3}) e^{\frac{3}{2}\beta(J_2 N_2 Q_2+J^\prime N_1Q_1)s^2+l_2\gamma s}d\textrm{s}}{\int_{-1}^{1} e^{\frac{3}{2}\beta(J_2 N_2 Q_2+J^\prime N_1Q_1)s^2+l_2\gamma s}d\textrm{s}},  \label{eq:consistent_2} \\
F_3 &= \frac{L}{N_1l_1} - \frac{\int_{-1}^{1} s e^{\frac{3}{2}\beta(J_1N_1Q_1+J^\prime N_2 Q_2)s^2+l_1\gamma s}d\textrm{s}}{\int_{-1}^{1} e^{\frac{3}{2}\beta(J_1N_1Q_1+J^\prime N_2 Q_2)s^2+l_1\gamma s}d\textrm{s}} \notag \\ 
-&\bar{l}\bar{N}\frac{\int_{-1}^{1} s e^{\frac{3}{2}\beta(J_2 N_2 Q_2+J^\prime N_1Q_1)s^2+l_2\gamma s}d\textrm{s}}{\int_{-1}^{1}  e^{\frac{3}{2}\beta(J_2 N_2 Q_2+J^\prime N_1Q_1)s^2+l_2\gamma s}d\textrm{s}}.     \label{eq:consistent_3}
\end{align}
The $p+1$ iteration value of $X_i$ is given by \begin{equation}
    \label{eq:newton}
X_i[p+1] = X_i[p] - \sum_{j}  (\mathbf{M}^{-1})_{ij}F_j(\bm{X}[p]),
\end{equation}
where $M_{ij}=\partial F_i(\bm{X})/\partial X_j$ evaluated at $\bm{X}[p]$ and $\mathbf{M}^{-1}$ denotes the inverse matrix. The iteration is stopped when $F_i = 0$ within a tolerance of $10^{-9}$. There may be more than one solution for some values of $\beta$ as discussed in the main text which leads to the possibility of a phase transition. In the multi-valued $T$ region for given $\kappa_1$, we initialize the iteration with three different $X_i$ to obtain the three different solutions. To find the solution at a given energy, we scan through a range of $T$ values, obtain the solutions of $X_i$ and select the one with the correct energy. Alternatively we may increase the parameters $\bm{X}=(Q_1,Q_2,\beta,\gamma)$ and add a fourth equation $F_4(\bm{X})$ of the energy constraint in the Newton's method.

\section{Maximum energy of two-component system} \label{app:max_E}

In the axisymmetric two-component system studied in this paper, the total energy is given by Eq.~\eqref{eq:Etot}. For radially nonoverlapping circular components $J'< \sqrt{J_1J_2}$, and the local extrema of $E$ with respect to $Q_1$ and $Q_2$ can be found by setting $\partial E/\partial Q_1 = \partial E/\partial Q_2 = 0$, which gives $Q_1 = Q_2 = 0$. The second derivative is negative definite, showing that this is a maximum point of $E$. 

For radially overlapping components on circular orbits, $\bar{a}=1$ and $J'^2=J_1J_2$, implying that
\begin{align}
    \label{eq:E_max}
  E=& -\frac34 J_1N_1^2(Q_1+\bar{M} Q_2)^2
\end{align}
where the maximum energy $E=0$ is attained if 
\begin{equation}
    \label{eq:Emax_condition}
    Q_1=-\bar{M} Q_2.
\end{equation}
For non-zero $Q_{1,2}$, this requires exactly one of $Q_{1,2}$ to take negative values, while the other to take a positive value. Generally the bounds on $Q_{1,2}$ are limited by the angular momentum of each component \cite{Roupas_2017}:
\begin{equation}
    \label{eq:Q_bound}
-\frac13 + \sa_{1,2}^2\leq Q_{1,2}\leq \frac 23,
\end{equation}
where $\sa_{1,2} = L_{1,2}/N_{1,2}l_{1,2}$. The total angular momentum constraint can be written in terms of the $L_{\rm norm}$ (defined in Eq.~\ref{eq:L_norm}) as 
\begin{equation}
    \label{eq:s2_s1}
    \sa_1 = L_{\rm norm} (1+\bar{M})-\bar{M}\sa_2.
\end{equation}

For $L_{\rm norm} \leq 1/\sqrt{3}$, it is possible to arrange the angular momenta such that both $\sa_{1,2} \leq 1/\sqrt{3}$. Hence both $Q_1$ and $Q_2$ can take negative values and the system can reach $E=0$ via satisfying Eq.~\eqref{eq:Emax_condition}.

Conversely, if $L_{\rm norm} > 1/\sqrt{3}$, we cannot have both $\sa_{1,2} \leq 1/\sqrt{3}$. But to reach $E=0$ Eqs.~\eqref{eq:Emax_condition} and \ref{eq:Q_bound} require that one of the components must satisfy $Q_{1,2}\leq 0$ and hence  $\sa_{1,2}\leq 1/\sqrt{3}$. Without loss of generality assume that $Q_1> 0$ and $Q_2\leq 0$, so that $\sa_1 > 1/\sqrt{3}$ and 
\begin{equation}
    \label{eq:Q_neg}
\sa_2 \leq \frac{1}{\sqrt{3}}.
\end{equation}
For Eq.~\eqref{eq:Emax_condition} to hold, we require 
\begin{equation}
    \label{eq:Q1_Q2_below_zero}
    0 = Q_1 + \bar{M} Q_2 \geq \min(Q_1)+\bar{M}\min(Q_2)
\end{equation}
where
\begin{equation}
    \label{eq:Q1Q2min}
    \rm min (Q_{1,2}) = -\frac 13 +\sa_{1,2} ^2.
\end{equation}
 Substituting Eqs.~\eqref{eq:s2_s1} and Eqs.~\eqref{eq:Q1Q2min}, the inequality \eqref{eq:Q1_Q2_below_zero} may be solved in terms of the variables $(L_{\rm norm},\bar{M})$ subject to the constraint of the inequality \eqref{eq:Q_neg}. The solution is $L_{\rm norm} \leq 1\sqrt{3}$ irrespective of the ratio of $\bar{M}$. This contradicts the fact that $L_{\rm norm} > 1/\sqrt{3}$. Hence $L_{\rm norm} \leq 1/\sqrt{3}$ is a necessary and sufficient condition for the maximum energy to reach $E=0$.
 
For $L_{\rm norm}\geq 1/\sqrt{3}$, Eq.~\eqref{eq:Emax_condition} cannot hold since both $Q_1$ and $Q_2$ can only take positive values. Substituting Eqs.~\eqref{eq:Q1Q2min} and \eqref{eq:s2_s1} into Eq.~\eqref{eq:E_max}, the energy may be expressed as a function of $(\bar{M},L_{\rm norm},\sa_2)$. Taking the derivative of $E$ with respect to $\langle s\rangle_2$ gives the maximum energy at a fixed set of $(\bar{M},L_{\rm norm})$:
\begin{align}
    &\frac{dE}{d\sa_2} = \bar{M}(1+\bar{M})^2(L_{\rm norm} - \sa_2) \notag \\
    & \left(   -1+3L_{\rm norm}^2(1+\bar{M})-6L_{\rm norm} \bar{M} \sa_2 + 3\bar{M} \sa_2^2     \right) \label{eq:dE_ds2},
\end{align}
where the second bracket vanishes when $\sa_1 = \sa_2 = L_{\rm norm}$ and the last bracket has no zeros at $L_{\rm norm} > 1/\sqrt{3}$. The maximum energy at $\sa_1 = \sa_2 = L_{\rm norm}$ is given by
\begin{equation}
    \label{eq:max_E_highL}
E_{\rm norm,max} = -\frac 14 (1-3L_{\rm norm}^2)^2.
\end{equation}
At the maximum energy, the order parameters are given by 
\begin{equation}
    \label{eq:Q_atE_max}
    Q_1=Q_2 = -\frac{1}{3} + L_{\rm norm}^2.
\end{equation}

\section{Plots of mass segregation without a dominant component} \label{app:plots}
\begin{figure*}
	\centering 
\includegraphics[scale = 0.6]{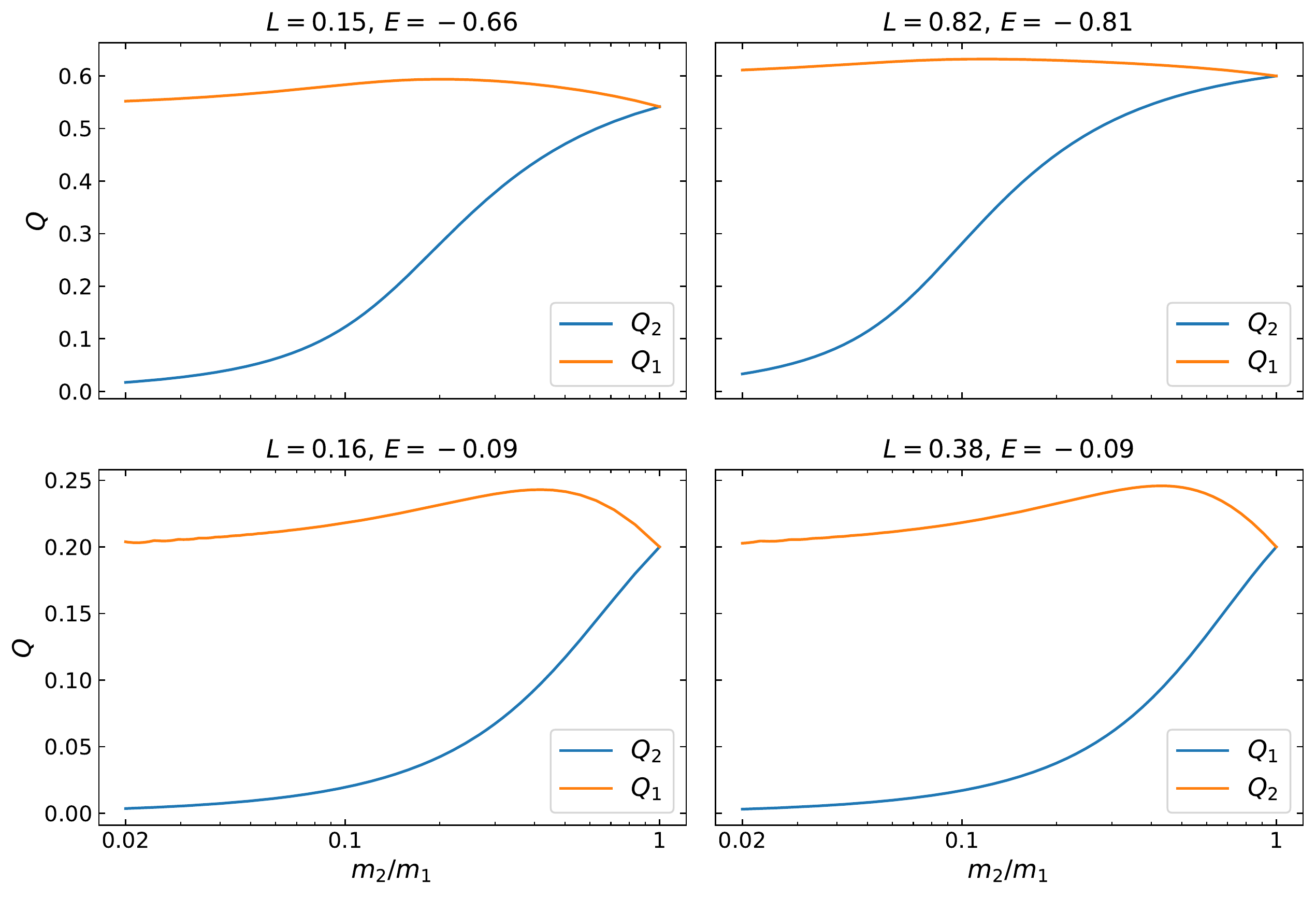}    
	\caption{The order parameter of both components as a function of mass for $a_2/a_1=N_2/N_1=1$. Different panels have different initial conditions parameterized by the conserved quantities $(L_{\rm norm},E_{\rm norm})$ as labelled (Eq.~\ref{eq:E_norm} and \ref{eq:L_norm}).
 } \label{fig:Q_m_two}
\end{figure*}
\begin{figure*}
    \centering
    \includegraphics[scale = 0.6]{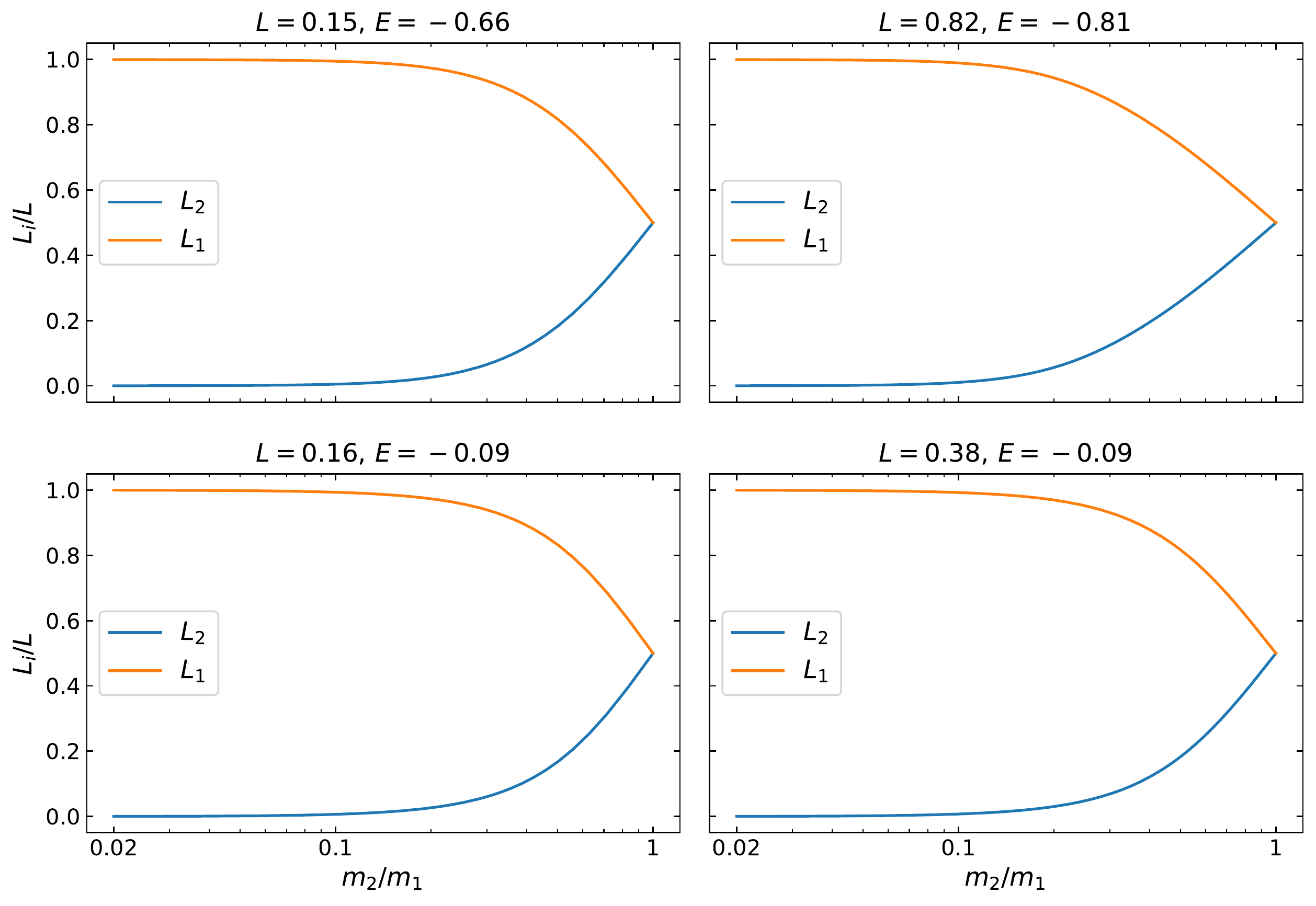}
    \caption{Similar to Figure~\ref{fig:Q_m_two} but showing the net angular momentum of each component relative to the total for a system with $a_2/a_1=N_2/N_1=1$.}
    \label{fig:L_m_two}
\end{figure*}
\begin{figure*}
	\centering 
        \includegraphics[scale = 0.6]{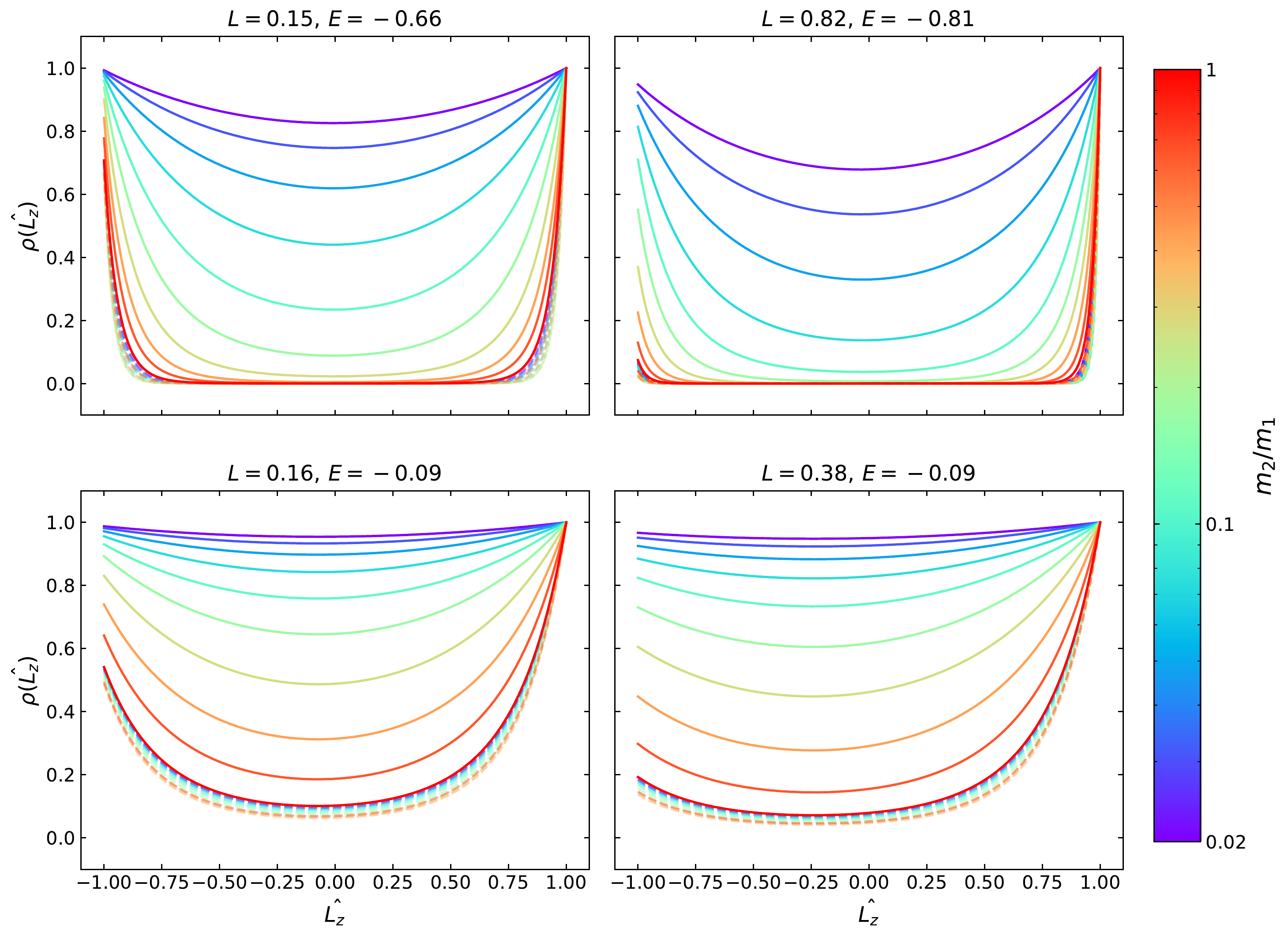}
	\caption{Similar to Figures~\ref{fig:lvl_m_bath} and \ref{fig:lvl_bath_spherical}, showing the distribution function of the  $\hat{L}_z$ component of the angular momentum vector for the less heavy component for fixed $a_2/a_1=N_2/N_1=1$ and different $m_2/m_1$ shown in the colorbar. The distribution function is plotted for 10 selected mass ratios separated by log scale. The dashed lines represent the distribution of the more massive component 1 using the same colour code. The top left plot has $L =0.15,E=-0.66$, the bottom left plot has $L=0.82,E=-0.81$, the top right plot has $L=0.16,E=-0.09$, the bottom right plot has $L=0.38,E=-0.09$. Note that $L$ and $E$ here stand for the normalised total angular momentum and energy as defined in Eq.~\eqref{eq:E_norm} and Eq.~\eqref{eq:L_norm}.} \label{fig:rho_m_two}
\end{figure*}
\begin{figure*}
	\centering 
    \includegraphics[scale =0.65]{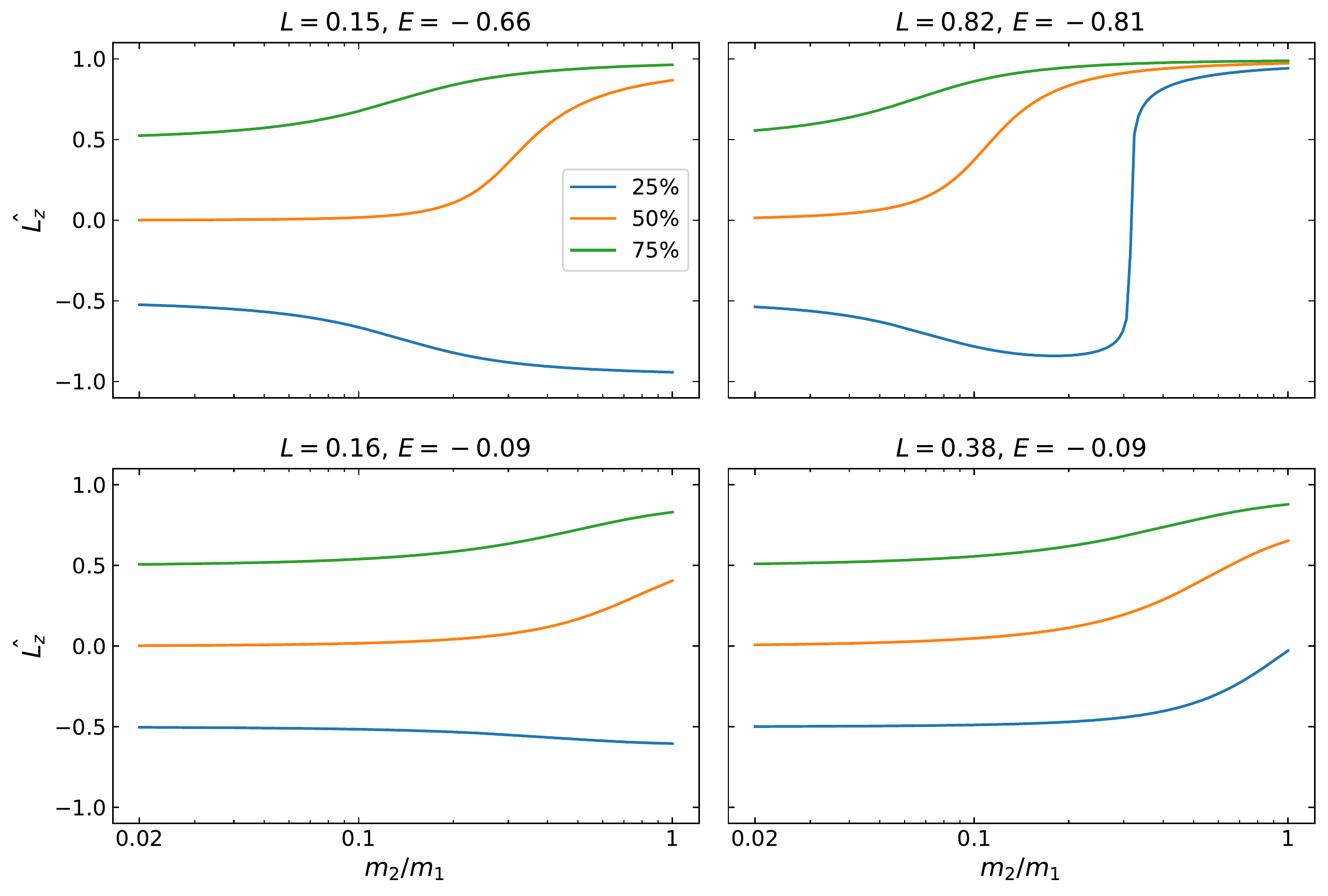}
	\caption{The cumulative distribution levels of $\hat{L}_z$ are plotted for different mass ratios for the less heavy stellar component $\mathcal{C}_1$. The plots show the value of $\hat{L}_z$ at which the cumulative distribution function reaches $25\%,50\%,75\%$. 
	Different panels show different normalised total angular momentum and total VRR energy as defined in Eqs.~\eqref{eq:E_norm} and \eqref{eq:L_norm}: $(L_{\rm norm},E_{\rm norm}) =(0.15,-0.66)$ top left, $(0.82,-0.81$ bottom left, $(0.16,-0.09)$ top right, $(0.38,-0.09)$ bottom right. } \label{fig:lvl_m_two}
\end{figure*}

\begin{figure}
    \centering
    \includegraphics[scale = 0.5]{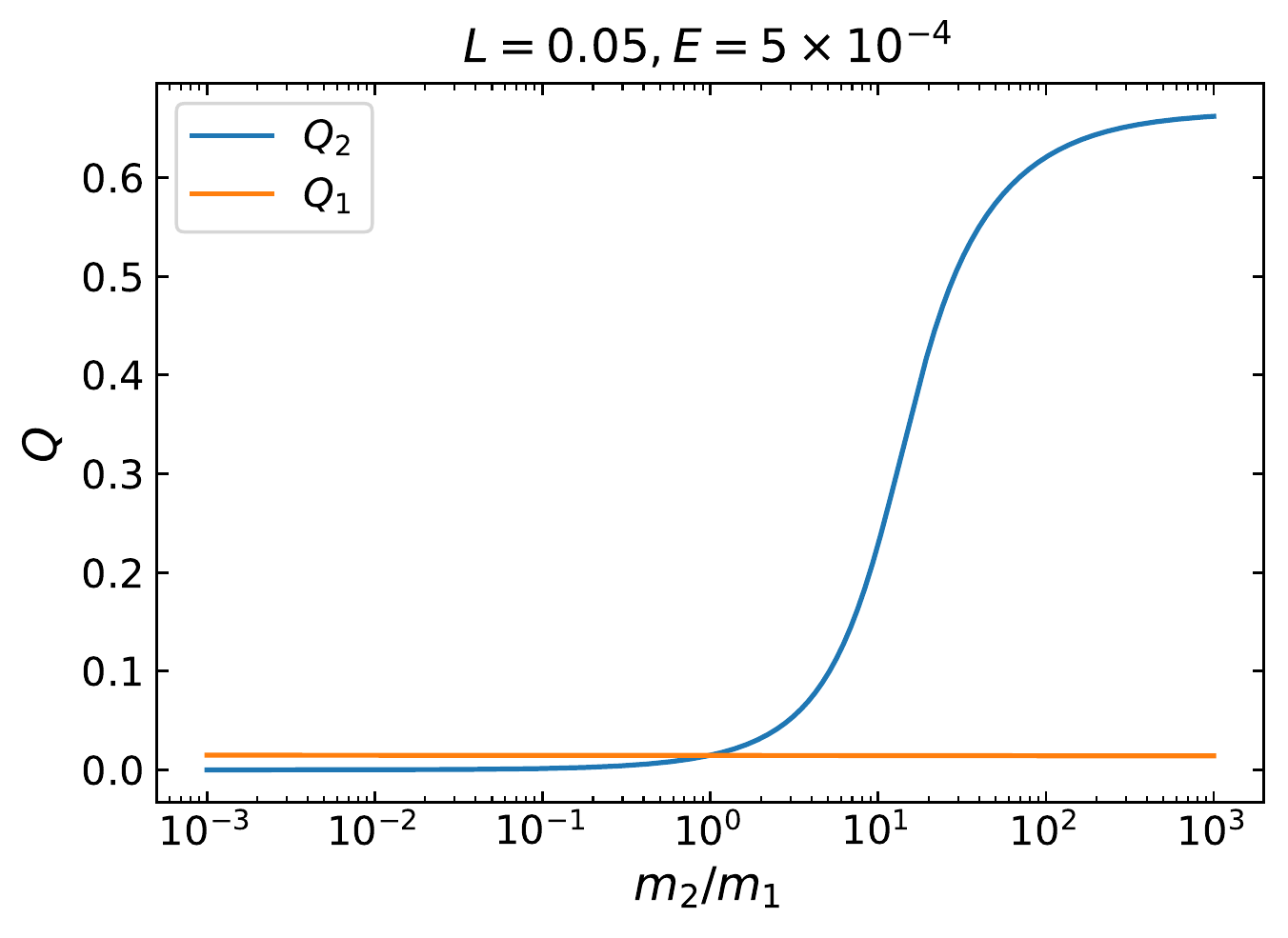}
    \caption{Similar to Figure~\ref{fig:Q_m_two} but showing the order parameter $Q_1$ and $Q_2$ at different mass ratios $m_2/m_1$ for nearly isotropic initial conditions with $(E_{\textrm{norm}}, L_{\textrm{norm}})  \approx ( 5\times 10^{-4}, 0.05)$ and $M_2/M_1 = 10^{-3}$ and $a_2/a_1 =1$.}
    \label{fig:two_sphere_QQ}
\end{figure}
\begin{figure}
    \centering
    \includegraphics[scale = 0.5]{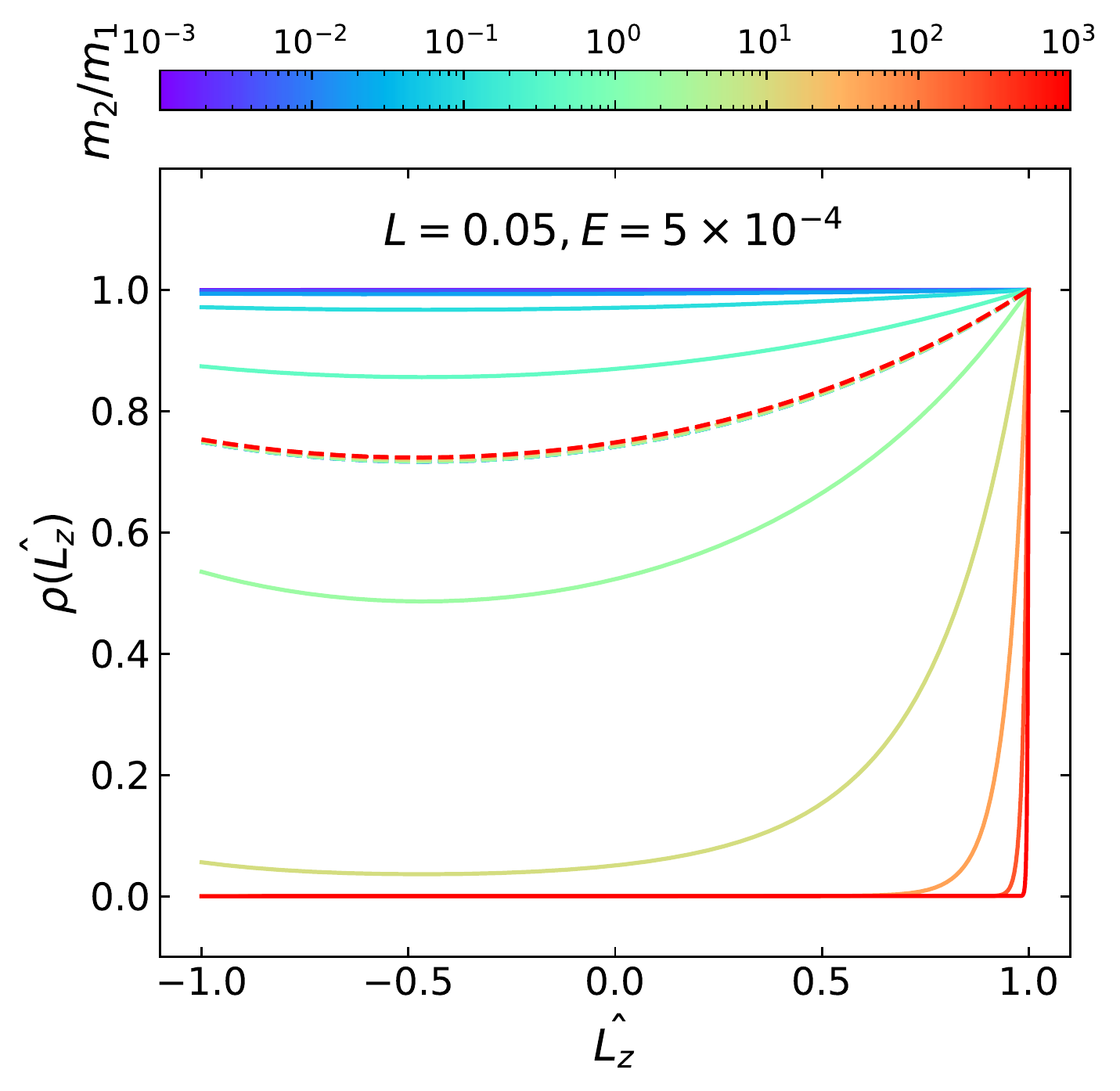}
    \caption{Similar to Figure~\ref{fig:rho_m_two} showing the distribution function of the normalised orbital angular momentum vector direction $\cos\theta$ of both components along the axis of symmetry but for nearly isotropic initial conditions with $(E_{\textrm{norm}}, L_{\textrm{norm}}) \approx (5\times 10^{-4},0.05)$ and $M_2/M_1 = 10^{-3}$ and $a_2/a_1 =1$. The solid curves correspond to the lighter total-mass component $\mathcal{C}_2$ while the dashed curves correspond to the heavier component $\mathcal{C}_1$.}
    \label{fig:two_sphere_selected}
\end{figure}
\begin{figure}
    \centering
    \includegraphics[scale = 0.5]{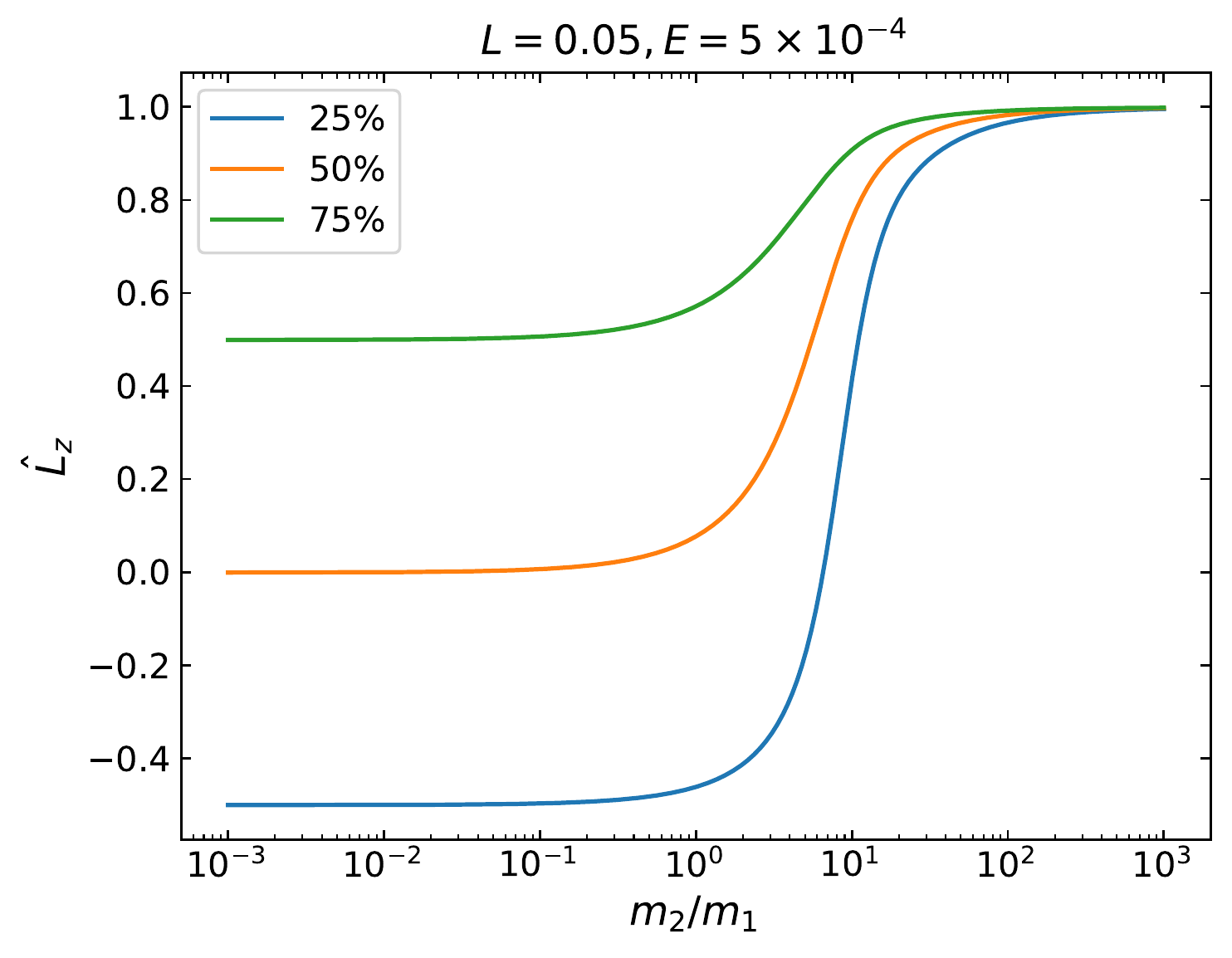}
    \caption{Similar to Figure~\ref{fig:lvl_m_two} showing the cumulative distribution levels of $\hat{L}_z$ as a function of mass ratio $m_2/m_1$ of the subdominant component $\mathcal{C}_2$ for the nearly isotropic initial conditions with $(E_{\textrm{norm}}, L_{\textrm{norm}}) \approx (5\times 10^{-4},0.05)$, $M_2/M_1 = 10^{-3}$ and $a_2/a_1 =1$.}
    \label{fig:two_sphere_lvl}
\end{figure}

\bibliographystyle{apsrev4-1}
\bibliography{references}

\begin{thebibliography}{46}%
\makeatletter
\providecommand \@ifxundefined [1]{%
 \@ifx{#1\undefined}
}%
\providecommand \@ifnum [1]{%
 \ifnum #1\expandafter \@firstoftwo
 \else \expandafter \@secondoftwo
 \fi
}%
\providecommand \@ifx [1]{%
 \ifx #1\expandafter \@firstoftwo
 \else \expandafter \@secondoftwo
 \fi
}%
\providecommand \natexlab [1]{#1}%
\providecommand \enquote  [1]{``#1''}%
\providecommand \bibnamefont  [1]{#1}%
\providecommand \bibfnamefont [1]{#1}%
\providecommand \citenamefont [1]{#1}%
\providecommand \href@noop [0]{\@secondoftwo}%
\providecommand \href [0]{\begingroup \@sanitize@url \@href}%
\providecommand \@href[1]{\@@startlink{#1}\@@href}%
\providecommand \@@href[1]{\endgroup#1\@@endlink}%
\providecommand \@sanitize@url [0]{\catcode `\\12\catcode `\$12\catcode
  `\&12\catcode `\#12\catcode `\^12\catcode `\_12\catcode `\%12\relax}%
\providecommand \@@startlink[1]{}%
\providecommand \@@endlink[0]{}%
\providecommand \url  [0]{\begingroup\@sanitize@url \@url }%
\providecommand \@url [1]{\endgroup\@href {#1}{\urlprefix }}%
\providecommand \urlprefix  [0]{URL }%
\providecommand \Eprint [0]{\href }%
\providecommand \doibase [0]{http://dx.doi.org/}%
\providecommand \selectlanguage [0]{\@gobble}%
\providecommand \bibinfo  [0]{\@secondoftwo}%
\providecommand \bibfield  [0]{\@secondoftwo}%
\providecommand \translation [1]{[#1]}%
\providecommand \BibitemOpen [0]{}%
\providecommand \bibitemStop [0]{}%
\providecommand \bibitemNoStop [0]{.\EOS\space}%
\providecommand \EOS [0]{\spacefactor3000\relax}%
\providecommand \BibitemShut  [1]{\csname bibitem#1\endcsname}%
\let\auto@bib@innerbib\@empty
\bibitem [{\citenamefont {{Genzel}}\ \emph {et~al.}(2010)\citenamefont
  {{Genzel}}, \citenamefont {{Eisenhauer}},\ and\ \citenamefont
  {{Gillessen}}}]{supmassivBH}%
  \BibitemOpen
  \bibfield  {author} {\bibinfo {author} {\bibfnamefont {R.}~\bibnamefont
  {{Genzel}}}, \bibinfo {author} {\bibfnamefont {F.}~\bibnamefont
  {{Eisenhauer}}}, \ and\ \bibinfo {author} {\bibfnamefont {S.}~\bibnamefont
  {{Gillessen}}},\ }\href {\doibase 10.1103/RevModPhys.82.3121} {\bibfield
  {journal} {\bibinfo  {journal} {Reviews of Modern Physics}\ }\textbf
  {\bibinfo {volume} {82}},\ \bibinfo {pages} {3121} (\bibinfo {year}
  {2010})},\ \Eprint {http://arxiv.org/abs/1006.0064} {arXiv:1006.0064
  [astro-ph.GA]} \BibitemShut {NoStop}%
\bibitem [{\citenamefont {{Neumayer}}\ \emph {et~al.}(2020)\citenamefont
  {{Neumayer}}, \citenamefont {{Seth}},\ and\ \citenamefont
  {{B{\"o}ker}}}]{Neumayer+2020}%
  \BibitemOpen
  \bibfield  {author} {\bibinfo {author} {\bibfnamefont {N.}~\bibnamefont
  {{Neumayer}}}, \bibinfo {author} {\bibfnamefont {A.}~\bibnamefont {{Seth}}},
  \ and\ \bibinfo {author} {\bibfnamefont {T.}~\bibnamefont {{B{\"o}ker}}},\
  }\href {\doibase 10.1007/s00159-020-00125-0} {\bibfield  {journal} {\bibinfo
  {journal} {\aapr}\ }\textbf {\bibinfo {volume} {28}},\ \bibinfo {eid} {4}
  (\bibinfo {year} {2020})},\ \Eprint {http://arxiv.org/abs/2001.03626}
  {arXiv:2001.03626 [astro-ph.GA]} \BibitemShut {NoStop}%
\bibitem [{\citenamefont {Bartko}\ \emph {et~al.}(2009)\citenamefont {Bartko},
  \citenamefont {Martins}, \citenamefont {Fritz}, \citenamefont {Genzel},
  \citenamefont {Levin}, \citenamefont {Perets}, \citenamefont {Paumard},
  \citenamefont {Nayakshin}, \citenamefont {Gerhard}, \citenamefont
  {Alexander}, \citenamefont {Dodds-Eden}, \citenamefont {Eisenhauer},
  \citenamefont {Gillessen}, \citenamefont {Mascetti}, \citenamefont {Ott},
  \citenamefont {Perrin}, \citenamefont {Pfuhl}, \citenamefont {Reid},
  \citenamefont {Rouan}, \citenamefont {Sternberg},\ and\ \citenamefont
  {Trippe}}]{Bartko_2009}%
  \BibitemOpen
  \bibfield  {author} {\bibinfo {author} {\bibfnamefont {H.}~\bibnamefont
  {Bartko}}, \bibinfo {author} {\bibfnamefont {F.}~\bibnamefont {Martins}},
  \bibinfo {author} {\bibfnamefont {T.~K.}\ \bibnamefont {Fritz}}, \bibinfo
  {author} {\bibfnamefont {R.}~\bibnamefont {Genzel}}, \bibinfo {author}
  {\bibfnamefont {Y.}~\bibnamefont {Levin}}, \bibinfo {author} {\bibfnamefont
  {H.~B.}\ \bibnamefont {Perets}}, \bibinfo {author} {\bibfnamefont
  {T.}~\bibnamefont {Paumard}}, \bibinfo {author} {\bibfnamefont
  {S.}~\bibnamefont {Nayakshin}}, \bibinfo {author} {\bibfnamefont
  {O.}~\bibnamefont {Gerhard}}, \bibinfo {author} {\bibfnamefont
  {T.}~\bibnamefont {Alexander}}, \bibinfo {author} {\bibfnamefont
  {K.}~\bibnamefont {Dodds-Eden}}, \bibinfo {author} {\bibfnamefont
  {F.}~\bibnamefont {Eisenhauer}}, \bibinfo {author} {\bibfnamefont
  {S.}~\bibnamefont {Gillessen}}, \bibinfo {author} {\bibfnamefont
  {L.}~\bibnamefont {Mascetti}}, \bibinfo {author} {\bibfnamefont
  {T.}~\bibnamefont {Ott}}, \bibinfo {author} {\bibfnamefont {G.}~\bibnamefont
  {Perrin}}, \bibinfo {author} {\bibfnamefont {O.}~\bibnamefont {Pfuhl}},
  \bibinfo {author} {\bibfnamefont {M.~J.}\ \bibnamefont {Reid}}, \bibinfo
  {author} {\bibfnamefont {D.}~\bibnamefont {Rouan}}, \bibinfo {author}
  {\bibfnamefont {A.}~\bibnamefont {Sternberg}}, \ and\ \bibinfo {author}
  {\bibfnamefont {S.}~\bibnamefont {Trippe}},\ }\href {\doibase
  10.1088/0004-637x/697/2/1741} {\bibfield  {journal} {\bibinfo  {journal}
  {\apj}\ }\textbf {\bibinfo {volume} {697}},\ \bibinfo {pages} {1741}
  (\bibinfo {year} {2009})}\BibitemShut {NoStop}%
\bibitem [{\citenamefont {{Yelda}}\ \emph {et~al.}(2014)\citenamefont
  {{Yelda}}, \citenamefont {{Ghez}}, \citenamefont {{Lu}}, \citenamefont
  {{Do}}, \citenamefont {{Meyer}}, \citenamefont {{Morris}},\ and\
  \citenamefont {{Matthews}}}]{yelda2014}%
  \BibitemOpen
  \bibfield  {author} {\bibinfo {author} {\bibfnamefont {S.}~\bibnamefont
  {{Yelda}}}, \bibinfo {author} {\bibfnamefont {A.~M.}\ \bibnamefont {{Ghez}}},
  \bibinfo {author} {\bibfnamefont {J.~R.}\ \bibnamefont {{Lu}}}, \bibinfo
  {author} {\bibfnamefont {T.}~\bibnamefont {{Do}}}, \bibinfo {author}
  {\bibfnamefont {L.}~\bibnamefont {{Meyer}}}, \bibinfo {author} {\bibfnamefont
  {M.~R.}\ \bibnamefont {{Morris}}}, \ and\ \bibinfo {author} {\bibfnamefont
  {K.}~\bibnamefont {{Matthews}}},\ }\href {\doibase
  10.1088/0004-637X/783/2/131} {\bibfield  {journal} {\bibinfo  {journal}
  {\apj}\ }\textbf {\bibinfo {volume} {783}},\ \bibinfo {eid} {131} (\bibinfo
  {year} {2014})},\ \Eprint {http://arxiv.org/abs/1401.7354} {arXiv:1401.7354
  [astro-ph.GA]} \BibitemShut {NoStop}%
\bibitem [{\citenamefont {{von Fellenberg}}\ \emph {et~al.}(2022)\citenamefont
  {{von Fellenberg}}, \citenamefont {{Gillessen}}, \citenamefont {{Stadler}},
  \citenamefont {{Baub{\"o}ck}}, \citenamefont {{Genzel}}, \citenamefont {{de
  Zeeuw}}, \citenamefont {{Pfuhl}}, \citenamefont {{Amaro Seoane}},
  \citenamefont {{Drescher}}, \citenamefont {{Eisenhauer}}, \citenamefont
  {{Habibi}}, \citenamefont {{Ott}}, \citenamefont {{Widmann}},\ and\
  \citenamefont {{Young}}}]{2022genzel}%
  \BibitemOpen
  \bibfield  {author} {\bibinfo {author} {\bibfnamefont {S.~D.}\ \bibnamefont
  {{von Fellenberg}}}, \bibinfo {author} {\bibfnamefont {S.}~\bibnamefont
  {{Gillessen}}}, \bibinfo {author} {\bibfnamefont {J.}~\bibnamefont
  {{Stadler}}}, \bibinfo {author} {\bibfnamefont {M.}~\bibnamefont
  {{Baub{\"o}ck}}}, \bibinfo {author} {\bibfnamefont {R.}~\bibnamefont
  {{Genzel}}}, \bibinfo {author} {\bibfnamefont {T.}~\bibnamefont {{de
  Zeeuw}}}, \bibinfo {author} {\bibfnamefont {O.}~\bibnamefont {{Pfuhl}}},
  \bibinfo {author} {\bibfnamefont {P.}~\bibnamefont {{Amaro Seoane}}},
  \bibinfo {author} {\bibfnamefont {A.}~\bibnamefont {{Drescher}}}, \bibinfo
  {author} {\bibfnamefont {F.}~\bibnamefont {{Eisenhauer}}}, \bibinfo {author}
  {\bibfnamefont {M.}~\bibnamefont {{Habibi}}}, \bibinfo {author}
  {\bibfnamefont {T.}~\bibnamefont {{Ott}}}, \bibinfo {author} {\bibfnamefont
  {F.}~\bibnamefont {{Widmann}}}, \ and\ \bibinfo {author} {\bibfnamefont
  {A.}~\bibnamefont {{Young}}},\ }\href {\doibase 10.3847/2041-8213/ac68ef}
  {\bibfield  {journal} {\bibinfo  {journal} {\apjl}\ }\textbf {\bibinfo
  {volume} {932}},\ \bibinfo {eid} {L6} (\bibinfo {year} {2022})},\ \Eprint
  {http://arxiv.org/abs/2205.07595} {arXiv:2205.07595 [astro-ph.GA]}
  \BibitemShut {NoStop}%
\bibitem [{\citenamefont {{Sz{\"o}lgy{\'e}n}}\ and\ \citenamefont
  {{Kocsis}}(2018)}]{Szolgyen_Kocsis2018}%
  \BibitemOpen
  \bibfield  {author} {\bibinfo {author} {\bibfnamefont {{\'A}.}~\bibnamefont
  {{Sz{\"o}lgy{\'e}n}}}\ and\ \bibinfo {author} {\bibfnamefont
  {B.}~\bibnamefont {{Kocsis}}},\ }\href {\doibase
  10.1103/PhysRevLett.121.101101} {\bibfield  {journal} {\bibinfo  {journal}
  {\prl}\ }\textbf {\bibinfo {volume} {121}},\ \bibinfo {eid} {101101}
  (\bibinfo {year} {2018})},\ \Eprint {http://arxiv.org/abs/1803.07090}
  {arXiv:1803.07090 [astro-ph.GA]} \BibitemShut {NoStop}%
\bibitem [{\citenamefont {{Gruzinov}}\ \emph {et~al.}(2020)\citenamefont
  {{Gruzinov}}, \citenamefont {{Levin}},\ and\ \citenamefont
  {{Zhu}}}]{Gruzinov+2020}%
  \BibitemOpen
  \bibfield  {author} {\bibinfo {author} {\bibfnamefont {A.}~\bibnamefont
  {{Gruzinov}}}, \bibinfo {author} {\bibfnamefont {Y.}~\bibnamefont {{Levin}}},
  \ and\ \bibinfo {author} {\bibfnamefont {J.}~\bibnamefont {{Zhu}}},\ }\href
  {\doibase 10.3847/1538-4357/abbfaa} {\bibfield  {journal} {\bibinfo
  {journal} {\apj}\ }\textbf {\bibinfo {volume} {905}},\ \bibinfo {eid} {11}
  (\bibinfo {year} {2020})},\ \Eprint {http://arxiv.org/abs/2007.08471}
  {arXiv:2007.08471 [astro-ph.GA]} \BibitemShut {NoStop}%
\bibitem [{\citenamefont {{Magnan}}\ \emph {et~al.}(2022)\citenamefont
  {{Magnan}}, \citenamefont {{Fouvry}}, \citenamefont {{Pichon}},\ and\
  \citenamefont {{Chavanis}}}]{magnan2021orbital}%
  \BibitemOpen
  \bibfield  {author} {\bibinfo {author} {\bibfnamefont {N.}~\bibnamefont
  {{Magnan}}}, \bibinfo {author} {\bibfnamefont {J.-B.}\ \bibnamefont
  {{Fouvry}}}, \bibinfo {author} {\bibfnamefont {C.}~\bibnamefont {{Pichon}}},
  \ and\ \bibinfo {author} {\bibfnamefont {P.-H.}\ \bibnamefont {{Chavanis}}},\
  }\href {\doibase 10.1093/mnras/stac1248} {\bibfield  {journal} {\bibinfo
  {journal} {\mnras}\ }\textbf {\bibinfo {volume} {514}},\ \bibinfo {pages}
  {3452} (\bibinfo {year} {2022})},\ \Eprint {http://arxiv.org/abs/2111.09011}
  {arXiv:2111.09011 [astro-ph.GA]} \BibitemShut {NoStop}%
\bibitem [{\citenamefont {{M{\'a}th{\'e}}}\ \emph {et~al.}(2023)\citenamefont
  {{M{\'a}th{\'e}}}, \citenamefont {{Sz{\"o}lgy{\'e}n}},\ and\ \citenamefont
  {{Kocsis}}}]{Mathe:2022azz}%
  \BibitemOpen
  \bibfield  {author} {\bibinfo {author} {\bibfnamefont {G.}~\bibnamefont
  {{M{\'a}th{\'e}}}}, \bibinfo {author} {\bibfnamefont {{\'A}.}~\bibnamefont
  {{Sz{\"o}lgy{\'e}n}}}, \ and\ \bibinfo {author} {\bibfnamefont
  {B.}~\bibnamefont {{Kocsis}}},\ }\href {\doibase 10.1093/mnras/stad016}
  {\bibfield  {journal} {\bibinfo  {journal} {\mnras}\ }\textbf {\bibinfo
  {volume} {520}},\ \bibinfo {pages} {2204} (\bibinfo {year} {2023})},\ \Eprint
  {http://arxiv.org/abs/2202.07665} {arXiv:2202.07665 [astro-ph.GA]}
  \BibitemShut {NoStop}%
\bibitem [{\citenamefont {{Lynden-Bell}}(1967)}]{1967lyn}%
  \BibitemOpen
  \bibfield  {author} {\bibinfo {author} {\bibfnamefont {D.}~\bibnamefont
  {{Lynden-Bell}}},\ }\href {\doibase 10.1093/mnras/136.1.101} {\bibfield
  {journal} {\bibinfo  {journal} {\mnras}\ }\textbf {\bibinfo {volume} {136}},\
  \bibinfo {pages} {101} (\bibinfo {year} {1967})}\BibitemShut {NoStop}%
\bibitem [{\citenamefont {{Lynden-Bell}}\ and\ \citenamefont
  {{Wood}}(1968)}]{1968lyn}%
  \BibitemOpen
  \bibfield  {author} {\bibinfo {author} {\bibfnamefont {D.}~\bibnamefont
  {{Lynden-Bell}}}\ and\ \bibinfo {author} {\bibfnamefont {R.}~\bibnamefont
  {{Wood}}},\ }\href {\doibase 10.1093/mnras/138.4.495} {\bibfield  {journal}
  {\bibinfo  {journal} {\mnras}\ }\textbf {\bibinfo {volume} {138}},\ \bibinfo
  {pages} {495} (\bibinfo {year} {1968})}\BibitemShut {NoStop}%
\bibitem [{\citenamefont {Campa}\ \emph {et~al.}(2014)\citenamefont {Campa},
  \citenamefont {Dauxois}, \citenamefont {Fanelli},\ and\ \citenamefont
  {Ruffo}}]{Campa2014}%
  \BibitemOpen
  \bibfield  {author} {\bibinfo {author} {\bibfnamefont {A.}~\bibnamefont
  {Campa}}, \bibinfo {author} {\bibfnamefont {T.}~\bibnamefont {Dauxois}},
  \bibinfo {author} {\bibfnamefont {D.}~\bibnamefont {Fanelli}}, \ and\
  \bibinfo {author} {\bibfnamefont {S.}~\bibnamefont {Ruffo}},\ }\href
  {\doibase 10.1093/acprof:oso/9780199581931.001.0001} {\emph {\bibinfo {title}
  {{Physics of Long-Range Interacting Systems}}}}\ (\bibinfo  {publisher}
  {Oxford University Press},\ \bibinfo {year} {2014})\BibitemShut {NoStop}%
\bibitem [{\citenamefont {{Rauch}}\ and\ \citenamefont
  {{Tremaine}}(1996)}]{RAUCH1996149}%
  \BibitemOpen
  \bibfield  {author} {\bibinfo {author} {\bibfnamefont {K.~P.}\ \bibnamefont
  {{Rauch}}}\ and\ \bibinfo {author} {\bibfnamefont {S.}~\bibnamefont
  {{Tremaine}}},\ }\href {\doibase 10.1016/S1384-1076(96)00012-7} {\bibfield
  {journal} {\bibinfo  {journal} {\na}\ }\textbf {\bibinfo {volume} {1}},\
  \bibinfo {pages} {149} (\bibinfo {year} {1996})},\ \Eprint
  {http://arxiv.org/abs/astro-ph/9603018} {arXiv:astro-ph/9603018 [astro-ph]}
  \BibitemShut {NoStop}%
\bibitem [{\citenamefont {{Roupas}}(2020)}]{Roupas2020}%
  \BibitemOpen
  \bibfield  {author} {\bibinfo {author} {\bibfnamefont {Z.}~\bibnamefont
  {{Roupas}}},\ }\href {\doibase 10.1088/1751-8121/ab5f7b} {\bibfield
  {journal} {\bibinfo  {journal} {Journal of Physics A Mathematical General}\
  }\textbf {\bibinfo {volume} {53}},\ \bibinfo {eid} {045002} (\bibinfo {year}
  {2020})},\ \Eprint {http://arxiv.org/abs/1910.05735} {arXiv:1910.05735
  [astro-ph.GA]} \BibitemShut {NoStop}%
\bibitem [{\citenamefont {{Kocsis}}\ and\ \citenamefont
  {{Tremaine}}(2011)}]{kocsis2011}%
  \BibitemOpen
  \bibfield  {author} {\bibinfo {author} {\bibfnamefont {B.}~\bibnamefont
  {{Kocsis}}}\ and\ \bibinfo {author} {\bibfnamefont {S.}~\bibnamefont
  {{Tremaine}}},\ }\href {\doibase 10.1111/j.1365-2966.2010.17897.x} {\bibfield
   {journal} {\bibinfo  {journal} {\mnras}\ }\textbf {\bibinfo {volume}
  {412}},\ \bibinfo {pages} {187} (\bibinfo {year} {2011})},\ \Eprint
  {http://arxiv.org/abs/1006.0001} {arXiv:1006.0001 [astro-ph.GA]} \BibitemShut
  {NoStop}%
\bibitem [{\citenamefont {{Touma}}\ and\ \citenamefont
  {{Tremaine}}(2014)}]{Touma_Tremaine2014}%
  \BibitemOpen
  \bibfield  {author} {\bibinfo {author} {\bibfnamefont {J.}~\bibnamefont
  {{Touma}}}\ and\ \bibinfo {author} {\bibfnamefont {S.}~\bibnamefont
  {{Tremaine}}},\ }\href@noop {} {\bibfield  {journal} {\bibinfo  {journal}
  {arXiv e-prints}\ ,\ \bibinfo {eid} {arXiv:1401.5534}} (\bibinfo {year}
  {2014})},\ \Eprint {http://arxiv.org/abs/1401.5534} {arXiv:1401.5534
  [astro-ph.GA]} \BibitemShut {NoStop}%
\bibitem [{\citenamefont {{Bar-Or}}\ and\ \citenamefont
  {{Alexander}}(2014)}]{BarOr_Alexander2014}%
  \BibitemOpen
  \bibfield  {author} {\bibinfo {author} {\bibfnamefont {B.}~\bibnamefont
  {{Bar-Or}}}\ and\ \bibinfo {author} {\bibfnamefont {T.}~\bibnamefont
  {{Alexander}}},\ }\href {\doibase 10.1088/0264-9381/31/24/244003} {\bibfield
  {journal} {\bibinfo  {journal} {Classical and Quantum Gravity}\ }\textbf
  {\bibinfo {volume} {31}},\ \bibinfo {eid} {244003} (\bibinfo {year}
  {2014})},\ \Eprint {http://arxiv.org/abs/1404.0351} {arXiv:1404.0351
  [astro-ph.GA]} \BibitemShut {NoStop}%
\bibitem [{\citenamefont {{Sridhar}}\ and\ \citenamefont
  {{Touma}}(2016)}]{Sridhar_Touma2016}%
  \BibitemOpen
  \bibfield  {author} {\bibinfo {author} {\bibfnamefont {S.}~\bibnamefont
  {{Sridhar}}}\ and\ \bibinfo {author} {\bibfnamefont {J.~R.}\ \bibnamefont
  {{Touma}}},\ }\href {\doibase 10.1093/mnras/stw543} {\bibfield  {journal}
  {\bibinfo  {journal} {\mnras}\ }\textbf {\bibinfo {volume} {458}},\ \bibinfo
  {pages} {4143} (\bibinfo {year} {2016})},\ \Eprint
  {http://arxiv.org/abs/1509.02401} {arXiv:1509.02401 [astro-ph.GA]}
  \BibitemShut {NoStop}%
\bibitem [{\citenamefont {{Roupas}}\ \emph {et~al.}(2017)\citenamefont
  {{Roupas}}, \citenamefont {{Kocsis}},\ and\ \citenamefont
  {{Tremaine}}}]{Roupas_2017}%
  \BibitemOpen
  \bibfield  {author} {\bibinfo {author} {\bibfnamefont {Z.}~\bibnamefont
  {{Roupas}}}, \bibinfo {author} {\bibfnamefont {B.}~\bibnamefont {{Kocsis}}},
  \ and\ \bibinfo {author} {\bibfnamefont {S.}~\bibnamefont {{Tremaine}}},\
  }\href {\doibase 10.3847/1538-4357/aa7141} {\bibfield  {journal} {\bibinfo
  {journal} {\apj}\ }\textbf {\bibinfo {volume} {842}},\ \bibinfo {eid} {90}
  (\bibinfo {year} {2017})},\ \Eprint {http://arxiv.org/abs/1701.03271}
  {arXiv:1701.03271 [astro-ph.GA]} \BibitemShut {NoStop}%
\bibitem [{\citenamefont {{Touma}}\ \emph {et~al.}(2019)\citenamefont
  {{Touma}}, \citenamefont {{Tremaine}},\ and\ \citenamefont
  {{Kazandjian}}}]{Touma+2019}%
  \BibitemOpen
  \bibfield  {author} {\bibinfo {author} {\bibfnamefont {J.}~\bibnamefont
  {{Touma}}}, \bibinfo {author} {\bibfnamefont {S.}~\bibnamefont {{Tremaine}}},
  \ and\ \bibinfo {author} {\bibfnamefont {M.}~\bibnamefont {{Kazandjian}}},\
  }\href {\doibase 10.1103/PhysRevLett.123.021103} {\bibfield  {journal}
  {\bibinfo  {journal} {\prl}\ }\textbf {\bibinfo {volume} {123}},\ \bibinfo
  {eid} {021103} (\bibinfo {year} {2019})},\ \Eprint
  {http://arxiv.org/abs/1907.01555} {arXiv:1907.01555 [astro-ph.GA]}
  \BibitemShut {NoStop}%
\bibitem [{\citenamefont {{Tremaine}}(2020{\natexlab{a}})}]{Tremaine2020}%
  \BibitemOpen
  \bibfield  {author} {\bibinfo {author} {\bibfnamefont {S.}~\bibnamefont
  {{Tremaine}}},\ }\href {\doibase 10.1093/mnras/stz3181} {\bibfield  {journal}
  {\bibinfo  {journal} {\mnras}\ }\textbf {\bibinfo {volume} {491}},\ \bibinfo
  {pages} {1941} (\bibinfo {year} {2020}{\natexlab{a}})},\ \Eprint
  {http://arxiv.org/abs/1911.04508} {arXiv:1911.04508 [astro-ph.GA]}
  \BibitemShut {NoStop}%
\bibitem [{\citenamefont {{Tremaine}}(2020{\natexlab{b}})}]{Tremaine2020b}%
  \BibitemOpen
  \bibfield  {author} {\bibinfo {author} {\bibfnamefont {S.}~\bibnamefont
  {{Tremaine}}},\ }\href {\doibase 10.1093/mnras/staa420} {\bibfield  {journal}
  {\bibinfo  {journal} {\mnras}\ }\textbf {\bibinfo {volume} {493}},\ \bibinfo
  {pages} {2632} (\bibinfo {year} {2020}{\natexlab{b}})},\ \Eprint
  {http://arxiv.org/abs/2002.05006} {arXiv:2002.05006 [astro-ph.GA]}
  \BibitemShut {NoStop}%
\bibitem [{\citenamefont {{Levin}}(2022)}]{Levin2022}%
  \BibitemOpen
  \bibfield  {author} {\bibinfo {author} {\bibfnamefont {Y.}~\bibnamefont
  {{Levin}}},\ }\href {\doibase 10.48550/arXiv.2211.12754} {\bibfield
  {journal} {\bibinfo  {journal} {arXiv e-prints}\ ,\ \bibinfo {eid}
  {arXiv:2211.12754}} (\bibinfo {year} {2022})},\ \Eprint
  {http://arxiv.org/abs/2211.12754} {arXiv:2211.12754 [astro-ph.GA]}
  \BibitemShut {NoStop}%
\bibitem [{\citenamefont {{Kocsis}}\ and\ \citenamefont
  {{Tremaine}}(2015)}]{Kocsis_Tremaine2015}%
  \BibitemOpen
  \bibfield  {author} {\bibinfo {author} {\bibfnamefont {B.}~\bibnamefont
  {{Kocsis}}}\ and\ \bibinfo {author} {\bibfnamefont {S.}~\bibnamefont
  {{Tremaine}}},\ }\href {\doibase 10.1093/mnras/stv057} {\bibfield  {journal}
  {\bibinfo  {journal} {\mnras}\ }\textbf {\bibinfo {volume} {448}},\ \bibinfo
  {pages} {3265} (\bibinfo {year} {2015})},\ \Eprint
  {http://arxiv.org/abs/1406.1178} {arXiv:1406.1178 [astro-ph.GA]} \BibitemShut
  {NoStop}%
\bibitem [{\citenamefont {{Fouvry}}\ \emph {et~al.}(2019)\citenamefont
  {{Fouvry}}, \citenamefont {{Bar-Or}},\ and\ \citenamefont
  {{Chavanis}}}]{Fouvry+2019}%
  \BibitemOpen
  \bibfield  {author} {\bibinfo {author} {\bibfnamefont {J.-B.}\ \bibnamefont
  {{Fouvry}}}, \bibinfo {author} {\bibfnamefont {B.}~\bibnamefont {{Bar-Or}}},
  \ and\ \bibinfo {author} {\bibfnamefont {P.-H.}\ \bibnamefont {{Chavanis}}},\
  }\href {\doibase 10.3847/1538-4357/ab2f78} {\bibfield  {journal} {\bibinfo
  {journal} {\apj}\ }\textbf {\bibinfo {volume} {883}},\ \bibinfo {eid} {161}
  (\bibinfo {year} {2019})},\ \Eprint {http://arxiv.org/abs/1812.07053}
  {arXiv:1812.07053 [astro-ph.GA]} \BibitemShut {NoStop}%
\bibitem [{\citenamefont {{Tak{\'a}cs}}\ and\ \citenamefont
  {{Kocsis}}(2018)}]{takacs2018}%
  \BibitemOpen
  \bibfield  {author} {\bibinfo {author} {\bibfnamefont {{\'A}.}~\bibnamefont
  {{Tak{\'a}cs}}}\ and\ \bibinfo {author} {\bibfnamefont {B.}~\bibnamefont
  {{Kocsis}}},\ }\href {\doibase 10.3847/1538-4357/aab268} {\bibfield
  {journal} {\bibinfo  {journal} {\apj}\ }\textbf {\bibinfo {volume} {856}},\
  \bibinfo {eid} {113} (\bibinfo {year} {2018})},\ \Eprint
  {http://arxiv.org/abs/1712.04449} {arXiv:1712.04449 [astro-ph.GA]}
  \BibitemShut {NoStop}%
\bibitem [{\citenamefont {{Hunter}}\ and\ \citenamefont
  {{Toomre}}(1969)}]{Hunter_Toomre1969}%
  \BibitemOpen
  \bibfield  {author} {\bibinfo {author} {\bibfnamefont {C.}~\bibnamefont
  {{Hunter}}}\ and\ \bibinfo {author} {\bibfnamefont {A.}~\bibnamefont
  {{Toomre}}},\ }\href {\doibase 10.1086/149908} {\bibfield  {journal}
  {\bibinfo  {journal} {\apj}\ }\textbf {\bibinfo {volume} {155}},\ \bibinfo
  {pages} {747} (\bibinfo {year} {1969})}\BibitemShut {NoStop}%
\bibitem [{\citenamefont {{Nelson}}\ and\ \citenamefont
  {{Tremaine}}(1995)}]{Nelson_Tremaine1995}%
  \BibitemOpen
  \bibfield  {author} {\bibinfo {author} {\bibfnamefont {R.~W.}\ \bibnamefont
  {{Nelson}}}\ and\ \bibinfo {author} {\bibfnamefont {S.}~\bibnamefont
  {{Tremaine}}},\ }\href {\doibase 10.1093/mnras/275.4.897} {\bibfield
  {journal} {\bibinfo  {journal} {\mnras}\ }\textbf {\bibinfo {volume} {275}},\
  \bibinfo {pages} {897} (\bibinfo {year} {1995})},\ \Eprint
  {http://arxiv.org/abs/astro-ph/9408068} {arXiv:astro-ph/9408068 [astro-ph]}
  \BibitemShut {NoStop}%
\bibitem [{\citenamefont {{Ulubay-Siddiki}}\ \emph {et~al.}(2009)\citenamefont
  {{Ulubay-Siddiki}}, \citenamefont {{Gerhard}},\ and\ \citenamefont
  {{Arnaboldi}}}]{Ulubay-Siddiki+2009}%
  \BibitemOpen
  \bibfield  {author} {\bibinfo {author} {\bibfnamefont {A.}~\bibnamefont
  {{Ulubay-Siddiki}}}, \bibinfo {author} {\bibfnamefont {O.}~\bibnamefont
  {{Gerhard}}}, \ and\ \bibinfo {author} {\bibfnamefont {M.}~\bibnamefont
  {{Arnaboldi}}},\ }\href {\doibase 10.1111/j.1365-2966.2009.15089.x}
  {\bibfield  {journal} {\bibinfo  {journal} {\mnras}\ }\textbf {\bibinfo
  {volume} {398}},\ \bibinfo {pages} {535} (\bibinfo {year} {2009})},\ \Eprint
  {http://arxiv.org/abs/0909.5333} {arXiv:0909.5333 [astro-ph.CO]} \BibitemShut
  {NoStop}%
\bibitem [{\citenamefont {{Batygin}}(2018)}]{Batygin18}%
  \BibitemOpen
  \bibfield  {author} {\bibinfo {author} {\bibfnamefont {K.}~\bibnamefont
  {{Batygin}}},\ }\href {\doibase 10.1093/mnras/sty162} {\bibfield  {journal}
  {\bibinfo  {journal} {\mnras}\ }\textbf {\bibinfo {volume} {475}},\ \bibinfo
  {pages} {5070} (\bibinfo {year} {2018})},\ \Eprint
  {http://arxiv.org/abs/1803.01258} {arXiv:1803.01258 [astro-ph.EP]}
  \BibitemShut {NoStop}%
\bibitem [{\citenamefont {{Panamarev}}\ and\ \citenamefont
  {{Kocsis}}(2022)}]{taras_2022_discs}%
  \BibitemOpen
  \bibfield  {author} {\bibinfo {author} {\bibfnamefont {T.}~\bibnamefont
  {{Panamarev}}}\ and\ \bibinfo {author} {\bibfnamefont {B.}~\bibnamefont
  {{Kocsis}}},\ }\href {\doibase 10.1093/mnras/stac3050} {\bibfield  {journal}
  {\bibinfo  {journal} {\mnras}\ }\textbf {\bibinfo {volume} {517}},\ \bibinfo
  {pages} {6205} (\bibinfo {year} {2022})},\ \Eprint
  {http://arxiv.org/abs/2207.06398} {arXiv:2207.06398 [astro-ph.GA]}
  \BibitemShut {NoStop}%
\bibitem [{\citenamefont {{Rasskazov}}\ and\ \citenamefont
  {{Kocsis}}(2019)}]{Rasskazov_Kocsis2019}%
  \BibitemOpen
  \bibfield  {author} {\bibinfo {author} {\bibfnamefont {A.}~\bibnamefont
  {{Rasskazov}}}\ and\ \bibinfo {author} {\bibfnamefont {B.}~\bibnamefont
  {{Kocsis}}},\ }\href {\doibase 10.3847/1538-4357/ab2c74} {\bibfield
  {journal} {\bibinfo  {journal} {\apj}\ }\textbf {\bibinfo {volume} {881}},\
  \bibinfo {eid} {20} (\bibinfo {year} {2019})},\ \Eprint
  {http://arxiv.org/abs/1902.03242} {arXiv:1902.03242 [astro-ph.HE]}
  \BibitemShut {NoStop}%
\bibitem [{\citenamefont {{Samsing}}\ \emph {et~al.}(2022)\citenamefont
  {{Samsing}}, \citenamefont {{Bartos}}, \citenamefont {{D'Orazio}},
  \citenamefont {{Haiman}}, \citenamefont {{Kocsis}}, \citenamefont {{Leigh}},
  \citenamefont {{Liu}}, \citenamefont {{Pessah}},\ and\ \citenamefont
  {{Tagawa}}}]{Samsing+2022}%
  \BibitemOpen
  \bibfield  {author} {\bibinfo {author} {\bibfnamefont {J.}~\bibnamefont
  {{Samsing}}}, \bibinfo {author} {\bibfnamefont {I.}~\bibnamefont {{Bartos}}},
  \bibinfo {author} {\bibfnamefont {D.~J.}\ \bibnamefont {{D'Orazio}}},
  \bibinfo {author} {\bibfnamefont {Z.}~\bibnamefont {{Haiman}}}, \bibinfo
  {author} {\bibfnamefont {B.}~\bibnamefont {{Kocsis}}}, \bibinfo {author}
  {\bibfnamefont {N.~W.~C.}\ \bibnamefont {{Leigh}}}, \bibinfo {author}
  {\bibfnamefont {B.}~\bibnamefont {{Liu}}}, \bibinfo {author} {\bibfnamefont
  {M.~E.}\ \bibnamefont {{Pessah}}}, \ and\ \bibinfo {author} {\bibfnamefont
  {H.}~\bibnamefont {{Tagawa}}},\ }\href {\doibase 10.1038/s41586-021-04333-1}
  {\bibfield  {journal} {\bibinfo  {journal} {\nat}\ }\textbf {\bibinfo
  {volume} {603}},\ \bibinfo {pages} {237} (\bibinfo {year}
  {2022})}\BibitemShut {NoStop}%
\bibitem [{\citenamefont {{Maier}}\ and\ \citenamefont
  {{Saupe}}(1958)}]{maier_saupe}%
  \BibitemOpen
  \bibfield  {author} {\bibinfo {author} {\bibfnamefont {W.}~\bibnamefont
  {{Maier}}}\ and\ \bibinfo {author} {\bibfnamefont {A.}~\bibnamefont
  {{Saupe}}},\ }\href {\doibase 10.1515/zna-1958-0716} {\bibfield  {journal}
  {\bibinfo  {journal} {Zeitschrift Naturforschung Teil A}\ }\textbf {\bibinfo
  {volume} {13}},\ \bibinfo {pages} {564} (\bibinfo {year} {1958})}\BibitemShut
  {NoStop}%
\bibitem [{\citenamefont {{Plischke}}\ and\ \citenamefont
  {{Bergersen}}(2006)}]{plischke2006}%
  \BibitemOpen
  \bibfield  {author} {\bibinfo {author} {\bibfnamefont {M.}~\bibnamefont
  {{Plischke}}}\ and\ \bibinfo {author} {\bibfnamefont {B.}~\bibnamefont
  {{Bergersen}}},\ }\href {\doibase 10.1142/5660} {\emph {\bibinfo {title}
  {{Equilibrium Statistical Physics (3RD Edition)}}}}\ (\bibinfo {year}
  {2006})\BibitemShut {NoStop}%
\bibitem [{\citenamefont {{Nayakshin}}(2005)}]{nayakshin2005_torus}%
  \BibitemOpen
  \bibfield  {author} {\bibinfo {author} {\bibfnamefont {S.}~\bibnamefont
  {{Nayakshin}}},\ }\href {\doibase 10.1111/j.1365-2966.2005.08913.x}
  {\bibfield  {journal} {\bibinfo  {journal} {\mnras}\ }\textbf {\bibinfo
  {volume} {359}},\ \bibinfo {pages} {545} (\bibinfo {year} {2005})},\ \Eprint
  {http://arxiv.org/abs/astro-ph/0411791} {arXiv:astro-ph/0411791 [astro-ph]}
  \BibitemShut {NoStop}%
\bibitem [{\citenamefont {{{\v{S}}ubr}}\ \emph {et~al.}(2009)\citenamefont
  {{{\v{S}}ubr}}, \citenamefont {{Schovancov{\'a}}},\ and\ \citenamefont
  {{Kroupa}}}]{2009subr_torus}%
  \BibitemOpen
  \bibfield  {author} {\bibinfo {author} {\bibfnamefont {L.~.}\ \bibnamefont
  {{{\v{S}}ubr}}}, \bibinfo {author} {\bibfnamefont {J.}~\bibnamefont
  {{Schovancov{\'a}}}}, \ and\ \bibinfo {author} {\bibfnamefont
  {P.}~\bibnamefont {{Kroupa}}},\ }\href {\doibase 10.1051/0004-6361:200811075}
  {\bibfield  {journal} {\bibinfo  {journal} {\aap}\ }\textbf {\bibinfo
  {volume} {496}},\ \bibinfo {pages} {695} (\bibinfo {year}
  {2009})}\BibitemShut {NoStop}%
\bibitem [{\citenamefont {{Smith}}\ and\ \citenamefont
  {{Wardle}}(2014)}]{smith2014_torus}%
  \BibitemOpen
  \bibfield  {author} {\bibinfo {author} {\bibfnamefont {I.~L.}\ \bibnamefont
  {{Smith}}}\ and\ \bibinfo {author} {\bibfnamefont {M.}~\bibnamefont
  {{Wardle}}},\ }\href {\doibase 10.1093/mnras/stt2092} {\bibfield  {journal}
  {\bibinfo  {journal} {\mnras}\ }\textbf {\bibinfo {volume} {437}},\ \bibinfo
  {pages} {3159} (\bibinfo {year} {2014})},\ \Eprint
  {http://arxiv.org/abs/1310.8429} {arXiv:1310.8429 [astro-ph.GA]} \BibitemShut
  {NoStop}%
\bibitem [{\citenamefont {{Hsieh}}\ \emph {et~al.}(2017)\citenamefont
  {{Hsieh}}, \citenamefont {{Koch}}, \citenamefont {{Ho}}, \citenamefont
  {{Kim}}, \citenamefont {{Tang}}, \citenamefont {{Wang}}, \citenamefont
  {{Yen}},\ and\ \citenamefont {{Hwang}}}]{2017torus_Hsieh}%
  \BibitemOpen
  \bibfield  {author} {\bibinfo {author} {\bibfnamefont {P.-Y.}\ \bibnamefont
  {{Hsieh}}}, \bibinfo {author} {\bibfnamefont {P.~M.}\ \bibnamefont {{Koch}}},
  \bibinfo {author} {\bibfnamefont {P.~T.~P.}\ \bibnamefont {{Ho}}}, \bibinfo
  {author} {\bibfnamefont {W.-T.}\ \bibnamefont {{Kim}}}, \bibinfo {author}
  {\bibfnamefont {Y.-W.}\ \bibnamefont {{Tang}}}, \bibinfo {author}
  {\bibfnamefont {H.-H.}\ \bibnamefont {{Wang}}}, \bibinfo {author}
  {\bibfnamefont {H.-W.}\ \bibnamefont {{Yen}}}, \ and\ \bibinfo {author}
  {\bibfnamefont {C.-Y.}\ \bibnamefont {{Hwang}}},\ }\href {\doibase
  10.3847/1538-4357/aa8329} {\bibfield  {journal} {\bibinfo  {journal} {\apj}\
  }\textbf {\bibinfo {volume} {847}},\ \bibinfo {eid} {3} (\bibinfo {year}
  {2017})},\ \Eprint {http://arxiv.org/abs/1708.08579} {arXiv:1708.08579
  [astro-ph.GA]} \BibitemShut {NoStop}%
\bibitem [{\citenamefont {{Girma}}\ and\ \citenamefont
  {{Loeb}}(2019)}]{Girma2019_IMBH}%
  \BibitemOpen
  \bibfield  {author} {\bibinfo {author} {\bibfnamefont {E.}~\bibnamefont
  {{Girma}}}\ and\ \bibinfo {author} {\bibfnamefont {A.}~\bibnamefont
  {{Loeb}}},\ }\href {\doibase 10.1093/mnras/sty2643} {\bibfield  {journal}
  {\bibinfo  {journal} {\mnras}\ }\textbf {\bibinfo {volume} {482}},\ \bibinfo
  {pages} {3669} (\bibinfo {year} {2019})},\ \Eprint
  {http://arxiv.org/abs/1807.02469} {arXiv:1807.02469 [astro-ph.GA]}
  \BibitemShut {NoStop}%
\bibitem [{\citenamefont {{Arca Sedda}}\ \emph {et~al.}(2019)\citenamefont
  {{Arca Sedda}}, \citenamefont {{Askar}},\ and\ \citenamefont
  {{Giersz}}}]{arca_sedda_2019_IMBH}%
  \BibitemOpen
  \bibfield  {author} {\bibinfo {author} {\bibfnamefont {M.}~\bibnamefont
  {{Arca Sedda}}}, \bibinfo {author} {\bibfnamefont {A.}~\bibnamefont
  {{Askar}}}, \ and\ \bibinfo {author} {\bibfnamefont {M.}~\bibnamefont
  {{Giersz}}},\ }\href@noop {} {\bibfield  {journal} {\bibinfo  {journal}
  {arXiv e-prints}\ ,\ \bibinfo {eid} {arXiv:1905.00902}} (\bibinfo {year}
  {2019})},\ \Eprint {http://arxiv.org/abs/1905.00902} {arXiv:1905.00902
  [astro-ph.GA]} \BibitemShut {NoStop}%
\bibitem [{\citenamefont {{Deme}}\ \emph {et~al.}(2020)\citenamefont {{Deme}},
  \citenamefont {{Meiron}},\ and\ \citenamefont {{Kocsis}}}]{Deme2020_IMBH}%
  \BibitemOpen
  \bibfield  {author} {\bibinfo {author} {\bibfnamefont {B.}~\bibnamefont
  {{Deme}}}, \bibinfo {author} {\bibfnamefont {Y.}~\bibnamefont {{Meiron}}}, \
  and\ \bibinfo {author} {\bibfnamefont {B.}~\bibnamefont {{Kocsis}}},\ }\href
  {\doibase 10.3847/1538-4357/ab7921} {\bibfield  {journal} {\bibinfo
  {journal} {\apj}\ }\textbf {\bibinfo {volume} {892}},\ \bibinfo {eid} {130}
  (\bibinfo {year} {2020})},\ \Eprint {http://arxiv.org/abs/1909.04678}
  {arXiv:1909.04678 [astro-ph.GA]} \BibitemShut {NoStop}%
\bibitem [{\citenamefont {{Sz{\"o}lgy{\'e}n}}\ \emph
  {et~al.}(2021)\citenamefont {{Sz{\"o}lgy{\'e}n}}, \citenamefont
  {{M{\'a}th{\'e}}},\ and\ \citenamefont {{Kocsis}}}]{szolgyen2021_IMBH}%
  \BibitemOpen
  \bibfield  {author} {\bibinfo {author} {\bibfnamefont {{\'A}.}~\bibnamefont
  {{Sz{\"o}lgy{\'e}n}}}, \bibinfo {author} {\bibfnamefont {G.}~\bibnamefont
  {{M{\'a}th{\'e}}}}, \ and\ \bibinfo {author} {\bibfnamefont {B.}~\bibnamefont
  {{Kocsis}}},\ }\href {\doibase 10.3847/1538-4357/ac13ab} {\bibfield
  {journal} {\bibinfo  {journal} {\apj}\ }\textbf {\bibinfo {volume} {919}},\
  \bibinfo {eid} {140} (\bibinfo {year} {2021})},\ \Eprint
  {http://arxiv.org/abs/2103.14042} {arXiv:2103.14042 [astro-ph.GA]}
  \BibitemShut {NoStop}%
\bibitem [{\citenamefont {{Chavanis}}(2006)}]{chavanis_review2006}%
  \BibitemOpen
  \bibfield  {author} {\bibinfo {author} {\bibfnamefont {P.~H.}\ \bibnamefont
  {{Chavanis}}},\ }\href {\doibase 10.1142/S0217979206035400} {\bibfield
  {journal} {\bibinfo  {journal} {International Journal of Modern Physics B}\
  }\textbf {\bibinfo {volume} {20}},\ \bibinfo {pages} {3113} (\bibinfo {year}
  {2006})}\BibitemShut {NoStop}%
\bibitem [{\citenamefont {{Chavanis}}(2002{\natexlab{a}})}]{Chavanis_2002}%
  \BibitemOpen
  \bibfield  {author} {\bibinfo {author} {\bibfnamefont {P.-H.}\ \bibnamefont
  {{Chavanis}}},\ }\href {\doibase 10.1103/PhysRevE.65.056123} {\bibfield
  {journal} {\bibinfo  {journal} {\pre}\ }\textbf {\bibinfo {volume} {65}},\
  \bibinfo {eid} {056123} (\bibinfo {year} {2002}{\natexlab{a}})},\ \Eprint
  {http://arxiv.org/abs/cond-mat/0109294} {arXiv:cond-mat/0109294
  [cond-mat.stat-mech]} \BibitemShut {NoStop}%
\bibitem [{\citenamefont {{Chavanis}}(2002{\natexlab{b}})}]{chavanis_vortices}%
  \BibitemOpen
  \bibfield  {author} {\bibinfo {author} {\bibfnamefont {P.-H.}\ \bibnamefont
  {{Chavanis}}},\ }in\ \href {\doibase 10.48550/arXiv.cond-mat/0212223} {\emph
  {\bibinfo {booktitle} {Dynamics and Thermodynamics of Systems with Long-Range
  Interactions}}},\ Vol.\ \bibinfo {volume} {602},\ \bibinfo {editor} {edited
  by\ \bibinfo {editor} {\bibfnamefont {T.}~\bibnamefont {{Dauxois}}}, \bibinfo
  {editor} {\bibfnamefont {S.}~\bibnamefont {{Ruffo}}}, \bibinfo {editor}
  {\bibfnamefont {E.}~\bibnamefont {{Arimondo}}}, \ and\ \bibinfo {editor}
  {\bibfnamefont {M.}~\bibnamefont {{Wilkens}}}}\ (\bibinfo {year} {2002})\
  pp.\ \bibinfo {pages} {208--289}\BibitemShut {NoStop}%
\end{thebibliography}%
\end{document}